\documentclass[a4paper,single column,14pt]{article}
\usepackage[colorlinks=true,linktocpage=true,linkcolor=blue,citecolor=blue]{hyperref}
\usepackage{epsfig,amsmath,amssymb,mathtools}
\usepackage[T1]{fontenc}
\usepackage{bm}% bold math
\usepackage{slashed}
\usepackage{cancel}
\usepackage{soul}
\usepackage{ulem}
\usepackage{float}
\usepackage{color}
\usepackage[mathscr]{eucal}

\newcommand{\G}{\mathcal{G}}
\newcommand{\Gc}{\text{\sout{$\mathcal{G}$}}}
\newcommand{\U}{\mathbb{U}}
\newcommand{\Uc}{\text{\sout{$\mathbb{U}$}}}
\newcommand{\B}{\mathbb{B}}
\newcommand{\Bc}{\text{\sout{$\mathbb{B}$}}}

\newcommand{\cll}{\text{\sout{$<$}}}
\newcommand{\crr}{\text{\sout{$>$}}}
\newcommand{\h}{\hspace{1.6mm}}
\usepackage{caption}

\begin{document}
	%\Large		
	%\maketitle
%\flushbottom
\begin{flushright}
	{\normalsize
		%June 21 2008}
}
\end{flushright}
\vskip 0.1in
\begin{center}
{\Large {\bf  Generalized relativistic second order magnetohydrodynamics: \\ A correlation function approach using Zubarev's nonequilibrium statistical operator }}
\end{center}
\vskip 0.1in
\begin{center}
Abhishek Tiwari$^\dag$\footnote{abhishek\_t@ph.iitr.ac.in}and Binoy Krishna Patra$^\dag$\footnote{binoy@ph.iitr.ac.in} 
\vskip 0.02in
{\small {\it $^\dag$ Department of Physics, Indian Institute of
		Technology Roorkee, Roorkee 247 667, India\\
} }
\end{center}
\vskip 0.01in
\addtolength{\baselineskip}{0.4\baselineskip} %wide line spacing
%opening

\section* {Abstract}
		
			We use total energy-momentum conservation and the Bianchi identity (magnetic-flux conservation) to construct second-order relativistic magnetohydrodynamics in a Zubarev's non-equilibrium statistical operator (NESO) framework. We obtain all dissipative tensors in the medium by focusing on a relativistic magnetized plasma that preserves parity and is symmetric to charge-conjugation. We also provide Kubo formulas for all transport coefficients that arise at second order. Moreover, we extend the NESO formalism to systematically take into account for nonlocal contributions.

	%
	%%%%%%%%%%%%%%%%%%%%%%
	
	%\maketitle
	
	%%%%%%%%%%%%%%%%%%%%%%%%%%%%%%%%%%%%%%%%%%%%%%%%%%%%%%

	\section{Introduction}\label{introdunction}

	Relativistic hydrodynamics has demonstrated exceptional effectiveness as a macroscopic framework for describing strongly interacting matter in nuclear~\cite{Gyulassy:2004zy}, astrophysical~\cite{Baiotti:2016qnr,Marti}, and high-energy contexts~\cite{Heinz:2013th}. In the realm of relativistic heavy-ion collisions at RHIC and the LHC, hydrodynamic models have successfully replicated the collective dynamics of the quark-gluon plasma (QGP)~\cite{Gale:2013da}, capturing features such as elliptic and higher-order flow harmonics, event-by-event fluctuations~\cite{Niemi}, and long-range correlations. These successes have solidified the QGP's status as an almost perfect fluid with remarkably low shear viscosity, showcasing the predictive capabilities of relativistic hydrodynamics in capturing the long-wavelength behavior of strongly coupled relativistic systems~\cite{Shuryak:2003xe,Gyulassy:2004zy,Romatschke:2009kr,Heinz:2013th,Jaiswal:2016hex,Hayata:2015lga}. Moreover, relativistic hydrodynamics has also found important applications in condensed-matter physics, especially in systems where quasiparticles display linear, Dirac-like dispersion relations and interact strongly enough to form a nearly ideal fluid~\cite{Gallagher,Narozhny:2022ncn,Gabbana:2019ydb,Jaiswal:2024urq}.

	Nevertheless, this otherwise persuasive scenario becomes considerably more complex when dynamic electromagnetic fields are introduced~\cite{Kharzeev:2024zzm,Tuchin:2014iua}. Their presence alters the conventional hydrodynamic equations, necessitating a broader framework known as relativistic magnetohydrodynamics (RMHD)~\cite{Hernandez:2017mch}. RMHD offers a macroscopic model for magnetized relativistic plasmas and forms the basis for cutting-edge simulations of neutron stars, binary mergers, and highly magnetized astrophysical environments~\cite{Anile,Duez:2005sf}. Over the past few decades, relativistic magnetohydrodynamics has seen significant theoretical advancements, including systematic gradient expansions, causal and stable constitutive relations, and fully covariant formulations of anisotropic transport in intense electromagnetic fields~\cite{Huang:2011dc,Shokri:2018qcu,Siddique:2019gqh,Denicol:2018rbw,Denicol:2019iyh,Molnar:2024fgy,Panda:2020zhr,Panda:2021pvq,Kushwah:2024zgd,Fang:2024skm,Singh:2024leo,Dash:2022xkz,Armas:2022wvb,Hoult:2024qph}.

	The standard derivation of RMHD starts with two fundamental equations: (i) the conservation of fluid energy-momentum influenced by electromagnetic fields and (ii) Maxwell's equations with sources~\cite{Huang:2011dc,Denicol:2019iyh,Panda:2020zhr,Panda:2021pvq}. This method involves breaking down the total energy-momentum tensor into distinct fluid and electromagnetic components, a nuanced process due to the need for consistent specification of interaction terms. Additionally, these equations do not appear solely in a conservative form and inherently include non-hydrodynamic (gapped) modes that diminish over time. This challenge stems from considering the electric field as a hydrodynamic variable, despite electric fields not being hydrodynamic degrees of freedom due to Debye screening. The typical solution is to assume infinite electrical conductivity, thus excluding the electric field from the hydrodynamic sector. While this approach is adequate for many ideal and resistive MHD applications, it creates a conceptual conflict: a framework based on the assumption of infinite conductivity cannot consistently incorporate dissipative corrections.

	Recent research has developed a formulation that relies exclusively on conserved densities and their fluxes by considering the magnetic field as a hydrodynamic variable. Unlike electric fields, magnetic fields are not subject to static screening. The Bianchi identity, which reflects the absence of magnetic monopoles, prevents the existence of a magnetic equivalent to Debye screening. This has led to a new approach to RMHD that is grounded in the conservation of total energy and momentum, along with the Bianchi identity (conservation of magnetic flux) as the core dynamic equations~\cite{Grozdanov:2016tdf,Armas:2018zbe,Hongo:2020qpv,Hattori:2022hyo}. A limited number of studies have initially examined RMHD using this framework at the first order in gradients~\cite{Hongo:2020qpv,Hattori:2022hyo}. However, it is well recognized that first-order relativistic dissipative theories are generally acausal; these challenges become even more complex when electromagnetic fields are involved. Therefore, it is crucial to develop a second-order RMHD theory based on this new formulation.

	Starting from this new foundational basis, where total energy-momentum conservation together with the Bianchi identity (magnetic-flux conservation) serve as the fundamental dynamical equations, we extend RMHD to second order in the gradient expansion using Zubarev’s nonequilibrium statistical operator (NESO) formalism. The NESO provides a microscopic basis for relativistic hydrodynamics by constructing a density operator that satisfies local constraints, such as the energy-momentum tensor and conserved charges, and by solving the Liouville equation with a retarded prescription~\cite{Zubarev1979derivation,Van1982maximum}. Through this framework, one obtains hydrodynamic constitutive relations that naturally incorporate quantum and statistical effects. The NESO formalism generalizes the Gibbs ensemble in a manner that preserves the local validity of the Gibbs relation. Additionally, the causal “switch-on’’ procedure for the irreversibility parameter selects the retarded response, fixes the arrow of time by incorporating irreversibility, and ensures consistency with the second law of thermodynamics. Moreover, linearization of the NESO leads to transport coefficients expressed in terms of equilibrium correlation functions. Systematic higher-order corrections follow from expanding the NESO beyond first order, generating relaxation times and nonlinear couplings, Israel-Stewart-type terms, encoded in memory kernels and multi-point correlators~\cite{HARUTYUNYAN2022168755}.

	This paper is organized as follows. In Sec.~II, we present the equations of motion for the energy-momentum tensor and for the dual electromagnetic field-strength tensor. The entropy current analysis for the first order is discussed in Sec.~III. Section~IV introduces the NESO formalism and derives the nonequilibrium contribution using the tensor decompositions of the energy-momentum tensor and the dual field-strength tensor. In Sec.~V, we revisit first-order RMHD and express the dissipative tensors in terms of their corresponding thermodynamic forces. Section~VI provides the complete set of second-order constitutive relations together with the evolution equations for all dissipative quantities. Finally, Sec.~VII summarizes our findings and offers concluding remarks.

	Throughout this paper, we employ the metric tensor 
	$\eta_{\mu\nu}={\rm diag}(1,-1,-1,-1)$. The projection operator transverse to the fluid velocity $u_\mu$ is 
	defined as $\Delta_{\mu\nu}=\eta_{\mu\nu}-u_\mu u_\nu$, with the 
	normalization condition for the fluid velocity given by $u_\mu 
	u^\mu=1$. Whereas, the projection operator transverse to fluid velocity and magnetic vector $b_\mu$ is defined as $\G_{\mu\nu}=\eta_{\mu\nu}-u_\mu u_\nu + b_\mu b_\nu$, with the condition $b_\mu b^\mu=-1$ and $b_\mu u^\mu=0$. In this paper, we work in a general frame and used the following projectors:
	\begin{align}
		\G^{\alpha\beta}_{\h\h \mu\nu} &= \frac{1}{2}\Big(\G^{\alpha}_{\h \mu} \G^{\beta}_{\h \nu}
		+ \G^{\alpha}_{\h \nu} \G^{\beta}_{\h \mu}
		- \G^{\alpha\beta} \G_{\mu\nu}\Big), \label{Gprojectos}\\
		\Gc^{\mu\nu}_{\h\h \alpha\beta} &= \frac{1}{2}\Big(\G^{\alpha}_{\h \mu} \G^{\beta}_{\h \nu}
		- \G^{\alpha}_{\h \nu} \G^{\beta}_{\h \mu}
		\Big),\label{Gcprojectos}\\
		\U^{\alpha\beta}_{\h\h \mu} &=  \frac{1}{2}\Big(u^{\alpha}\mathcal{G}^{\beta}_{\h \mu}+u^{\beta}\mathcal{G}^{\alpha}_{\h \mu}\Big),\label{noncutprojectorU}\\
		\B^{\alpha\beta}_{\h\h \mu} &=  \frac{1}{2}\Big(b^{\alpha}\mathcal{G}^{\beta}_{\h \mu}+b^{\beta}\mathcal{G}^{\alpha}_{\h \mu}\Big)\label{noncutprojectorB}\\
		\Uc^{\alpha\beta}_{\h\h \mu} &=  \frac{1}{2}\Big(u^{\alpha}\mathcal{G}^{\beta}_{\h \mu}-u^{\beta}\mathcal{G}^{\alpha}_{\h \mu}\Big),\label{cutprojectorU}\\
		\Bc^{\alpha\beta}_{\h\h \mu} &= \frac{1}{2}\Big(b^{\alpha}\mathcal{G}^{\beta}_{\h \mu}-b^{\beta}\mathcal{G}^{\alpha}_{\h \mu}\Big). \label{cutprojectorB}
	\end{align}
In addition, other shorthand notations used in this paper are given by,
	\begin{align}
		A^{<\mu\nu>} &= \G_{\h \h \alpha\beta}^{\mu\nu} A^{\alpha\beta}, ~~ A^{\cll\mu\nu\crr} = \Gc_{\h \h \alpha\beta}^{\mu\nu} A^{\alpha\beta},\label{shorthand1}\\
		A^{(\mu\nu)} &= \frac{1}{2}(A^{\mu\nu}+A^{\nu\mu}),~~A^{[\mu\nu]} = \frac{1}{2}(A^{\mu\nu}-A^{\nu\mu}).\label{shorthand2}
	\end{align}

\section{Equations of motion}
	
	To derive the framework of relativistic magnetohydrodynamics, we begin by identifying the fundamental conserved quantities of the system.
	For our case, the total energy-momentum of the system and the magnetic flux are the conserve quantities for the system of interest and their conservation equations are given by 
	\begin{align}
		\partial_\mu T^{\mu\nu}&=0, \label{con_T}\\
		\partial_\mu \tilde{F}^{\mu\nu}&=0 .\label{con_F}
	\end{align}
	Here, $T^{\mu\nu}$ is the total energy momentum of the system that includes matter part as well as the contribution from the EM field. Whereas $\tilde{F}^{\mu\nu}$ is the dual of the electromagnetic field strength tensor $F^{\mu\nu}$.
	
	The next step is to perform a tensorial decomposition of the energy-momentum tensor $T^{\mu\nu}$ and the dual electromagnetic field tensor $\tilde{F}^{\mu\nu}$ along fluid four-velocity $u^\mu$, magnetic field vector $b^\mu$ and $\G^{\mu\nu}$. The most general tensor structure of $T^{\mu\nu}$ and $\tilde{F}^{\mu\nu}$ for a system invariant under the symmetries of charge conjugation and parity, is given by~\cite{Hattori:2022hyo}
	\begin{align}
		T^{\mu\nu} &= \overbracket[0.7pt][0.25cm]{\varepsilon\,u^{\mu}u^{\nu}
		+ p_{\parallel}\, b^{\mu}b^{\nu}
		- p_{\perp}\, \G^{\mu\nu} }^{\text{}~\mathcal{O}(\partial^0)}
		+ \, \Pi_{\parallel}\, b^{\mu}b^{\nu}-\Pi_{\perp}\, \G^{\mu\nu}
		+ h^{\mu}u^{\nu} + h^{\nu}u^{\mu}
		+ f^{\mu}b^{\nu} + f^{\nu}b^{\mu}
		+ \pi^{\mu\nu}_{\perp} , \label{T_decom} \\
		\tilde F^{\mu\nu}
		&=  \underbracket[0.7pt][0.25cm]{ B\,(b^{\mu}u^{\nu}-b^{\nu}u^{\mu})}_{\text{}~\mathcal{O}(\partial^0)} 
			 \;-\; (l^{\mu}b^{\nu}-l^{\nu}b^{\mu})
		\;+\; g^{\mu}u^{\nu} - g^{\nu}u^{\mu} \;+\; m^{\mu\nu} . \label{F_decom}
	\end{align}
	Here, $\varepsilon$ is total energy density, $B$ is the magnetic flux density, $p_\parallel$ and $p_\perp$ are the total pressure densities in the direction parallel and perpendicular to magnetic field, respectively. In general, the tensors $\Pi_\parallel$, $\Pi_\perp$, $h^\mu$, $f^\mu$, $l^\mu$, $g^\mu$, $\pi_\perp^{\mu\nu}$, and $m^{\mu\nu}$ contain the dissipative corrections up to all orders. They satisfies the conditions: $u_\mu h^\mu=0$, $b_\mu h^\mu=0$, $u_\mu f^\mu=0$, $b_\mu f^\mu=0$, $\pi_\perp^{\mu\nu}=\pi_\perp^{\nu\mu}$, $\pi^\mu_{\perp\mu}=0$, $u_\mu \pi_\perp^{\mu\nu}=0$, $b_\mu \pi_\perp^{\mu\nu}=0$, $u_\mu l^\mu=0$, $b_\mu l^\mu=0$, $u_\mu g^\mu=0$, $b_\mu g^\mu=0$, $m^{\mu\nu}=-m^{\nu\mu}$, $u_\mu m^{\mu\nu}=0$, and $b_\mu m^{\mu\nu}=0$. In this formalism, it is important to emphasize that the magnetic field is treated as an $\mathcal{O}(\partial^0)$ quantity, whereas the electric field appears only at higher order in the gradient expansion. Consequently, the electric field does not enter the ideal part of $\tilde{F}^{\mu\nu}$. Instead, it emerges as a higher-order effect generated by the magnetic field through the dissipative contributions to $\tilde{F}^{\mu\nu}$, which give rise to the induced electric field.

		Since the objective of this work is to derive the second-order relativistic magnetohydrodynamics, it is sufficient to determine the equations of motion up to second order in the hydrodynamic gradient expansion. To achieve this, we introduce the first-order corrections $T_{1}^{\mu\nu}$ and $\tilde{F}_1^{\mu\nu}$, which respectively contain all terms of order $\mathcal{O}(\partial^1)$ in the expansions of $T^{\mu\nu}$ and $\tilde{F}^{\mu\nu}$. Their expressions are given by
	\begin{align}
		T_1^{\mu\nu} &= \Pi_{1\parallel }\, b^{\mu}b^{\nu}-\Pi_{1\perp}\, \G^{\mu\nu}
		+ h_1^{\mu}u^{\nu} + h_1^{\nu}u^{\mu}
		+ f_1^{\mu}b^{\nu} + f_1^{\nu}b^{\mu}
		+ \pi^{\mu\nu}_{1\perp} , \label{T_1}\\
		\tilde{F}_1^{\mu\nu} &=  -\; (l_1^{\mu}b^{\nu}-l_1^{\nu}b^{\mu})
		\;+\; g_1^{\mu}u^{\nu} - g_1^{\nu}u^{\mu} \;+\; m_1^{\mu\nu} . \label{F_1}
	\end{align}
	Before proceeding to derive the equations of motion, we first write down the thermodynamic relations in the presence of the magnetic flux $B^\mu$~\cite{Huang:2011dc,Hattori:2022hyo}:
	\begin{align}
		ds &= \beta\, d\varepsilon - \mathcal{H}_\alpha\, dB^\alpha = \beta\, d\varepsilon - \beta {H}_\alpha\, dB^\alpha , \label{ds_thermo}\\
		s &= \beta\,\varepsilon + \beta\, p_{\perp} - B \mathcal{H} , \label{s_thermo}\\
		\beta d p_{\perp} &= -(\varepsilon+p_{\perp})\, d\beta + B^\alpha d\mathcal{H}_\alpha . \label{dP_thermo}
	\end{align}
	Here, $s$ is the entropy density, $\beta$ denotes the inverse temperature, 
$B_\mu = B\, b_\mu$ represents the magnetic flux vector, 
$H_\mu = -H\, b_\mu$ corresponds to the magnetic field vector, 
and $\mathcal{H}_\mu = \beta H_\mu$ defines the reduced magnetic field vector, 
which later appears as the Lagrange multiplier associated with the dual field strength tensor in the density operator. 
In addition, the longitudinal ($p_\parallel$) and transverse ($p_\perp$) pressures are related through~\cite{Hattori:2022hyo}
\begin{align}
	p_{\parallel}=p_{\perp}-B H	 . \label{p_relation}
\end{align}
The thermodynamic relations~\eqref{ds_thermo},~\eqref{s_thermo}, and~\eqref{dP_thermo} connect quantities such as energy density, pressure, entropy, temperature, and magnetic field density through the equation of state and Gibbs-Duhem relations. These equations are needed to obtain the entropy current.

	To obtain the equations of motion, we use the conservation equations and calculate their projection along $u_\nu$, $b_\nu$, and $\G^{\alpha}_{\nu}$. This decomposition separates the dynamics into components aligned with and orthogonal to the magnetic field. Thus, by projecting $u_\nu$, $b_\nu$, and $\G^{\alpha}_{\h\nu}$ onto Eq.~\eqref{con_T}, and employing the tensor structure of $T^{\mu\nu}$ up to first-order in gradients, we obtain
	\begin{align}
		u_{\nu}\partial_{\mu}T^{\mu\nu}&= u_{\nu}\partial_{\mu}T_0^{\mu\nu}+u_{\nu}\partial_{\mu}T_1^{\mu\nu}=0 ,\nonumber\\
		\Rightarrow\;	D\varepsilon
		&= -\varepsilon\,\theta 
		- p_{\parallel}\,\theta_{\parallel} - p_{\perp}\, \theta_\perp - u_{\nu}\partial_{\mu}T_1^{\mu\nu} , \label{udelT} \\
		% - (p_{\parallel}+\Pi_{\parallel})\, b^{\mu}b^{\nu}\,\partial_{\mu}u_{\nu} + (p_{\perp}+\Pi_{\perp})\, \G^{\mu\nu}(\partial_{\mu}u_{\nu})
		%+ \partial_{\mu}h^{\mu} - h^{\nu}Du_{\nu}
		%- f^{\mu}b^{\nu}\partial_{\mu}u_{\nu} - f^{\nu}b^{\mu}\partial_{\mu}u_{\nu} \nonumber\\
		% &\quad
		%- \pi^{\mu\nu}_{\perp}(\partial_{\mu}u_{\nu}) 
		b_{\nu}\,\partial_{\mu}T^{\mu\nu}
		&= b_{\nu}\,\partial_{\mu}T_{0}^{\mu\nu}
		+ b_{\nu}\,\partial_{\mu}T_1^{\mu\nu} =0 ,\nonumber\\[0.25em]
		\Rightarrow\;
		(p_{\perp}+\varepsilon)\, b_{\nu} D u^{\nu}
		&= b^{\mu}\partial_{\mu}p_{\parallel}
		- B H\,\partial_{\mu}b^{\mu}
		- b_{\nu}\,\partial_{\mu}T_1^{\mu\nu} ,\label{bdelT}\\
		\G^{\alpha}_{\h\nu}\,\partial_{\mu}T^{\mu\nu}
		&= \G^{\alpha}_{\h \nu}\,\partial_{\mu}T_{0}^{\mu\nu}
		+ \G^{\alpha}_{\h \nu}\,\partial_{\mu}T_1^{\mu\nu} =0 ,\nonumber\\[0.25em]
		\Rightarrow\;
		(\varepsilon+p_{\perp})\, D u^{\alpha}
		&= \G^{\alpha\mu} \partial_\mu p_{\perp}
		+ B H\, b^{\mu}\partial_{\mu}b^{\alpha} \nonumber\\
		&- (\varepsilon+p_{\perp})\, b^{\alpha} b_{\nu} D u^{\nu} - B H\, u^{\alpha}\theta_{\parallel}
		- \G^{\alpha}_{\h \nu}\,\partial_{\mu}T_1^{\mu\nu}  .\label{GdelT}
	\end{align}
	Here, we have defined $\theta \equiv \partial_\mu u^\mu$, $\theta_\parallel \equiv-b^\mu b^\nu \partial_\mu u_\nu$, $\theta_\perp \equiv \mathcal{G}^{\mu\nu}\partial_\mu u_\nu$, and the comoving derivative $D\equiv u^\mu\partial_\mu$. Similarly, by projecting $u_\nu$, $b_\nu$, and $\G^{\alpha}_{\h\nu}$ onto Eq.~\eqref{con_F}, we obtain
	\begin{align}
		u_{\nu}\,\partial_{\mu}\tilde F^{\mu\nu}
		&= u_{\nu}\,\partial_{\mu}\tilde F_{0}^{\mu\nu}
		+ u_{\nu}\,\partial_{\mu}\tilde F_{1}^{\mu\nu}=0 , \nonumber\\[0.25em]
		\Rightarrow\;
		b_{\nu} D u^{\nu}
		&= -\frac{1}{B}\,\Big[\,\partial_{\mu}\!\left(B b^{\mu}\right)
		+ u_{\nu}\,\partial_{\mu}\tilde F_{1}^{\mu\nu}\,\Big] ,\label{udelF}\\
		b_{\nu}\,\partial_{\mu}\tilde F^{\mu\nu}
		&= b_{\nu}\,\partial_{\mu}\tilde F_{0}^{\mu\nu}
		+ b_{\nu}\,\partial_{\mu}\tilde F_{1}^{\mu\nu}=0 ,\nonumber\\
	\Rightarrow\; DB
		&= - B\theta +B\, \theta_{\parallel} 
		- b_{\nu}\,\partial_{\mu}\tilde F_{1}^{\mu\nu} \,  ,\label{bdelF}\\
		\G^{\alpha}_{\h \nu}\,\partial_{\mu}\tilde F^{\mu\nu}
		&= \G^{\alpha}_{\h \nu}\,\partial_{\mu}\tilde F_{0}^{\mu\nu}
		+ \G^{\alpha}_{\h \nu}\,\partial_{\mu}\tilde F_{1}^{\mu\nu} =0 , \nonumber\\[0.25em]
		\Rightarrow\;
		B\,D b^{\alpha}
		&= B\,u^{\alpha}u_{\nu} D b^{\nu}
		+ B\,b^{\mu}\,\partial_{\mu}u^{\alpha}
		- B\,b^{\alpha}\theta_{\parallel}
		+ \G^{\alpha}_{\h \nu}\,\partial_{\mu}\tilde F_{1}^{\mu\nu} .\label{GdelF}
	\end{align}
	By combining Eqs.~\eqref{bdelT} and~\eqref{udelF}, one can evaluate the four-divergence of the magnetic flux $B^\mu$, which in turn facilitates the simplification of the comoving derivatives of $u^\mu$ and $b^\mu$. To further simplify Eqs.~\eqref{bdelT} and~\eqref{udelF}, we make use of the thermodynamic relation. Starting from Eq.~\eqref{bdelT}, we first use the relation~\eqref{p_relation} to express Eq.~\eqref{bdelT} in terms of $p_\perp$. 
Subsequently, by applying the thermodynamic relation~\eqref{dP_thermo}, Eq.~\eqref{bdelT} can be rewritten as
	\begin{align}
		(p_{\perp}+\varepsilon)\, b_{\nu} D u^{\nu} 
		&= \beta^{-1}\Big[-(\varepsilon+p_{\perp})+BH\Big]b^{\mu} \partial_\mu\beta - H\,\partial_{\mu}B^{\mu}
		- b_{\nu}\partial_{\mu}T_1^{\mu\nu} \, .
	\end{align}
Further, we substitute the value of $b_\nu D u^\nu$ from~\eqref{udelF}, and obtain the four divergence of magnetic flux vector, given by	
	\begin{align}
		-\frac{1}{B}\,(\,p_{\perp}+\varepsilon\,)\,
		\Big[\,\partial_{\mu}B^{\mu}+u_{\nu}\,\partial_{\mu}\tilde F_{1}^{\mu\nu}\,\Big]
		&= \Big[-(\varepsilon+p_{\perp})+BH\Big]\,
		\beta^{-1}\, b^{\mu}\partial_{\mu}\beta
		- H\,\partial_{\mu}B^{\mu}
		- b_{\nu}\,\partial_{\mu}T_{1}^{\mu\nu} ,\nonumber\\[0.35em]
		\Rightarrow\quad
		\partial_{\mu}B^{\mu}
		&= \beta^{-1}\,B^\mu\,\partial_{\mu}\beta
		+ \frac{1}{-(p_{\perp}+\varepsilon)+BH}\,
		\Big[ -\,B_{\nu}\,\partial_{\mu}T_{1}^{\mu\nu}
		+ (p_{\perp}+\varepsilon)\,u_{\nu}\partial_{\mu}\tilde F_{1}^{\mu\nu} \Big] . \label{delB}
	\end{align}
Next, we substitute the value of $(p_{\perp}+\varepsilon)\, b_{\nu} D u^{\nu}$ from~\eqref{bdelT} in~\eqref{GdelT} and make use of~\eqref{delB}. We obtain
\begin{align}
		D u^{\alpha}
		= -\beta^{-1}\nabla^{\alpha}\beta
		- \frac{1}{(\varepsilon+p_{\perp})}
		\Big[\,2\beta^{-1} B_{\nu}\,\nabla^{[\nu}\mathcal{H}^{\alpha]}
		+ u^{\alpha} B H\,\theta_{\parallel}\,\Big]
		+ \frac{H}{K}\, b^{\alpha} u_{\nu}\partial_{\mu}\tilde F_{1}^{\mu\nu}
		- \frac{1}{(\varepsilon+p_{\perp})}
		\Big[\frac{\varepsilon+p_{\perp}}{K}\,b^{\alpha} b_{\nu}
		+ \G^{\alpha}_{\h \nu}\Big]\partial_{\mu}T_{1}^{\mu\nu} .\label{Du}
\end{align}
Here, we define $K = BH - (p_{\perp}+\varepsilon)$. Similarly, we substitute the value of $u_{\nu} D b^{\nu}$ from~\eqref{udelF} in~\eqref{GdelF} and make use of~\eqref{delB}. We obtain
\begin{align}
	D b^{\alpha}
	= \beta^{-1} u^{\alpha}  b^\mu\,\partial_{\mu}\beta
	+  b^{\mu}\partial_{\mu}u^{\alpha}
	-  b^{\alpha}\theta_{\parallel}
	- \frac{u^{\alpha}  b_{\nu}}{K}\,\partial_{\mu}T_{1}^{\mu\nu}
	+ \frac{1}{B}\Big[\frac{B H\,u^{\alpha}u_{\nu}}{K}+\G^{\alpha}_{\h \nu}\Big]\partial_{\mu}\tilde F_{1}^{\mu\nu}  . \label{Db}
\end{align}

	\section{Entropy current analysis}

	With the help of the thermodynamic relations~\eqref{ds_thermo},~\eqref{s_thermo}, and~\eqref{dP_thermo}, we can express the entropy current $S^\mu$. 
	Alternatively, the entropy current can be derived from first principle using the quantum statistical density operator. 
	The form of entropy current in the presence of an external magnetic field, up to first order in gradients, is given by
	\begin{align}
		S^{\mu} &= s u^{\mu} + \beta_{\nu}\,T_1^{\mu\nu} + \mathcal{H}_{\nu}\,\tilde F_1^{\mu\nu} . \label{Smu}
    \end{align}	
    According to the second law of thermodynamics, the divergence of the entropy current i.e., the entropy production rate in local rest frame, must satisfy
    \begin{align}
    	\partial_\mu S^\mu \ge 0 , \label{delSge}
    \end{align}
    for any dissipative evolution. Using~\eqref{Smu}, the divergence of entropy current can be written as
    \begin{align}
    	\partial_{\mu}S^{\mu}
    	&= T_{1}^{\mu\nu}(\partial_{\mu}\beta_{\nu})
    	+ \tilde F_{1}^{\mu\nu}(\partial_{\mu}\mathcal{H}_{\nu}) + \Big[\partial_{\mu}(s u^{\mu})
    	- (\partial_{\mu}\tilde F_{0}^{\mu\nu})\,\mathcal{H}_{\nu}
    	- (\partial_{\mu}T_{0}^{\mu\nu})\,\beta_{\nu}\Big]. \label{delS1}
    \end{align}
     Here, $T_{0}^{\mu\nu}$ and $\tilde{F}_{0}^{\mu\nu}$ denote the zeroth-order part of $T^{\mu\nu}$ and $\tilde{F}^{\mu\nu}$, respectively. 
    In simplifying Eq.~\eqref{delS1}, we have used the conservation laws \eqref{con_T} and \eqref{con_F}. 
    Employing the thermodynamic identities \eqref{ds_thermo}, \eqref{s_thermo}, and \eqref{dP_thermo}, together with Eqs.~\eqref{udelT} and \eqref{bdelF}, one finds that the terms inside the square brackets in Eq.~\eqref{delS1} vanish. 
    Therefore,
    \begin{equation}
    	\partial_{\mu}S^{\mu} = T_{1}^{\mu\nu}(\partial_{\mu}\beta_{\nu})
    	+ \tilde F_{1}^{\mu\nu}(\partial_{\mu}\mathcal{H}_{\nu}) . \label{delS2}
    \end{equation}
    Employing~\eqref{T_1} and~\eqref{F_1}, we obtain
    \begin{align}
    	\partial_{\mu}S^{\mu}
    	&= -\beta\,\Pi_{1\parallel}\,\theta_{\parallel}
    	- \beta\,\Pi_{1\perp}\,\theta_{\perp}
    	+ \pi_{1\perp}^{\mu\nu}\,\G^{\alpha\beta}_{\h\h \mu\nu} \partial_\alpha \beta_\beta 
    	+ m_1^{\mu \nu} \Gc^{\alpha\beta}_{\h\h \mu\nu} \partial_\alpha \mathcal{H}_\beta \nonumber\\
    	&\quad + 2 h_1^\mu \, \U^{\alpha\beta}_{\h\h \mu} \partial_\alpha \beta_\beta + 2 f_1^\mu \, \B^{\alpha\beta}_{\h\h \mu} \partial_\alpha \beta_\beta - 2 g_1^\mu \, \Uc^{\alpha\beta}_{\h\h \mu} \partial_\alpha \mathcal{H}_\beta + 2 l_1^\mu \, \Bc^{\alpha\beta}_{\h\h \mu} \partial_\alpha \mathcal{H}_\beta , \label{delS3}
    \end{align}
    where, we used the projectors defined in~\eqref{Gprojectos}-\eqref{cutprojectorB}
    and employ the relations~\eqref{proonh}-\eqref{proonm} from Appendix~\ref{cal_C}.

  We now examine the parity properties of the dissipative tensors appearing in Eq.~\eqref{delS3}. 
  Since both $\partial_\mu$ and $\beta_\mu (u_\mu)$ transform as proper vectors and acquire a minus sign under a parity transformation, the combination $\partial_\mu \beta_\nu$ is parity even. In contrast, the structure $\partial_\mu \mathcal{H}_\nu$ is parity odd due to the presence of the pseudovector $b^\mu$, which is itself parity even. Using Eqs.~\eqref{Gprojectos}-\eqref{cutprojectorB}, we therefore deduce that the thermodynamic forces $\theta_\parallel$, $\theta_\perp$, $\B^{\alpha\beta}_{\h\h \mu} \partial_\alpha \beta_\beta$, $\Uc^{\alpha\beta}_{\h\h \mu} \partial_\alpha \mathcal{H}_\beta$, and $\G^{\alpha\beta}_{\h\h \mu\nu} \partial_\alpha \beta_\beta$ are parity even, whereas $\Gc^{\alpha\beta}_{\h\h \mu\nu} \partial_\alpha \mathcal{H}_\beta$, $\U^{\alpha\beta}_{\h\h \mu} \partial_\alpha \beta_\beta$, and $ \Bc^{\alpha\beta}_{\h\h \mu} \partial_\alpha \mathcal{H}_\beta$ are parity odd. Because $\partial_\mu S^\mu$ is a proper scalar and therefore parity is even, hence every term contributing to Eq.~\eqref{delS3} must also be of even-parity for consistency. Consequently, the dissipative tensors that couple to even-parity thermodynamic forces must themselves be even parity, whereas those coupled to odd-parity thermodynamic forces must have odd parity. It follows that $\Pi_\parallel$, $\Pi_\perp$, $f^\mu$, $g^\mu$, and $\pi_\perp^{\mu\nu}$ are parity even, while $m^{\mu\nu}$, $h^\mu$, and $l^\mu$ are parity odd. These parity assignments will play an important role in determining the allowed correlation functions in the subsequent sections.

 With these parity assignments established, the next step in refining Eq.~\eqref{delS3} involves simplifying the terms proportional to $h_1^\mu$ and $g_1^\mu$. Using the relations~\eqref{UtoBcrelation} and~\eqref{UctoBrelation} from Appendix~\ref{important_relation}, we can write
\begin{align}
	2 h_1^\mu \, \U^{\alpha\beta}_{\h\h \mu} \partial_\alpha \beta_\beta &=\frac{-2B}{\varepsilon+p_\perp}\,h_1^\mu \, \Bc^{\alpha\beta}_{\h\h \mu} \partial_\alpha \mathcal{H}_\beta , \label{eq5}\\
	2 g_1^\mu \, \Uc^{\alpha\beta}_{\h\h \mu} \partial_\alpha \mathcal{H}_\beta &= -2 H g_1^\mu \, \B^{\alpha\beta}_{\h\h \mu} \partial_{\alpha} \beta_\beta  . \label{eq6}
\end{align}
Using~\eqref{eq5} and~\eqref{eq6}, we can rewrite~\eqref{delS3} as
\begin{align}
	\partial_{\mu}S^{\mu}
	&= -\,\Pi_{1\parallel}\,\bar{\theta}_{\parallel}
	- \,\Pi_{1\perp}\,\bar{\theta}_{\perp}
	+ \pi_{1\perp}^{\mu\nu}\,\bar{\sigma}_{\perp \mu\nu} 
	+ m_1^{\mu \nu} \bar{\chi}_{ \mu\nu}   +  \, \mathscr{K}_1^\mu \, \bar{\mathscr{Y}}_\mu  +  \mathscr{J}_1^\mu  \, \bar{\mathscr{X}}_\mu . \label{delSfinal}
\end{align}
Here, $ \mathscr{K}_1^\mu$ and $ \mathscr{J}_1^\mu$ are the first order correction to the currents $\mathscr{K}^\mu$ and $\mathscr{J}^\mu$, respectively and they are given by
\begin{align}
	\mathscr{K}^\mu  &= f^\mu+ H g^\mu, \\
		\mathscr{J}^\mu &= l^\mu -\frac{B}{\varepsilon+p_\perp} h^\mu,
\end{align}
respectively. Also, we use $\bar{\theta}_\parallel=\beta \theta_\parallel$, $\bar{\theta}_\perp=\beta \theta_\perp$, and define the thermodynamic forces:  $ \bar{\sigma}_{\perp \mu\nu}\equiv\G^{\alpha\beta}_{\h\h \mu\nu}\,\partial_{\alpha}\beta_{\beta}$, $\bar{\chi}_{ \mu\nu}\equiv\Gc^{\alpha\beta}_{\h\h \mu\nu}\,\partial_{\alpha} \mathcal{H}_{\beta}$, $\bar{\mathscr{X}}_\mu\equiv2 \, \Bc^{\alpha\beta}_{\h\h \mu} \partial_\alpha \mathcal{H}_\beta$, and $\bar{\mathscr{Y}}_\mu\equiv2 \,\B^{\alpha\beta}_{\h\h \mu} \partial_\alpha \beta_\beta$. 

To ensure the semi-positivity of Eq.~\eqref{delSfinal}, we impose linear constitutive relations that express the dissipative tensors in terms of the thermodynamic forces.
Guided by Curie’s symmetry principle that an effect cannot possess more symmetries than its cause. Accordingly, the dissipative tensors can be expressed only as linear combinations of thermodynamic forces possessing the same tensorial rank and parity. Thus, we obtain
\begin{align}
	\Pi_{1\parallel}
	&= -\bar{\zeta}_{\parallel}\bar{\theta}_{\parallel}
	-\bar{\zeta}_{\times} \bar{\theta}_{\perp},\label{pi1par}\\
		\Pi_{1\perp}
	&= -\bar{\zeta}_{\perp}\bar{\theta}_{\perp}
	-\bar{\zeta}_{\times}'\bar{\theta}_{\parallel},\label{pi1perp}\\
	\pi_{1\perp}^{\mu\nu}&=2 \bar{\eta}_\perp \bar{\sigma}_\perp^{\mu\nu},\label{pi1munu}\\
	m_{1}^{\mu\nu}&=2 \bar{\rho}_\perp \bar{\chi}^{\mu\nu},\label{m1munu}\\
	\mathscr{K}_1^\mu &= \bar{\eta}_\parallel \bar{\mathscr{Y}}^\mu,\label{k1mu}\\
	\mathscr{J}_1^\mu &= \bar{\rho}_\parallel \bar{\mathscr{X}}^\mu \label{j1mu}.
\end{align}
As we can see there is no cross term between $\mathscr{K}_1^\mu$ and $\mathscr{J}_1^\mu$ because they are of different parity. 
This construction yields the associated transport coefficients $\bar{\zeta}_{\parallel}$, $\bar{\zeta}_{\perp}$, $\bar{\eta}_\parallel$, $\bar{\eta}_\perp$, $\bar{\rho}_\parallel$, and $\bar{\rho}_\perp$, which are constrained to be nonnegative. However, for a charge conjugate symmetric plasma $\bar{\zeta}_\times=\bar{\zeta}'_\times$ and follow the inequality $\bar{\zeta}_\parallel \bar{\zeta}_\perp-\bar{\zeta}^2_\times\geq 0$~\cite{Hattori:2022hyo}.

	%%%%%%%%%%%%%%%%%%%%%%%%%%%%%%%%%%%%%%%%%%%%%%%%%%%%%%%%%%%%%%%%%%%%%%%%%%%%%%%%%%%%%%
	%%%%%%%%%%%%%%%%%%%%%%%%%%%%%%%%%%%%%%%%%%%%%%%%%%%%%%%%%%%%%%%%%%%%%%%%%%%%%%%%%%%%%%

	%%%%%%%%%%%%%%%%%%%%%%%%%%%%%%%%%%%%%%%%%%%%%%%%%%%%%%%%%%%%%%%%%%%%%%%%%%%%%%%%%%%%%%
	%%%%%%%%%%%%%%%%%%%%%%%%%%%%%%%%%%%%%%%%%%%%%%%%%%%%%%%%%%%%%%%%%%%%%%%%%%%%%%%%%%%%%%

	\section{Nonequilibrium statistical operator}\label{NESO}
	%%%%%%%%%%%%%%%%%%%%%%%%%%%%%%%%%%%%%%%%%%%%%%%%%%%%%%%%%%%%%%%%%%%%%%%%%%%
	%%%%%%%%%%%%%%%%%%%%%%%%%%%%%%%%%%%%%%%%%%%%%%%%%%%%%%%%%%%%%%%%%%%%%%%%%%%
	We consider the system in hydrodynamic regime, wherein thermodynamic parameters can be defined locally (local thermodynamic equilibrium). 
	For a system in which the magnetic flux is treated as a conserved quantity, the nonequilibrium statistical operator $\hat{\rho}(t)$ is defined as~\cite{Hongo:2020qpv}
	\begin{equation}
		\hat{\rho}(t) = Q^{-1}\exp\Big[ -\int d^3x \hat{Z}(\vec{x},t)\Big], ~\text{with}~~Q = {\rm Tr}\exp\Big[ -\int d^3x \hat{Z}(\vec{x},t)\Big],
	\end{equation}
	where
	\begin{equation}
		\hat{Z}(\vec{x},t)=\epsilon \int_{-\infty}^t dt_1 e^{\epsilon(t_1-t)}\Big[  \beta_{\nu}(\vec x,t_1)\,\hat{T}^{0\nu}(\vec x,t_1)
		+ \mathcal{H}_{\nu}(\vec x,t_1)\,\hat{\tilde F}^{0\nu}(\vec x,t_1) \Big].
	\end{equation}
	Here, $\epsilon$ denotes a small infinitesimal parameter that enforces irreversibility throughout the calculation, and it is set to zero at the end. 
	The tensors $\hat{T}^{\mu\nu}$ and $\hat{\tilde{F}}^{\mu\nu}$ are operator-valued function. Zubarev's 
	prescription for taking the statistical average of any operator $\hat{O}(\vec{x},t)$ is given by~\cite{Zubarev1979derivation}
	\begin{equation}
		\left< \hat{O}(\vec{x},t)\right> = \lim_{\epsilon \to 0^+} \lim_{V\to\infty}{\rm Tr}\left[\hat{\rho}(t)\hat{O}(\vec{x},t) \right].
	\end{equation}
	The statistical averages of the operators $\hat{T}^{\mu\nu}$ and $\hat{\tilde{F}}^{\mu\nu}$ reproduce the corresponding hydrodynamic
	values in Eqs.~\eqref{T_decom} and~\eqref{F_decom}, i.e. $\big<\hat{T}^{\mu\nu}\big>=T^{\mu\nu}$ and $\big<\hat{\tilde{F}}^{\mu\nu}\big>=\tilde{F}^{\mu\nu}$. 
	A principal advantage of the NESO formalism is that it gives a clean separation into equilibrium and nonequilibrium parts. 
	This follows by integrating $\hat{Z}(\vec{x},t)$ by-parts, which yields the density operator
	\begin{equation}
		\hat{\rho}(t)=Q^{-1}\exp{(-\hat{\mathcal{A}}+\hat{\mathcal{B}})},
	\end{equation}
	where
\begin{align}
	\hat{\mathcal{A}}(t) &= \int d^3x \left[ \beta_{\nu}(\vec x,t)\,\hat{T}^{0\nu}(\vec x,t)
	+ \mathcal{H}_{\nu}(\vec x,t)\,\hat{\tilde F}^{0\nu}(\vec x,t) \right],\label{A} \\[0.25em]
	\hat{\mathcal{B}}(t) &= \int d^3x \int_{-\infty}^{t} dt_1\; e^{\epsilon (t_1-t)}\, \hat{\mathcal{C}}(\vec x,t_1), \label{B} \\[0.25em]
	\hat{\mathcal{C}}(\vec x,t) &= \hat T^{\mu\nu}(\vec x,t)\,\partial_{\mu}\beta_{\nu}(\vec x,t)
	+ \hat{\tilde F}^{\mu\nu}(\vec x,t)\,\partial_{\mu}\mathcal{H}_{\nu}(\vec x,t). \label{C}
\end{align}
Here, $\hat{\mathcal{A}}(t)$ denotes the equilibrium contribution and $\hat{\mathcal{B}}(t)$ the nonequilibrium part. We used the conservation laws to write the nonequilibrium part $\hat{\mathcal{B}}(t)$ in a covariant form. 
Treating $\hat{\mathcal{B}}$ as a perturbation, the $\hat{\rho}(t)$ can be expanded around local equilibrium, thereby 
systematically incorporating deviations around local equilibrium. 
Consequently, the statistical average of any operator, expanded up-to second order about the local equilibrium, can be written as~\cite{Hosoya:1983id,HARUTYUNYAN2022168755}:
	\begin{equation}
		\left< \hat{O}(x)\right> = \left< \hat{O}(x)\right>_l + \int d^4 x_1 \Big( \hat{O}(x), \hat{\mathcal{C}}(x_1) \Big) + \int d^4 x_1 d^4 x_2 \Big( \hat{O}(x), \hat{\mathcal{C}}(x_1), \hat{\mathcal{C}}(x_2) \Big) +\cdots \label{staAve},
	\end{equation}
	here, $\int d^4 x_1 = \int d^3 x_1 \int_{-\infty}^t dt_1 
	e^{\epsilon(t_1-t)}$ and the statistical averaging in local equilibrium is denoted
	by the subscript $l$. The two-point and three-point 
	correlation functions are defined by
	\begin{align}
		\Big( \hat{O}(x), \hat{X}(x_1) \Big) &= \int_0^1 d\tau \Big< O(x) \left[ X_\tau(x_1)-\left< X_\tau(x_1) \right>_l \right] \Big>_l, \label{twopointdef}\\
		\Big( \hat{O}(x), \hat{X}(x_1), \hat{Y}(x_2) \Big) &=\frac{1}{2} \int_0^1 d\tau \int_0^1 d\lambda \Big< \tilde{T} \Big\{\hat{O}(x) \Big[ \hat{X}_\lambda (x_1) \hat{Y}_\tau(x_2)-\langle \hat{X}_\lambda (x_1)\rangle_l \hat{Y}_\tau(x_2)\nonumber\\
		&~~~-\hat{X}_\lambda (x_1) \langle \hat{Y}_\tau(x_2)\rangle_l-\langle \tilde{T} \hat{X}_\lambda (x_1) \hat{Y}_\tau(x_2)\rangle_l + 2 \langle \hat{X}_\lambda (x_1)\rangle_l \langle \hat{Y}_\tau(x_2)\rangle_l  \Big] \Big\} \Big>_l,\label{threepointdef}
	\end{align}
	respectively. Here, $X_\tau = e^{-\tau A} X e^{\tau A}$ and 
	$\tilde{T}\{ X_\lambda Y_\tau \}$ is the anti-chronological 
	time-ordering operator with respect to variable $\tau$ and 
	$\lambda$. Also, observing~\eqref{threepointdef}, we infer that the three-point correlation function follows the symmetry relation:
	\begin{align}
		\int d^4 x_1 d^4 x_2 \Big( \hat{O}(x), \hat{X}(x_1), \hat{Y}(x_2) \Big) = \int d^4 x_1 d^4 x_2 \Big( \hat{O}(x), \hat{Y}(x_1), \hat{X}(x_2) \Big). \label{threepointsymmetry}
	\end{align}

		Furthermore, to develop a second-order theory and to obtain the evolution equations for the dissipative tensors, $\Pi_{\parallel},\, \Pi_{\perp},\, \mathscr{K}^{\mu},\, \mathscr{J}^{\mu},\, \pi_{\perp}^{\mu\nu},$ and $m^{\mu\nu}$ up to second order in the gradient expansion within the NESO framework, it is necessary to express the statistical average in Eq.~\eqref{staAve} explicitly in terms of hydrodynamic gradients. To this end, we must systematically extract all first- and second-order contributions arising from the two-point and three-point correlation functions. For this purpose, consider an integral of the form 
	\begin{align}
		\int d^4 x_1 \Big( \hat{O}(x), \hat{\mathcal{C}}(x_1) \Big). \label{C_int}
	\end{align}
	We may factorize each term in $\hat{\mathcal{C}}$ into two components, writing $\hat{\mathcal{C}}= \hat{\mathcal{C}}_{oper}\mathcal{C}_{hydro}$
	where $\hat{\mathcal{C}}_{oper}$ denotes the operator-valued part and $\mathcal{C}_{hydro}$ represents the hydrodynamic fields. Performing a Taylor expansion of the hydrodynamic fields appearing in $\hat{\mathcal{C}}(x_1)$ about the point $x_1 = x$, while keeping all operator-valued quantities in $\hat{\mathcal{C}}(x_1)$ evaluated at $x_1$, we may write
	\begin{align}
		\hat{\mathcal{C}}(x_1) &= \hat{\mathcal{C}}_{oper}(x_1)\mathcal{C}_{hydro}(x_1) \nonumber\\
		&= \hat{\mathcal{C}}_{oper}(x_1)\left[\hat{\mathcal{C}}(x_1)\right]_x+\hat{\mathcal{C}}_{oper}(x_1)(x_1-x)^\lambda \left[\frac{\partial}{\partial x_1^\lambda}\mathcal{C}_{hydro}(x_1)\right]_x +\cdots .\label{tayexC}
	\end{align}
	Here, the notation $[...]_x$ indicates that values inside are evaluated at $x$.
	This expansion allows us to factor out all $x$-dependent hydrodynamic quantities from the integral in~\eqref{C_int}. Apart from that, the dominant contribution to the integral~\eqref{C_int} comes from $|\vec{x}_1-\vec{x}|$ which encodes the correlation length, of order the microscopic mean free path $\lambda$.  Consequently, $(x_1-x)^{\lambda}\,\partial/\partial x_1^{\lambda}\sim \lambda/L \equiv \mathrm{Kn}$, with $L$ being the macroscopic scale ($\mathrm{Kn}$ is the Knudsen number). Therefore, the second term in~\eqref{tayexC} is one order high in gradient ordering compared to the first term. The first term in \eqref{tayexC}, in which  the hydrodymamic fields are evaluated at $x_1=x$, is called ``local'' contribution. All the higher-order terms are correspondingly ``nonlocal'' contributions.

	Thus, we first decompose $\hat{\mathcal{C}}$ as $\hat{\mathcal{C}}=\hat{\mathcal{C}}_F+\hat{\mathcal{C}}_S$	and then perform a Taylor expansion of the hydrodynamic fields to extract the local and nonlocal contributions, as prescribed in Eq.~\eqref{tayexC}. Here, $\hat{\mathcal{C}}_{F}$ denotes the part of $\hat{\mathcal{C}}$ that contains hydrodynamic fields which are first order in hydrodynamic gradients, while $\hat{\mathcal{C}}_{S}$ denotes the part of $\hat{\mathcal{C}}$ that contains hydrodynamic fields which are second order in hydrodynamic gradients. In this way, the statistical average of any operator is systematically organized order-by-order in derivatives of the hydrodynamic fields.
	\begin{equation}\label{gradient_expansion}
		\left< \hat{O}(x)\right> = \left< \hat{O}(x)\right>_l + \left< \hat{O}(x)\right>_1 + \left< \hat{O}(x)\right>_2,
	\end{equation}
	where $\left< \hat{O}(x)\right>_1$ and $\left< 
	\hat{O}(x)\right>_2$ are the first-order and second-order 
	gradient terms, respectively. The first-order contribution to the statistical average emerges by extracting the local contribution from two-point correlation function. The 
	resulting expression is given by 
	\begin{equation}
		\left< \hat{O}(x)\right>_1 = \int d^4 x_1 \Big( \hat{O}(x), \hat{\mathcal{C}}_F(x_1) \Big)\Big|_{local}.\label{FOC}
	\end{equation}
	 Whereas, the second-order contribution to the statistical average 
	can originate from three possible ways, given by
	\begin{equation}
		\left< \hat{O}(x)\right>_2 = \left< \hat{O}(x)\right>_2^{(2),NL} + \left< \hat{O}(x)\right>_2^{(2),ET} +\left< \hat{O}(x)\right>_2^{(3)} ,\label{SOC}
	\end{equation}
	where,
	\begin{align}
		\left< \hat{O}(x)\right>_2^{(2),NL} &=  \int d^4 x_1 \Big( \hat{O}(x), [\partial_\lambda \hat{\mathcal{C}}(x_1)]_{x_1=x} \Big)(x_1-x)^\lambda, \label{nlc}\\
		\left< \hat{O}(x)\right>_2^{(2),ET} &= \int d^4 x_1 \Big( \hat{O}(x), \hat{\mathcal{C}}_S(x_1) \Big)\Big|_{local}, \label{etc}\\
		\left< \hat{O}(x)\right>_2^{(3)}  &= \int d^4 x_1 d^4 x_2 \Big( \hat{O}(x), \hat{\mathcal{C}}_F(x_1), \hat{\mathcal{C}}_F(x_2) \Big)\Big|_{local}. \label{tpc}
	\end{align}
	Here, the term $\big< \hat{O}(x)\big>_2^{(2),NL}$ contains the nonlocal corrections arising from two-point correlation functions and $[\partial_\lambda \hat{\mathcal{C}}(x_1)]_{x_1=x}$ signifies that we differentiate all hydrodynamic variables (not the operators) appearing in $\hat{\mathcal{C}}$ with respect to $x_1$, and then set $x_1=x$, while the operators remain evaluated at $x_1$. $\big< 
	\hat{O}(x)\big>_2^{(2),ET}$ contains the local corrections from two-point correlators driven by extended thermodynamic forces ($\hat{\mathcal{C}}_S$), 
	and $\big< \hat{O}(x)\big>_2^{(3)}$ comprises the local contributions from three-point correlation functions.

	In order to evaluate the second-order correction, we observe from Eqs.~\eqref{nlc}-\eqref{tpc} that both $\hat{\mathcal{C}}$ and $\partial_\lambda \hat{\mathcal{C}}$ are required. We therefore compute these quantities one by one. To obtain $\hat{\mathcal{C}}$ in terms of gradients of hydrodynamic quantities, our strategy is to begin with a tensor decomposition of the operators $\hat{T}^{\mu\nu}$ and $\hat{\tilde{F}}^{\mu\nu}$:
	\begin{align}
		\hat T^{\mu\nu} &= \hat\varepsilon\,u^{\mu}u^{\nu}
			+ \hat p_{\parallel}\, b^{\mu}b^{\nu}
			- \hat p_{\perp}\, \G^{\mu\nu} 
		+ \hat h^{\mu}u^{\nu} + \hat h^{\nu}u^{\mu}
		+ \hat f^{\mu}b^{\nu} + \hat f^{\nu}b^{\mu}
		+ \hat \pi^{\mu\nu}_{\perp} ,\label{To_decom} \\
		\hat{\tilde F}^{\mu\nu}
		&=  \hat B\,(b^{\mu}u^{\nu}-b^{\nu}u^{\mu}) 
		\;-\; (\hat l^{\mu}b^{\nu}-\hat l^{\nu}b^{\mu})
		\;+\; \hat g^{\mu}u^{\nu} - \hat g^{\nu}u^{\mu} \;+\; \hat m^{\mu\nu}. \label{Fo_decom}
	\end{align}
	All operators, except the pressures $\hat{p}_{\parallel}$ and $\hat{p}_{\perp}$, can be directly matched to their hydrodynamic counterparts by taking statistical averages with the full nonequilibrium statistical operator:
	\begin{align}
	\varepsilon \equiv \big\langle \hat{\varepsilon} \big\rangle,\quad
	B \equiv \big\langle \hat{B} \big\rangle,\quad
	h^\mu \equiv \big\langle \hat{h}^\mu \big\rangle,\quad
	f^\mu \equiv \big\langle \hat{f}^\mu \big\rangle,\quad
	l^\mu \equiv \big\langle \hat{l}^\mu \big\rangle,\quad
	g^\mu \equiv \big\langle \hat{g}^\mu \big\rangle,\quad
	\pi^{\mu\nu} \equiv \big\langle \hat{\pi}^{\mu\nu} \big\rangle,\quad
	m^{\mu\nu} \equiv \big\langle \hat{m}^{\mu\nu} \big\rangle.
	\end{align}
	In contrast, the statistical average of the total longitudinal and transverse pressure operators yield the equilibrium pressure plus the corresponding bulk viscous corrections:
	\begin{align}
	\big\langle \hat{p}_{\parallel} \big\rangle \equiv p_{\parallel} + \Pi_{\parallel}, 
	\qquad
	\big\langle \hat{p}_{\perp} \big\rangle \equiv p_{\perp} + \Pi_{\perp}. \label{bulk_to_eq_relation}
	\end{align}
	All above operators present in~\eqref{To_decom} and~\eqref{Fo_decom} can be expressed as suitable projections of $\hat{T}^{\mu\nu}$ and $\hat{\tilde{F}}^{\mu\nu}$.
	\begin{align}
		\hat \varepsilon &= u_\alpha u_\beta\, \hat T^{\alpha\beta},~~\hat p_{\perp} = -\frac{1}{2}\, \G_{\alpha\beta}\, \hat T^{\alpha\beta},~~\hat p_{\parallel} = b_\alpha b_\beta\, \hat T^{\alpha\beta},\\
		\hat h^\mu &= \U_{\alpha\beta}^{\h\h\mu}\, \hat T^{\alpha\beta},~~\hat f^\mu = -\, \B_{\alpha\beta}^{\h\h\mu}\, \hat T^{\alpha\beta},~~\hat \pi_{\perp}^{\mu\nu} = \G^{\mu\nu}_{\h\h \alpha\beta} \, \hat T^{\alpha\beta},\\
		\hat B &= u_\alpha b_\beta \,\hat{\tilde F}^{\alpha\beta},~~\hat g^\mu = -\, \Uc_{\alpha\beta}^{\h\h\mu}\,\hat{\tilde F}^{\alpha\beta},~~\hat l^\mu = -\,\Bc_{\alpha\beta}^{\h\h\mu}\,\hat{\tilde F}^{\alpha\beta},~~\hat m^{\mu\nu} = \Gc^{\mu\nu}_{\h\h\alpha\beta}\,\hat{\tilde F}^{\alpha\beta}.
	\end{align}
where the projectors are defined in~\eqref{Gprojectos}-\eqref{cutprojectorB}. In order to write $\hat{\mathcal{C}}$ in first- and second-order hydrodynamic gradients, we substitute the decomposition of $\hat T^{\mu \nu}$ and $\hat{\tilde F}^{\mu\nu}$ from~\eqref{To_decom} and~\eqref{Fo_decom} in~\eqref{C}. We obtain	
	\begin{align}
		\hat{\mathcal{C}} &= \hat\varepsilon\, D\beta - \hat B\, D \mathcal{H}
		- \beta\,\hat p_{\parallel}\,\theta_{\parallel}
		- \beta\,\hat p_{\perp}\,\theta_{\perp}
		- \hat B\, \mathcal{H}\, \theta_{\parallel}\nonumber\\
		&\quad + 2 \hat h^\mu \, \U^{\alpha\beta}_{\h\h \mu} \partial_\alpha \beta_\beta + 2 \hat f^\mu \, \B^{\alpha\beta}_{\h\h \mu} \partial_\alpha \beta_\beta - 2 \hat g^\mu \, \Uc^{\alpha\beta}_{\h\h \mu} \partial_\alpha \mathcal{H}_\beta + 2 \hat l^\mu \, \Bc^{\alpha\beta}_{\h\h \mu} \partial_\alpha \mathcal{H}_\beta \nonumber \\
		&~~~+ \hat \pi_{\perp}^{\mu\nu}\,\G^{\alpha\beta}_{\h\h \mu\nu} \partial_\alpha \beta_\beta 
		+ \hat m^{\mu \nu} \Gc^{\alpha\beta}_{\h\h \mu\nu} \partial_\alpha \mathcal{H}_\beta . \label{c_solve}
	\end{align}
Here, the first five terms in $\hat{\mathcal{C}}$ are simplified using $\beta_{\nu}=\beta\,u_{\nu}$ and $\mathcal{H}_{\nu}=-\mathcal{H}\,b_{\nu}$. 
In this work, we take two independent thermodynamic variables: the total energy density $\varepsilon$ and the magnetic flux density $B$---all other thermodynamic quantities can be treated as functions of $(\varepsilon,B)$. 
Consequently, the comoving derivative of any thermodynamic variable follows from the chain rule. 
For example, for $\beta\equiv\beta(\varepsilon,B)$ one finds
\begin{align}
	D\beta
	= \left.\frac{\partial \beta}{\partial \varepsilon}\right|_{B}\, D\varepsilon
	+ \left.\frac{\partial \beta}{\partial B}\right|_{\varepsilon}\, DB  .
\end{align}
Similarly, we write $D\mathcal{H}$ via the chain rule, and then substitute $D\varepsilon$ and $DB$ from Eqs.~\eqref{udelT} and~\eqref{bdelF}, respectively. 
In this way, the first five terms in $\hat{\mathcal{C}}$ can be simplified and written as (See Eq.~\eqref{Cscalar} Appendix~\ref{cal_C})
\begin{align}
		\hat\varepsilon D\beta - \hat B D H
		- \beta \hat p_{\parallel}\theta_{\parallel}
		- \beta \hat p_{\perp}\theta_{\perp}
		- \hat B H \theta_{\parallel}
		= -\beta \theta_{\perp} \hat P_{\perp}
		- \beta \theta_{\parallel} \hat P_{\parallel}
		+ \hat\beta^{*}\big[ -u_{\nu}\partial_{\mu}T_{1}^{\mu\nu} \big]
		- \hat{\mathcal{H}}^{*}\big[ -b_{\nu}\partial_{\mu}\tilde F_1^{\mu\nu} \big].
	\end{align}
Here, we defined	
\begin{align}
	\hat P_{\perp} &=\hat p_{\perp} - \hat p_{\perp}^*,~~\hat P_{\parallel} =\hat p_{\parallel} - \hat p_{\parallel}^* , \\
	\hat{p}_\perp^* &=  \hat\varepsilon\,\frac{\partial p_{\perp}}{\partial \varepsilon}
	+ \hat B\,\frac{\partial p_{\perp}}{\partial B},~~ \hat{p}_\parallel^* =  \hat\varepsilon\,\frac{\partial p_{\parallel}}{\partial \varepsilon}
	+ \hat B\,\frac{\partial p_{\parallel}}{\partial B} , \\
	\hat\beta^{*} &= \hat\varepsilon\,\frac{\partial \beta}{\partial \varepsilon}
	+ \hat B\,\frac{\partial \beta}{\partial B},~~
	\hat{\mathcal{H}}^{*} = \hat\varepsilon\,\frac{\partial \mathcal{H}}{\partial \varepsilon}
	+ \hat B\,\frac{\partial \mathcal{H}}{\partial B}. 
\end{align}

Furthermore, the structures $\U^{\alpha\beta}_{\h\h \mu} \partial_\alpha \beta_\beta$ and $\Uc^{\alpha\beta}_{\h\h \mu} \partial_\alpha \mathcal{H}_\beta$ contain $D u_\mu$ and $D b_\mu$, respectively. Therefore, each contribution can be decomposed into first- and second-order pieces (using eq.~\eqref{Du} and~\eqref{Db}). The first-order parts of $\U^{\alpha\beta}_{\h\h \mu} \partial_\alpha \beta_\beta$ and $\Uc^{\alpha\beta}_{\h\h \mu} \partial_\alpha \mathcal{H}_\beta$ can then be expressed as projections of $\Bc^{\alpha\beta}_{\h\h\mu}$ onto $\partial_\alpha \mathcal{H}_\beta$ and of $\B^{\alpha\beta}_{\h\h\mu}$ onto $\partial_\alpha u_\beta$, respectively (See Eq.~\eqref{UtoBcrelation} and~\eqref{UctoBrelation} in Appendix~\ref{important_relation}). Accordingly, we decompose Eq.~\eqref{c_solve} into two parts, collecting thermodynamic forces of first order and of second order in the hydrodynamic gradient expansion. We obtain
\begin{align}
	\hat{\mathcal C}_F
	&= -\,\bar{\theta}_{\perp}\,\hat{P}_{\perp}
	-\,\bar{\theta}_{\parallel}\,\hat{P}_{\parallel}
	+ \hat{\mathscr{K}}^{\mu}\,\bar{\mathscr{Y}}_\mu
	+ \hat{\mathscr{J}}^{\mu}\, \bar{\mathscr{X}}_\mu
	+ \hat \pi_{\perp}^{\mu\nu} \bar{\sigma}_{\perp \mu\nu}
	+ \hat m^{\mu\nu} \bar{\chi}_{ \mu\nu}, \label{CF}\\
	\hat{\mathcal C}_S
	&= \hat\beta^{*}\big[-u_\nu \partial_\mu T_1^{\mu\nu}\big]
	- \hat{\mathcal{H}}^{*}\big[-b_\nu \partial_\mu \tilde F_1^{\mu\nu}\big]
	+ \hat h^{\mu}\,\beta\,(D u_\mu)_2
	+ \hat g^{\mu}\,\mathcal{H}\,(D b_\mu)_2 \label{CS}.
\end{align}	
Inferring to~\eqref{Du} and~\eqref{Db}, the second order contribution $(Du_\mu)_2$ and $(Db_\mu)_2$ in~\eqref{CS} is given by
\begin{align}
	(Du_\mu)_2 &=  \frac{H}{K}\, b_{\mu} u_{\beta}\partial_{\alpha}\tilde F_{1}^{\alpha\beta}
	- \frac{1}{(\varepsilon+p_{\perp})}
	\Big[\frac{\varepsilon+p_{\perp}}{K}\,b_{\mu} b_{\beta}
	+ \G_{\mu \beta}\Big]\partial_{\alpha}T_{1}^{\alpha\beta} , \\
(Db_\mu)_2 &= - \frac{u_{\mu}  b_{\beta}}{K}\,\partial_{\alpha}T_{1}^{\alpha\beta}
	+ \frac{1}{B}\Big[\frac{B H\,u_{\mu}u_{\beta}}{K}+\G_{\mu \beta}\Big]\partial_{\alpha}\tilde F_{1}^{\alpha\beta}, 
\end{align}	
respectively.

Before proceeding to the evaluation of $\partial_\lambda \hat{\mathcal{C}}$
for the nonlocal contributions, we emphasize that earlier derivations of second-order relativistic hydrodynamics based on Zubarev’s nonequilibrium statistical operator formalism incorporated nonlocal corrections only for couplings between tensors of the same rank~\cite{HARUTYUNYAN2022168755,Tiwari:2024trl,She:2024rnx}. 
This was refined in Ref.~\cite{Harutyunyan:2025fgs}, where it was shown that at the nonlocal level Curie’s principle cannot be applied naively, and mixed-rank couplings are allowed. 
Accordingly, Ref.~\cite{Harutyunyan:2025fgs} included such couplings by Taylor-expanding the thermodynamic forces that appear in $\hat{\mathcal{C}}(x_1)$ (specifically $\partial_\mu u_\nu$ and $\partial_\mu\alpha$ only with $\alpha=\beta\mu$, in their work), thereby generates the acceleration terms in the evolution equations of dissipative tensors. However, $\hat{T}^{\mu\nu}$ and $\hat{\tilde{F}}^{\mu\nu}$ possess a tensor structure (\eqref{To_decom} and~\eqref{Fo_decom}) that include explicit hydrodynamic fields ($u^\mu$ and $b^\mu$) in addition to operator-valued components. Therefore a complete extraction of nonlocal corrections must also account for all such fields. Accordingly, in this work, apart from the thermodynamic forces $\partial_\mu\beta_\nu$ and $\partial_\mu \mathcal{H}_\nu$, we also included the hydrodynamic field appearing in $\hat{T}^{\mu\nu}$ and $\hat{\tilde{F}}^{\mu\nu}$ via tensor structure~\eqref{To_decom} and~\eqref{Fo_decom}, while doing the expansion~\eqref{tayexC}, thereby yielding an improvement in the nonlocal corrections.

For the nonlocal corrections, our goal is to Taylor-expand all hydrodynamic variables (but not operators) that appear in $\hat{\mathcal{C}}(x_1)$ about $x_1=x$ and then identify the coefficient of $(x_1-x)^\lambda$. 
Equivalently, this amounts to differentiating the hydrodynamic variables in $\hat{\mathcal{C}}(x_1)$ with respect to $x_1$ and setting $x_1=x$. 
Implementing this procedure by substituting Eqs.~\eqref{T_decom} and~\eqref{F_decom} into Eq.~\eqref{C}, we obtain
		\begin{align}
		\partial_\lambda \hat{\mathcal{C}}
		&= \hat{\varepsilon}\, \partial_\lambda (u^\mu u^\nu \partial_\mu \beta_\nu)
		+ \hat{p}_\parallel\, \partial_\lambda (b^\mu b^\nu \partial_\mu \beta_\nu)
		- \hat{p}_\perp\,  \partial_\lambda(\G^{\mu\nu}  \partial_\mu \beta_\nu)
		\nonumber\\
		&\quad
		+ \hat{h}^\mu \partial_\lambda (u^\nu \partial_\mu \beta_\nu)
		+ \hat{h}^\nu \partial_\lambda (u^\mu \partial_\mu \beta_\nu)
		+ \hat{f}^\mu \partial_\lambda (b^\nu \partial_\mu \beta_\nu)
		+ \hat{f}^\nu \partial_\lambda (b^\mu \partial_\mu \beta_\nu)
		\nonumber\\
		&\quad
		+ \hat{\pi}_{\perp}^{\mu\nu}\, \partial_\lambda \partial_\mu \beta_\nu
		+ \hat{B}\, \partial_\lambda 
		\big[(b^\mu u^\nu - b^\nu u^\mu)\partial_\mu \mathcal{H}_\nu \big]
		\nonumber\\
		&\quad
		- \hat{l}^\mu \partial_\lambda (b^\nu \partial_\mu \mathcal{H}_\nu)
		+ \hat{l}^\nu \partial_\lambda (b^\mu \partial_\mu \mathcal{H}_\nu)
		+ \hat{g}^\mu \partial_\lambda (u^\nu \partial_\mu \mathcal{H}_\nu)
		- \hat{g}^\nu \partial_\lambda (u^\mu \partial_\mu \mathcal{H}_\nu)
		\nonumber\\
		&\quad
		+ \hat{m}^{\mu\nu}\, \partial_\lambda \partial_\mu \mathcal{H}_\nu .\label{delC1}
	\end{align}
Using the relations collected in Appendix~A, Eq.~\eqref{delC1} simplifies to
\begin{align}
	\partial_\lambda \hat{\mathcal{C}}
	=& -\hat{P}_\parallel \partial_\lambda \bar{\theta}_\parallel  - \hat{P}_\perp \partial_\lambda \bar{\theta}_\perp +  \bar{\theta}_\parallel [\hat \varepsilon \,\partial_\lambda \gamma_\parallel + \hat B \,\partial_\lambda \phi_\parallel]  + \bar{\theta}_\perp [\hat \varepsilon \,\partial_\lambda \gamma_\perp + \hat B \,\partial_\lambda \phi_\perp]  \nonumber\\\
	&+ \hat{h}^\mu[ \partial_\lambda (2\U^{\alpha\beta}_{\h\h\mu}\partial_\alpha \beta_\beta) +2 D\beta \partial_\lambda u_\mu ] + \hat{f}^\mu[ \partial_\lambda (2\B^{\alpha\beta}_{\h\h\mu}\partial_\alpha \beta_\beta) +2\bar{\theta}_\parallel \partial_\lambda b_\mu ] \nonumber\\
	&+ \hat{\pi}_\perp^{\mu\nu} [\partial_\lambda \bar{\sigma}_{\perp\mu\nu} + \partial_\lambda u_\nu (2\U^{\alpha\beta}_{\h\h\mu}\partial_\alpha \beta_\beta) -  \partial_\lambda b_\nu (2\B^{\alpha\beta}_{\h\h\mu}\partial_\alpha \beta_\beta) ] \nonumber \\
	& + \hat l^\mu [\partial_\lambda (2\Bc^{\alpha\beta}_{\h\h\mu}\partial_\alpha \mathcal{H}_\beta) -(D\mathcal{H}+H\bar{\theta}_\parallel)\partial_\lambda u_\mu] - \hat{g}^\mu [\partial_\lambda (2\Uc^{\alpha\beta}_{\h\h\mu}\partial_\alpha \mathcal{H}_\beta) +(D\mathcal{H}+H\bar{\theta}_\parallel)\partial_\lambda b_\mu] \nonumber\\
	& + \hat{m}^{\mu\nu} [ \partial_\lambda \bar{\chi}_{\mu\nu} + \partial_\lambda b_\nu (2\Bc^{\alpha\beta}_{\h\h\mu}\partial_\alpha \mathcal{H}_\beta) -  \partial_\lambda u_\nu (2\Uc^{\alpha\beta}_{\h\h\mu}\partial_\alpha \mathcal{H}_\beta) ] . \label{delC2}
\end{align}
Here we defined, 
\begin{align}
	\gamma_{\parallel} &= \frac{\partial p_\parallel}{\partial \varepsilon},~ \gamma_{\perp} = \frac{\partial p_\perp}{\partial \varepsilon},\\
	\phi_{\parallel} &= \frac{\partial p_\parallel}{\partial B},~ \phi_{\perp} = \frac{\partial p_\perp}{\partial B}.
\end{align}
Finally, rewriting everything in terms of thermodynamic forces with the help of~\eqref{UtoBcrelation} and~\eqref{UctoBrelation}, we arrive at
\begin{align}
		\partial_\lambda \hat{\mathcal{C}}
		=&-\hat{P}_\parallel \partial_\lambda \bar{\theta}_\parallel  - \hat{P}_\perp \partial_\lambda \bar{\theta}_\perp +  \bar{\theta}_\parallel [\hat \varepsilon \,\partial_\lambda \gamma_\parallel + \hat B \,\partial_\lambda \phi_\parallel]  + \bar{\theta}_\perp [\hat \varepsilon \,\partial_\lambda \gamma_\perp + \hat B \,\partial_\lambda \phi_\perp]  \nonumber\\
		&+ \hat{\mathscr{J}}^\mu \partial_\lambda \bar{\mathscr{X}}_\mu +\hat{h}^\mu [-\bar{\mathscr{X}}_\mu \partial_\lambda B' +2DB \partial_\lambda u_\mu] - \hat{l}^\mu (D\mathcal{H}+H\bar{\theta}_\parallel)\partial_\lambda u_\mu \nonumber\\
		&+ \hat{\pi}_\perp^{\mu\nu} [\partial_\lambda \bar{\sigma}_{\perp\mu\nu} -B'\bar{\mathscr{X}}_\mu \partial_\lambda u_\nu -  \bar{\mathscr{Y}}_\mu \partial_\lambda b_\nu  ] \nonumber \\
		&+ \hat{\mathscr{K}}^\mu \partial_\lambda \bar{\mathscr{Y}}_\mu +2\hat{f}^\mu \bar{\theta}_\parallel \partial_\lambda b_\mu -\hat{g}^\mu [-\bar{\mathscr{Y}}_\mu \partial_\lambda H + (D\mathcal{H}+H\bar{\theta}_\parallel) \partial_\lambda b_\mu] \nonumber \\
		& + \hat{m}^{\mu\nu} [ \partial_\lambda \bar{\chi}_{\mu\nu} + \bar{\mathscr{X}}_\mu \partial_\lambda b_\nu  + H \bar{\mathscr{Y}}_\mu  \partial_\lambda u_\nu  ]  . \label{delC3}
\end{align}
Here, we defined $B'=\frac{B}{(\varepsilon + p_\perp)}$. 

We will employ~\eqref{delC3} to evaluate the nonlocal (NL) corrections in the subsequent sections. 
The complete derivation of $\partial_\lambda \hat{\mathcal{C}}$ is provided in Appendix~\eqref{delC_cal}.

	\section{First order magnetohydrodynamics}\label{FOMHD}

First-order relativistic hydrodynamics derived using Zubarev’s nonequilibrium statistical operator (NESO) method was previously developed in Ref.~\cite{Huang:2011dc}, where Kubo formulas for the transport coefficients of first-order magnetohydrodynamics were obtained. However, that derivation followed the conventional formulation in which the energy-momentum tensors of the fluid and the electromagnetic field are treated separately, rather than starting from a fully conserved set of equations. More recently, first-order relativistic magnetohydrodynamics has been formulated entirely in terms of conserved quantities by employing the conservation of the total energy-momentum tensor together with magnetic-flux conservation (the Bianchi identity)~\cite{Hongo:2020qpv,Hattori:2022hyo}. In this section, we provide a brief review of first-order magnetohydrodynamics within this modern framework.

We begin by evaluating the bulk viscous pressure parallel to the magnetic field, $\Pi_{\parallel}$. 
According to Eq.~\eqref{fullbulkparallel}, the first-order correction to $\Pi_{\parallel}$ is given by
\begin{align*}
		\Pi_{1\parallel}(x) &= \big< \hat P_{\parallel}(x) \big>_{1} = \int d^4 x_1\, \big( \hat P_{\parallel}(x),\, \hat{\mathcal{C}}_F(x_1) \big)\big|_{local}.
\end{align*}
		Here, we employ Eq.~\eqref{gradient_expansion} and~\eqref{FOC} to express $\big< \hat{P}_{\parallel} \big>_{1}$ in terms of two-point correlation function.
		By Curie’s symmetry principle, the scalar operator $\hat{P}_{\parallel}$ couples only to the operators of the same rank and parity; consequently, within $\hat{\mathcal{C}}_F$ (Eq.~\eqref{CF}) it can couple only to $\hat{P}_{\parallel}$ and $\hat{P}_{\perp}$. Hence,	
\begin{align} 
		\Pi_{1\parallel}(x)
		&= \int d^4 x_1\, \big( \hat P_{\parallel}(x),\, \hat P_{\parallel}(x_1) \big)\,[-\bar{\theta}_{\parallel}(x_1)]\big|_{local}
		+ \int d^4 x_1\, \big( \hat P_{\parallel}(x),\, \hat P_{\perp}(x_1) \big)\,[-\bar{\theta}_{\perp}(x_1)]\big|_{local}  . \label{pical} 
\end{align}
We next perform a Taylor expansion of $\bar{\theta}_{\parallel}(x_1)$ about $x_1=x$:
\begin{equation}\label{tayextau}
	\bar{\theta}_{\parallel}(x_1) = \bar{\theta}_{\parallel}(x)+(x_1-x)^\lambda \frac{\partial}{\partial x_1^\lambda}\left[\bar{\theta}_{\parallel}(x_1)\right]\bigg|_x +\cdots .
\end{equation}
As discussed in Sec~\ref{NESO}, $(x_1-x)^{\lambda}\,\partial/\partial x_1^{\lambda}\sim \lambda/L \equiv \mathrm{Kn}$ ($\mathrm{Kn}$ is the Knudsen number). Since $\theta_{\parallel}$ is already first order in gradients, this term scales as $\mathcal{O}(\mathrm{Kn}^2)$ (second order in the hydrodynamic gradients) and is neglected in first-order hydrodynamics. Hence, extracting the local terms in~\eqref{pical}, we obtain
	\begin{align}	
			\Pi_{1\parallel}(x)&= -\bar{\theta}_{\parallel}(x)\,\bar{\zeta}_{\parallel}(x)
		-\bar{\theta}_{\perp}(x)\,\bar{\zeta}_{\times}(x), \label{pi1parZ}
	\end{align}
where, we defined
	\begin{align}
		\bar{\zeta}_{\parallel}(x) &= \int d^4 x_1\, \big( \hat P_{\parallel}(x),\, \hat P_{\parallel}(x_1) \big), \\
		\bar{\zeta}_{\times}(x) &= \int d^4 x_1\, \big( \hat P_{\parallel}(x),\, \hat P_{\perp}(x_1) \big),
	\end{align}
by comparing~\eqref{pi1parZ} to~\eqref{pi1par}. Similarly, we can obtain 	
	\begin{align}
		\Pi_{1\perp}(x)
		&= -\bar{\theta}_{\perp}(x)\,\bar{\zeta}_{\perp}(x)
		-\bar{\theta}_{\parallel}(x)\,\bar{\zeta}_{\times}'(x), \label{pi1perpZ}
	\end{align}
	where the transport coefficients are defined by
	\begin{align}
		\bar{\zeta}_{\perp}(x) &=  \int d^4 x_1\, \big( \hat P_{\perp}(x),\, \hat P_{\perp}(x_1) \big), \\
		\bar{\zeta}_{\times}'(x) &= \int d^4 x_1\, \big( \hat P_{\perp}(x),\, \hat P_{\parallel}(x_1) \big).
	\end{align}
For a charge-conjugate-symmetric plasma the cross bulk transport coefficients $\bar{\zeta}_{\times}=\bar{\zeta}_{\times}'$~\cite{Hongo:2020qpv}.

Furthermore, in order to obtain the first-order corrections to the statistical averages of the remaining dissipative tensors, we use Eqs.~\eqref{gradient_expansion} and~\eqref{FOC} to express these corrections in terms of two-point correlation functions. 
Invoking Eq.~\eqref{CF} together with Curie’s principle, only correlators of same rank and parity survive, so the first-order corrections are given by
\begin{align}
	\mathscr{K}_1^\mu(x)=\left<\hat{\mathscr{K}}^\mu(x)\right>_1
	&= \int d^4 x_1\, \left( \hat{\mathscr{K}}^\mu(x),\, \hat{\mathscr{K}}^\alpha(x_1) \right) \bar{\mathscr{Y}}_\alpha (x_1) , \label{K1}\\
	\mathscr{J}_1^\mu(x)=\left<\hat{\mathscr{J}}^\mu(x)\right>_1
	&= \int d^4 x_1\, \left( \hat{\mathscr{J}}^\mu(x),\, \hat{\mathscr{J}}^\alpha(x_1) \right)  \bar{\mathscr{X}}_\alpha (x_1)  , \label{J1}\\
	\pi_{1\perp}^{\mu\nu} (x)=\left<\hat \pi_{\perp}^{\mu\nu} (x)\right>_1
	&= \int d^4 x_1\, \left( \hat \pi_{\perp}^{\mu\nu} (x),\, \hat \pi_{\perp}^{\alpha\beta}(x_1) \right) \bar\sigma_{\perp\alpha\beta}(x_1) , \label{pi1}\\
	m_1^{\mu\nu} (x)=\left<\hat m^{\mu\nu} (x)\right>_1
	&= \int d^4 x_1\, \left( \hat m^{\mu\nu} (x),\, m^{\alpha\beta}(x_1) \right) \bar{\chi}_{ \alpha\beta} (x_1) . \label{m1}
\end{align}
Here we have used the fact that the statistical average of dissipative tensors vanishes in local equilibrium, 
$\langle \hat{\mathscr{K}}^\mu \rangle_{l}=\langle \hat{\mathscr{J}}^\mu \rangle_{l}=0$, $\left<\hat \pi_{\perp}^{\mu\nu} (x)\right>_l=0$, and $\left<\hat m^{\mu\nu} (x)\right>_l=0$. Similar to the calculation of bulk viscous pressure, we expand the thermodynamic forces in~\eqref{K1}-\eqref{m1} around $x_1=x$ and keep the leading order terms, also we write the correlation functions in terms of traces and the corresponding projection operators, given by
\begin{align}
\left( \hat{\mathscr{K}}^\mu(x),\, \hat{\mathscr{K}}^\alpha(x_1) \right) &= \frac{1}{2} \, \G^{\mu\alpha}\left( \hat{\mathscr{K}}^\nu(x),\, \hat{\mathscr{K}}_\nu(x_1) \right) , \label{kuboK1}\\
	\left( \hat{\mathscr{J}}^\mu(x),\, \hat{\mathscr{J}}^\alpha(x_1) \right) &= \frac{1}{2} \, \G^{\mu\alpha}\left( \hat{\mathscr{J}}^\nu(x),\, \hat{\mathscr{J}}_\nu(x_1) \right)  , \label{kuboJ1} \\
 \left( \hat \pi_{\perp}^{\mu\nu} (x),\, \hat \pi_{\perp}^{\alpha\beta}(x_1) \right)&= \frac{1}{2} \, \G^{\mu\nu\alpha\beta} \Big( \hat \pi_{\perp}^{\rho\sigma}(x),\, \hat \pi_{\perp\rho\sigma}(x_1) \Big) , \label{kubopi1}\\
\left( \hat m^{\mu\nu} (x),\, m^{\alpha\beta}(x_1) \right) &= \Gc^{\mu\nu\alpha\beta} \Big( \hat m^{\rho\sigma}(x),\, \hat m_{\rho\sigma}(x_1) \Big)  .  \label{kubom1}
\end{align}
Thus we obtain, 
\begin{align}
	\mathscr{K}_1^\mu(x)=\left<\hat{\mathscr{K}}^\mu(x)\right>_1
	&= \bar{\eta}_\parallel(x) \,\bar{\mathscr{Y}}^\mu (x)  , \label{transK1}\\
	\mathscr{J}_1^\mu(x)=\left<\hat{\mathscr{J}}^\mu(x)\right>_1
	&= \bar{\rho}_\parallel(x) \,\bar{\mathscr{X}}^\mu (x)  , \label{transJ1}\\
	\pi_{1\perp}^{\mu\nu} (x)=\left<\hat \pi_{\perp}^{\mu\nu} (x)\right>_1
	&= 2 \,\bar{\eta}_\perp (x)\, \bar\sigma_{\perp}^{\mu\nu}(x)  , \label{transpi1}\\
	m_1^{\mu\nu} (x)=\left<\hat m^{\mu\nu} (x)\right>_1
	&= 	2 \, \bar{\rho}_\perp (x)\, \bar{\chi}^{\mu\nu} (x) , \label{transm1}
\end{align}
where the transport coefficients are defined by
\begin{align}
	\bar{\eta}_\parallel (x) &= \frac{1}{2}\int d^4 x_1\, \left( \hat{\mathscr{K}}^\nu(x),\, \hat{\mathscr{K}}_\nu(x_1) \right) , \\
	\bar{\rho}_\parallel (x) &= \frac{1}{2}\int d^4 x_1\, \left( \hat{\mathscr{J}}^\nu(x),\, \hat{\mathscr{J}}_\nu(x_1) \right) , \\
	\bar{\eta}_\perp (x)&= \frac{1}{4}\int d^4 x_1\, \Big( \hat \pi_{\perp}^{\rho\sigma}(x),\, \hat \pi_{\perp\rho\sigma}(x_1) \Big) , \\
	\bar{\rho}_\perp (x)&=  \frac{1}{2}\int d^4 x_1\, \Big( \hat m^{\rho\sigma}(x),\, \hat m_{\rho\sigma}(x_1) \Big) .
\end{align}

Here, we like to emphasize that the relations~\eqref{kuboK1}-\eqref{kubom1} are valid only to leading order in the gradient expansion. 
In principle, $u^{\mu}$ and $b^{\mu}$ vary in spacetime, so the projectors at $x$ and $x_1$ need not coincide. 
However, within linear response, when evaluating Kubo correlators \eqref{kuboK1}-\eqref{kubom1} in first order hydrodynamic theory, we can observe that the thermodynamic forces in \eqref{K1}-\eqref{m1} are already first order in gradients. 
Therefore, we may approximate $u^{\mu}(x)\simeq u^{\mu}(x_1)$ and $b^{\mu}(x)\simeq b^{\mu}(x_1)$ in the  leading order approximations, and the projectors in \eqref{kuboK1}-\eqref{kubom1} need not be tied to a specific point. 
However, if one goes beyond first order, e.g., to compute second order hydrodynamic corrections, one must keep distinct projectors $\G^{\mu\nu}(x)$ and $\G^{\mu\nu}(x_1)$ and Taylor-expand $\G^{\mu\nu}(x_1)$ around $x_1=x$. 
The resulting higher-order terms generate nonlocal contributions and new couplings between thermodynamic forces and dissipative tensors. 
These effects are naturally incorporated in the nonlocal correction, Eq.~\eqref{nlc}, via the evaluation of $\big[\partial_\lambda \hat{\mathcal{C}}(x_1)\big]_{x_1=x}$.

	%%%%%%%%%%%%%%%%%%%%%%%%%%%%%%%%%%%%%%%%%%%%%%%%%%%%%%%%%%%%%%%%%%%%%%%%%%%%%%%%%%%%%%
	%%%%%%%%%%%%%%%%%%%%%%%%%%%%%%%%%%%%%%%%%%%%%%%%%%%%%%%%%%%%%%%%%%%%%%%%%%%%%%%%%%%%%%

	%%%%%%%%%%%%%%%%%%%%%%%%%%%%%%%%%%%%%%%%%%%%%%%%%%%%%%%%%%%%%%%%%%%%%%%%%%%%%%%%%%%%%%
	%%%%%%%%%%%%%%%%%%%%%%%%%%%%%%%%%%%%%%%%%%%%%%%%%%%%%%%%%%%%%%%%%%%%%%%%%%%%%%%%%%%%%
	
	%

	\section{Second order magnetohydrodynamics}

	In the conventional way, second-order magnetohydrodynamics has been derived in both the non-resistive~\cite{Denicol:2018rbw} and resistive~\cite{Denicol:2019iyh} limits from the Boltzmann-Vlasov equation using the method of moments. Second-order formulations in the non-resistive and resistive limits have also been obtained within the relaxation-time approximation in Refs.~\cite{Panda:2020zhr,Panda:2021pvq}. In this section, we employ Zubarev’s nonequilibrium statistical operator formalism to derive second-order relativistic magnetohydrodynamics based entirely on the underlying conserved quantities and symmetries.

	\subsection{Second-order correction and evolution equation for shear stress tensor ($\hat{\pi}^{\mu\nu}_\perp$)}
	
	\subsubsection{Nonlocal correction from the two-point correlation function: $\big< \hat{\pi}_{\perp}^{\mu\nu} (x)\big>^{\text{(2),NL}}_2$}
	Using~\eqref{nlc}, the nonlocal correction is given by
	\begin{align}
		\Big< \hat{\pi}_{\perp}^{\mu\nu} (x)\Big>^{\text{(2),NL}}_2
		&=\int d^4 x_1 \Big( \hat{\pi}_{\perp\,\mu\nu} (x), [\partial_\lambda \hat{\mathcal{C}}(x_1)]_{x_1=x} \Big)(x_1-x)^\lambda. \label{pinl1}
	\end{align}
	We now employ Eq.~\eqref{delC3} to evaluate $\big[\partial_\lambda \hat{\mathcal{C}}(x_1)\big]_{x_1=x}$ by setting all hydrodynamic variables to $x_1=x$ while leaving the operators at $x_1$. Invoking Curie’s symmetry principle, that correlations can exist only between operators of the same rank and parity, we find that only the term containing $\hat{\pi}_{\perp}^{\mu\nu}$ in 
	$\big[\partial_\lambda \hat{\mathcal{C}}(x_1)\big]_{x_1=x}$ contributes to the integral in Eq.~\eqref{pinl1}. This yields
\begin{align}
	\Big< \hat{\pi}_{\perp}^{\mu\nu} (x)\Big>^{\text{(2),NL}}_2
	&=\int d^4 x_1 \Big( \hat{\pi}_{\perp}^{\mu\nu} (x), \hat{\pi}_\perp^{\alpha\beta}(x_1) [\partial_\lambda \bar{\sigma}_{\perp\alpha\beta} -B'\bar{\mathscr{X}}_\alpha \partial_\lambda u_\beta -  \bar{\mathscr{Y}}_\alpha \partial_\lambda b_\beta  ]_x  \Big)(x_1-x)^\lambda \nonumber\\
	&= \big[\partial_\lambda \bar{\sigma}_{\perp\alpha\beta}(x) -B'(x)\bar{\mathscr{X}}_\alpha (x) \partial_\lambda u_\beta (x) -  \bar{\mathscr{Y}}_\alpha (x) \partial_\lambda b_\beta (x) \big]\int d^4 x_1 \Big( \hat{\pi}_{\perp}^{\mu\nu} (x), \hat{\pi}_\perp^{\alpha\beta}(x_1) \Big)(x_1-x)^\lambda  .  \label{pinl2}
\end{align}
Since the integral in Eq.~\eqref{pinl2} already contains a factor that is second order in hydrodynamics gradients, we may safely approximate the Kubo correlators using Eq.~\eqref{kubopi1}. 
Accordingly, Eq.~\eqref{pinl2} simplifies to
\begin{align}
	\Big< \hat{\pi}_{\perp}^{\mu\nu} (x)\Big>^{\text{(2),NL}}_2
	&=   2\,\G^{\mu\nu\alpha\beta}\big[\partial_\lambda \bar{\sigma}_{\perp\alpha\beta}(x) -B'(x)\bar{\mathscr{X}}_\alpha (x) \partial_\lambda u_\beta (x) -  \bar{\mathscr{Y}}_\alpha (x) \partial_\lambda b_\beta (x) \big]\, \mathfrak{a}_\pi^\lambda (x)  , \label{pinl3}
\end{align}
where 
\begin{align}
	\mathfrak{a}_\pi^\lambda (x)= \frac{1}{4}\int d^4 x_1   \Big( \hat \pi_{\perp}^{\rho\sigma}(x),\, \hat \pi_{\perp\rho\sigma}(x_1) \Big)(x_1-x)^\lambda  .
\end{align}
Here, $\mathfrak{a}_\pi^\lambda$ can be further simplified with the help of~\eqref{freq_dep_trans} in Appendix~\ref{correlators}. We get
\begin{align}
	\mathfrak{a}_\pi^\lambda&= i u^\lambda \lim_{\omega \to 0}\frac{d}{d \omega}\bar{\eta}_{{}_{\pi\pi}}(\omega) ,
\end{align}
where $\bar{\eta}_{{}_{\pi\pi}}(\omega)$ is the frequency dependent transport coefficient, given by
\begin{align}
	\bar{\eta}_{{}_{\pi\pi}}(\omega)=\frac{1}{4}\int d^4 x_1  e^{i\omega(t-t_1)} \Big( \hat \pi_{\perp}^{\rho\sigma}(x),\, \hat \pi_{\perp\rho\sigma}(x_1) \Big) .
\end{align}
With this, Eq.~\eqref{pinl3} takes the simplified form,
\begin{align}
		\big< \hat{\pi}_{\perp}^{\mu\nu}\big>^{\text{(2),NL}}_2 &=2\big[{}^<D\bar{\sigma}_{\perp}^{\mu\nu>} -B'\bar{\mathscr{X}}^{<\mu} D u^{\nu>} -  \bar{\mathscr{Y}}^{<\mu} D b^{\nu>}\big]\mathscr{W}_{\pi\pi} . \label{final_shear_nonlocal}
\end{align}
Here, for brevity, we omitted the explicit arguments. Also, we have defined
\begin{align}
	\mathscr{W}_{\pi\pi} = i\lim_{\omega \to 0}\frac{d}{d \omega}\bar{\eta}_{{}_{\pi\pi}}(\omega) .
\end{align}

	%%%%%%%%%%%%%%%%%%%%%%%%%%%%%%%%%%%%%%%%%%%%%%%%%%%%%%%%%%%%%%%%%%%%%%%%%%%%%

	\subsubsection{Local correction from the two-point correlation function using extended thermodynamic forces: $\big< \hat{\pi}_{\perp}^{\mu\nu} (x)\big>^{\text{(2),ET}}_2$}\label{secPi3P}

Since the extended thermodynamic forces, $\hat{\mathcal{C}}_{S}$, contains no rank-two operators, Curie’s symmetry principle implies that it cannot couple to the rank-two tensor operator $\hat{\pi}_{\perp}^{\mu\nu}$. 
Consequently, the corresponding two-point correlation vanishes:
\begin{align}
\Big< \hat{\pi}_{\perp}^{\mu\nu} (x)\Big>^{\text{(2),ET}}_2 =	\int d^4x_1\,\big(\hat{\pi}_{\perp}^{\mu\nu}(x),\,\hat{\mathcal{C}}_{S}(x_1)\big)=0. \label{final_shear_extended}
\end{align}

	%%%%%%%%%%%%%%%%%%%%%%%%%%%%%%%%%%%%%%%%%%%%%%%%%%%%%%%%%%%%%%%%%%%%%%%%%%%%%%

\subsubsection{Local correction from the three-point correlation function: $\big< \hat{\pi}_{\perp}^{\mu\nu} (x)\big>^{\text{(3)}}_2$}

Unlike the two-point correlators—where, by Curie’s principle, only operators of the same tensorial rank couple—three-point correlation functions can admit couplings among tensors of different ranks. 
Such mixed-rank couplings are allowed provided the correlator can be decomposed into irreducible tensors that reproduce the symmetry, parity, and exchange properties of the operators entering the correlator.

With these symmetry considerations in hand, we now turn to the evaluation of the three-point correlators. Using~\eqref{tpc} and~\eqref{CF}, we get
\begin{align}
		\Big< \hat{\pi}_{\perp}^{\mu\nu} (x)\Big>^{\text{(3)}}_2 = 	\int d^4x_1 d^4x_2\,\big(\hat{\pi}_{\perp}^{\mu\nu}(x),\,\hat{\mathcal{C}}_{F}(x_1), \,\hat{\mathcal{C}}_{F}(x_2)\big) \big|_{local} . \label{pi3local}
\end{align}
To avoid confusion, we use 
\begin{align}
	\hat{\mathcal{C}}_{F}(x_1) &= -\bar{\theta}_{\perp}\,\hat{P}_{\perp}
	-\bar{\theta}_{\parallel}\,\hat{P}_{\parallel}
	+ \hat{\mathscr{K}}^{\rho}\,\bar{\mathscr{Y}}_\rho
	+ \hat{\mathscr{J}}^{\rho}\, \bar{\mathscr{X}}_\rho
	+ \hat \pi_{\perp}^{\rho\sigma} \bar{\sigma}_{\perp \rho\sigma}
	+ \hat m^{\rho\sigma} \bar{\chi}_{ \rho\sigma} , \\
	\hat{\mathcal{C}}_{F}(x_2) &= -\bar{\theta}_{\perp}\,\hat{P}_{\perp}
	-\bar{\theta}_{\parallel}\,\hat{P}_{\parallel}
	+ \hat{\mathscr{K}}^{\alpha}\,\bar{\mathscr{Y}}_\alpha
	+ \hat{\mathscr{J}}^{\alpha}\, \bar{\mathscr{X}}_\alpha
	+ \hat \pi_{\perp}^{\alpha\beta} \bar{\sigma}_{\perp \alpha\beta}
	+ \hat m^{\alpha\beta} \bar{\chi}_{\alpha\beta} .
\end{align}
As discussed in the Section~\ref{FOMHD}, we extract the local terms by evaluating the hydrodynamic quantities present in $\hat{\mathcal{C}}_{F}(x_1)$ and $\hat{\mathcal{C}}_{F}(x_2)$ at $x_1=x$ and $x_2=x$, respectively. As a result, each term in Eq.~\eqref{pi3local} carries a second-order gradient term extracted from the hydrodynamic variables in $\hat{\mathcal{C}}_{F}$. 
Consequently, all Kubo correlators can be evaluated to leading order as mention in section~\ref{FOMHD}, and we can use the projectors (irreducible tensors) to write a tensor decomposition without specifying a particular spacetime point. 
In this approximation, a general covariant tensor decomposition for a sixth-rank Kubo correlator, transverse to both fluid velocity and magnetic field, takes the form
\begin{align}
	\Big( \hat{X}^{\mu\nu}(x), \hat{Y}^{\rho\sigma}(x_1),\hat{Z}^{\alpha\beta}(x_2) \Big)  &= a_1\,\G^{\mu\nu}\G^{\rho\alpha}\G^{\sigma\beta} +a_2 \, \G^{\mu\nu}\G^{\rho\beta}\G^{\sigma\alpha} +b_1 \, \G^{\rho\sigma}\G^{\mu\alpha}\G^{\nu\beta} +b_2\, \G^{\rho\sigma}\G^{\mu\beta}\G^{\nu\alpha}\nonumber\\
	&~~~ +c_1 \,\G^{\alpha\beta}\G^{\mu\rho}\G^{\nu\sigma} +c_2 \, \G^{\alpha\beta}\G^{\mu\sigma}\G^{\nu\rho} +d\, \G^{\mu\nu}\G^{\rho\sigma}\G^{\alpha\beta} \nonumber\\
	&~~~ +e_1 \,\G^{\mu\rho}\G^{\nu\alpha}\G^{\sigma\beta} +e_2\, \G^{\mu\rho}\G^{\nu\beta}\G^{\sigma\alpha} +e_3\, \G^{\mu\sigma}\G^{\nu\alpha}\G^{\rho\beta} +e_4\, \G^{\mu\sigma}\G^{\nu\beta}\G^{\rho\alpha} \nonumber\\
	&~~~ +e_5\, \G^{\mu\alpha}\G^{\nu\rho}\G^{\sigma\beta} +e_6\, \G^{\mu\alpha}\G^{\nu\sigma}\G^{\rho\beta} +e_7\, \G^{\mu\beta}\G^{\nu\rho}\G^{\sigma\alpha} +e_8\, \G^{\mu\beta}\G^{\nu\sigma}\G^{\rho\alpha}, \label{general_six_decom}
\end{align}
with scalar coefficients $a_i$, $b_i$, $c_i$, $d$, and $e_i$ fixed by the symmetries of the operators. 

Likewise, a general decomposition for fourth-rank Kubo correlators of type 
$\big( \hat{X}^{\mu\nu}(x), \hat{Y}^{\rho\sigma}(x_1), \hat{z}(x_2) \big)$
can be written as
\begin{align}
	A_1\, \G^{\mu\nu}\G^{\rho\sigma}+ A_2\,\G^{\mu\rho}\G^{\nu\sigma} + A_3\,\G^{\mu\sigma}\G^{\nu\rho} , \label{forthranklinear}
\end{align}
where $A_i$ are scalar functions determined by the underlying symmetries of the operators. A similar decomposition can also be written for the correlation function of type $\big( \hat{X}^{\mu\nu}(x), \hat{V}_1^{\rho}(x_1), \hat{V}_2^{\alpha}(x_2) \big)$. Additionally, the only available transverse object is the rank-two projector 
$\G^{\mu\nu}=g^{\mu\nu}-u^{\mu}u^{\nu}+b^{\mu}b^{\nu}$, 
which is orthogonal to both fluid velocity and magnetic field vector. 
Since one cannot construct a rank-three tensor from $\G^{\mu\nu}$ alone that preserves the required transversality and symmetry properties. Therefore, the correlators of the form 
$\big( \hat{X}^{\mu\nu}(x), \hat{V}_1^{\rho}(x_1), \hat{z}(x_2) \big)$ 
vanish.

Having established these constraints, we use the symmetry properties of the relevant dissipative tensors 
($\hat{\pi}_{\perp}^{\mu\nu}=\hat{\pi}_{\perp}^{\nu\mu}$, $\hat{\pi}_{\perp\,\mu}^{\ \ \mu}=0$, 
$\hat{m}^{\mu\nu}=-\hat{m}^{\nu\mu}$) to determine the coefficients 
$a_{1},a_{2},b_{1},b_{2},c_{1},c_{2}$, $e_{1}\text{--}e_{8}$, and $A_{1}\text{--}A_{3}$.

Substituting Eq.~\eqref{CF} into Eq.~\eqref{pi3local}, and imposing the constraint that only correlators with overall parity $+1$ survive, we obtain the following six nonvanishing correlators:
\begin{align}
	\big( \hat\pi_{\perp}^{\mu\nu}(x),\, \hat{\mathscr{J}}^\rho(x_1), \, \hat{\mathscr{J}}^\alpha(x_2) \big) &= \frac{1}{2}\G^{\mu\nu\rho\alpha} \big( \hat\pi_{\perp}^{\lambda\eta}(x),\, \hat{\mathscr{J}}_\lambda(x_1), \, \hat{\mathscr{J}}_\eta(x_2) \big) , \\
	\big( \hat\pi_{\perp}^{\mu\nu}(x),\, \hat{\mathscr{K}}^\rho(x_1), \, \hat{\mathscr{K}}^\alpha(x_2) \big) &= \frac{1}{2}\G^{\mu\nu\rho\alpha}\big( \hat\pi_{\perp}^{\lambda\eta}(x),\, \hat{\mathscr{K}}_\lambda(x_1), \, \hat{\mathscr{K}}_\eta(x_2) \big) , \\	
	\big( \hat\pi_{\perp}^{\mu\nu}(x),\, \hat\pi_{\perp}^{\rho\sigma}(x_1), \, \hat{P}_\perp(x_2) \big) &=\frac{1}{2}\G^{\mu\nu\rho\sigma} \big( \hat\pi_{\perp}^{\lambda\eta}(x),\, \hat\pi_{\perp\lambda\eta}(x_1), \, \hat{P}_\perp(x_2) \big) , \\
	\big( \hat\pi_{\perp}^{\mu\nu}(x),\, \hat{P}_\perp(x_1), \hat\pi_{\perp}^{\alpha\beta}(x_2) \,  \big) &= \frac{1}{2}\G^{\mu\nu\alpha\beta}\big( \hat\pi_{\perp}^{\lambda\eta}(x),\, \hat{P}_\perp(x_1), \, \hat\pi_{\perp\lambda\eta}(x_2) \big) , \\
	\big( \hat\pi_{\perp}^{\mu\nu}(x),\, \hat\pi_{\perp}^{\rho\sigma}(x_1), \, \hat{P}_\parallel(x_2) \big) &= \frac{1}{2}\G^{\mu\nu\rho\sigma}\big( \hat\pi_{\perp}^{\lambda\eta}(x),\, \hat\pi_{\perp\lambda\eta}(x_1), \, \hat{P}_\parallel(x_2) \big) , \\
	\big( \hat\pi_{\perp}^{\mu\nu}(x),\, \hat{P}_\parallel(x_1), \hat\pi_{\perp}^{\alpha\beta}(x_2) \,  \big) &=\frac{1}{2}\G^{\mu\nu\alpha\beta} \big( \hat\pi_{\perp}^{\lambda\eta}(x),\, \hat{P}_\parallel(x_1), \, \hat\pi_{\perp\lambda\eta}(x_2) \big) .
\end{align}
The correlators $\big( \hat\pi_{\perp}^{\mu\nu}(x),\, \hat\pi_{\perp}^{\rho\sigma}(x_1),\, \hat\pi_{\perp}^{\alpha\beta}(x_2) \big)$ and $\big( \hat\pi_{\perp}^{\mu\nu}(x),\, \hat m^{\rho\sigma}(x_1),\, \hat m^{\alpha\beta}(x_2) \big)$ also follow the parity constraint but they vanish due to their particular tensor structure. The complete derivation is provided in Appendix~\ref{three_point_correlatos}. 
Collecting all the nonvanishing correlators, we arrive at
\begin{align}
	\big\langle \hat{\pi}_{\perp}^{\mu\nu}(x)\big\rangle^{\text{(3)}}_{2} = \mathfrak{b}_1^{\mu\nu} (x) + \mathfrak{b}_2^{\mu\nu} (x) + \mathfrak{b}_3^{\mu\nu} (x) +\mathfrak{b}_4^{\mu\nu} (x) + \mathfrak{b}_5^{\mu\nu} (x) +\mathfrak{b}_6^{\mu\nu} (x) .
\end{align} 
Here,
\begin{align} 
	\mathfrak{b}_1^{\mu\nu}(x)&=\frac{1}{2}\G^{\mu\nu\rho\alpha}\,\bar{\mathscr{X}}_\rho(x)\,\bar{\mathscr{X}}_\alpha(x)\int d^4x_1 \,d^4x_2 \,	 \big( \hat\pi_{\perp}^{\lambda\eta}(x),\, \hat{\mathscr{J}}_\lambda(x_1), \, \hat{\mathscr{J}}_\eta(x_2) \big) , \\
	\mathfrak{b}_2^{\mu\nu}(x)&=\frac{1}{2}\G^{\mu\nu\rho\alpha}\, \bar{\mathscr{Y}}_\rho(x)\,\bar{\mathscr{Y}}_\alpha(x)\int d^4x_1 \,d^4x_2 \,	 \big( \hat\pi_{\perp}^{\lambda\eta}(x),\, \hat{\mathscr{K}}_\lambda(x_1), \, \hat{\mathscr{K}}_\eta(x_2) \big) , \\	
	\mathfrak{b}_3^{\mu\nu}(x)&=-\frac{1}{2}\, \bar{\sigma}^{\mu\nu}_{\perp}(x)\,\bar{\theta}_\perp(x)\int d^4x_1 \,d^4x_2 \,	\big( \hat\pi_{\perp}^{\lambda\eta}(x),\, \hat\pi_{\perp\lambda\eta}(x_1), \, \hat{P}_\perp(x_2) \big) , \\
	\mathfrak{b}_4^{\mu\nu}(x)&=-\frac{1}{2}\,\bar{\sigma}^{\mu\nu}_{\perp}(x)\,\bar{\theta}_\perp(x)\int d^4x_1 \,d^4x_2 \,	\big( \hat\pi_{\perp}^{\lambda\eta}(x),\, \hat{P}_\perp(x_1), \hat\pi_{\perp\lambda\eta}(x_2) \,  \big) , \\
	\mathfrak{b}_5^{\mu\nu}(x)&=-\frac{1}{2}\,\bar{\sigma}^{\mu\nu}_{\perp}(x)\,\bar{\theta}_\parallel(x)\int d^4x_1 \,d^4x_2 \,	\big( \hat\pi_{\perp}^{\lambda\eta}(x),\, \hat\pi_{\perp\lambda\eta}(x_1), \, \hat{P}_\parallel(x_2) \big) , \\
	\mathfrak{b}_6^{\mu\nu}(x)&=-\frac{1}{2}\,\bar{\sigma}^{\mu\nu}_{\perp}(x)\,\bar{\theta}_\parallel(x)\int d^4x_1 \,d^4x_2 \,	\big( \hat\pi_{\perp}^{\lambda\eta}(x),\, \hat{P}_\parallel(x_1), \hat\pi_{\perp\lambda\eta}(x_2) \,  \big).
\end{align}
Inferring to~\eqref{threepointsymmetry}, we can conclude that $\mathfrak{b}_3^{\mu\nu}=\mathfrak{b}_4^{\mu\nu}$ and $\mathfrak{b}_5^{\mu\nu}=\mathfrak{b}_6^{\mu\nu}$. Thus we obtain
\begin{align}
	\big\langle \hat{\pi}_{\perp}^{\mu\nu}\big\rangle^{\text{(3)}}_{2} = \bar{\mathscr{X}}^{<\mu}\,\bar{\mathscr{X}}^{\nu>}\, \bar{\eta}_{{}_{\pi\mathscr{J}\mathscr{J}}}
	+ \bar{\mathscr{Y}}^{<\mu}\,\bar{\mathscr{Y}}^{\nu>}\, \bar{\eta}_{{}_{\pi\mathscr{K}\mathscr{K}}}
	- 2\,\bar{\sigma}^{\mu\nu}_{\perp}\,\bar{\theta}_\perp \, \bar{\eta}_{{}_{\pi\pi P_\perp}}
	- 2\,\bar{\sigma}^{\mu\nu}_{\perp}\,\bar{\theta}_\parallel \, \bar{\eta}_{{}_{\pi\pi P_\parallel}} . \label{final_shear_three_point}
\end{align} 
Here, for brevity, we omitted explicit arguments. The transport coefficients appearing in~\eqref{final_shear_three_point} are defined by
\begin{align}
    \bar{\eta}_{{}_{\pi\mathscr{J}\mathscr{J}}}(x) &=\frac{1}{2}\int d^4x_1 \,d^4x_2 \,	 \big( \hat\pi_{\perp}^{\lambda\eta}(x),\, \hat{\mathscr{J}}_\lambda(x_1), \, \hat{\mathscr{J}}_\eta(x_2) \big) , \\
    \bar{\eta}_{{}_{\pi\mathscr{K}\mathscr{K}}}(x) &=\frac{1}{2}\int d^4x_1 \,d^4x_2 \,	 \big( \hat\pi_{\perp}^{\lambda\eta}(x),\, \hat{\mathscr{K}}_\lambda(x_1), \, \hat{\mathscr{K}}_\eta(x_2) \big) , \\
    \bar{\eta}_{{}_{\pi\pi P_\perp}}(x) &=\frac{1}{2}\int d^4x_1 \,d^4x_2 \,	\big( \hat\pi_{\perp}^{\lambda\eta}(x),\, \hat\pi_{\perp\lambda\eta}(x_1), \, \hat{P}_\perp(x_2) \big) , \\
     \bar{\eta}_{{}_{\pi\pi P_\parallel}}(x) &=\frac{1}{2}\int d^4x_1 \,d^4x_2 \,	\big( \hat\pi_{\perp}^{\lambda\eta}(x),\, \hat\pi_{\perp\lambda\eta}(x_1), \, \hat{P}_\parallel(x_2) \big).
\end{align}

\subsubsection{Full second order expression and the evolution equation for the shear stress tensor}

Combining the first-order correction from~\eqref{transpi1} and second-order expression from~\eqref{final_shear_nonlocal},~\eqref{final_shear_extended}, and~\eqref{final_shear_three_point}, we get the full expression of shear stress tensor up to second order, given by
\begin{align}
	\pi_\perp^{\mu\nu} =& \, 2 \,\bar{\eta}_\perp \bar\sigma_{\perp}^{\mu\nu} + 2\big[{}^<D\bar{\sigma}_{\perp}^{\mu\nu>} -B'\bar{\mathscr{X}}^{<\mu} D u^{\nu>} -  \bar{\mathscr{Y}}^{<\mu} D b^{\nu>}\big]\mathscr{W}_{\pi\pi} \nonumber\\
	& +  \bar{\mathscr{X}}^{<\mu}\,\bar{\mathscr{X}}^{\nu>}\, \bar{\eta}_{{}_{\pi\mathscr{J}\mathscr{J}}}
	+ \bar{\mathscr{Y}}^{<\mu}\,\bar{\mathscr{Y}}^{\nu>}\, \bar{\eta}_{{}_{\pi\mathscr{K}\mathscr{K}}}
	- 2\,\bar{\sigma}^{\mu\nu}_{\perp}\,\bar{\theta}_\perp \, \bar{\eta}_{{}_{\pi\pi P_\perp}}
	- 2\,\bar{\sigma}^{\mu\nu}_{\perp}\,\bar{\theta}_\parallel \, \bar{\eta}_{{}_{\pi\pi P_\parallel}} . \label{fullpi}
\end{align}

In order to derive the evolution equation for $\pi^{\mu\nu}_\perp$, we employ the first-order Navier-Stokes relation~\eqref{transpi1} to approximate the term ${}^<D\bar{\sigma}_{\perp}^{\mu\nu>}$, replacing $2\bar{\sigma}_\perp^{\mu\nu}$ with $\pi_\perp^{\mu\nu}\bar{\eta}_\perp^{-1}$. This leads to
\begin{align}
	2 \mathscr{W}_{\pi\pi} {}^<D\bar{\sigma}_{\perp}^{\mu\nu>}  \simeq \frac{ \mathscr{W}_{\pi\pi}}{\bar{\eta}_\perp} \big[ \dot{\pi}^{\mu\nu}_\perp - \pi^{\mu\nu}_\perp D \ln \bar{\eta}_\perp \big] . \label{nav_appro_pi}
\end{align}
Here, we have defined $\dot{\pi}^{\mu\nu}_\perp = \G^{\mu\nu}_{\h\h\alpha\beta}D \pi^{\alpha\beta}_\perp$. We observe that a characteristic timescale naturally emerges at this stage. We therefore define
\begin{align}
	 \frac{ \mathscr{W}_{\pi\pi}}{\bar{\eta}_\perp} = -\tau_\pi . \label{timepi}
\end{align}
 Substituting~\eqref{nav_appro_pi} and~\eqref{timepi} in~\eqref{fullpi}, we obtain the evolution equation for $\pi^{\mu\nu}_\perp$, given by
 \begin{align}
 	\pi_\perp^{\mu\nu} +\tau_\pi \dot{\pi}^{\mu\nu}_\perp =& \, 2 \,\bar{\eta}_\perp \bar\sigma_{\perp}^{\mu\nu} + \tau_\pi \pi^{\mu\nu}_\perp D \ln \bar{\eta}_\perp - 2\big[ B'\bar{\mathscr{X}}^{<\mu} D u^{\nu>} +  \bar{\mathscr{Y}}^{<\mu} D b^{\nu>}\big]\mathscr{W}_{\pi\pi} \nonumber\\
 	& +  \bar{\mathscr{X}}^{<\mu}\,\bar{\mathscr{X}}^{\nu>}\, \bar{\eta}_{{}_{\pi\mathscr{J}\mathscr{J}}}
 	+ \bar{\mathscr{Y}}^{<\mu}\,\bar{\mathscr{Y}}^{\nu>}\, \bar{\eta}_{{}_{\pi\mathscr{K}\mathscr{K}}}
 	- 2\,\bar{\sigma}^{\mu\nu}_{\perp}\,\bar{\theta}_\perp \, \bar{\eta}_{{}_{\pi\pi P_\perp}}
 	- 2\,\bar{\sigma}^{\mu\nu}_{\perp}\,\bar{\theta}_\parallel \, \bar{\eta}_{{}_{\pi\pi P_\parallel}}  . \label{evopi}
 \end{align}

\subsection{Second-order correction and evolution equation for $\hat m^{\mu\nu}$}

\subsubsection{Nonlocal correction from the two-point correlation function: $\big< \hat{m}^{\mu\nu} (x)\big>^{\text{(2),NL}}_2$}

Using~\eqref{nlc}, the nonlocal correction to the statistical average of $\hat m^{\mu\nu}$ is given by
\begin{align}
	\Big< \hat{m}^{\mu\nu} (x)\Big>^{\text{(2),NL}}_2
	&=\int d^4 x_1 \Big( \hat{m}^{\mu\nu} (x), [\partial_\lambda \hat{\mathcal{C}}(x_1)]_{x_1=x} \Big)(x_1-x)^\lambda. \label{mnl1}
\end{align}
Following the prescription for evaluating $\big[\partial_\lambda \hat{\mathcal{C}}(x_1)\big]_{x_1=x}$ and using Curie’s symmetry principle, which implies that $\hat{m}^{\mu\nu}$ couples only to $\hat{m}^{\mu\nu}$ in Eq.~\eqref{delC3}, we obtain
\begin{align}
	\Big< \hat{m}^{\mu\nu} (x)\Big>^{\text{(2),NL}}_2
	&=\big[ \partial_\lambda \bar{\chi}_{\alpha\beta} (x) + \bar{\mathscr{X}}_\alpha (x) \partial_\lambda b_\beta (x) + H (x)\bar{\mathscr{Y}}_\alpha (x) \partial_\lambda u_\beta (x) \big] \int d^4 x_1 \Big( \hat{m}^{\mu\nu} (x), \hat{m}^{\alpha\beta} (x_1) \Big)(x_1-x)^\lambda. \label{mnl2}
\end{align}
Using~\eqref{kubom1}, we further simplify~\eqref{mnl2} as:
\begin{align}
		\Big< \hat{m}^{\mu\nu} (x)\Big>^{\text{(2),NL}}_2
		&=2\,\Gc^{\mu\nu\alpha\beta}\big[ \partial_\lambda \bar{\chi}_{\alpha\beta} (x) + \bar{\mathscr{X}}_\alpha (x) \partial_\lambda b_\beta (x) + H (x)\bar{\mathscr{Y}}_\alpha (x) \partial_\lambda u_\beta (x) \big] \mathfrak{a}_m^\lambda (x), \label{mnl3}
\end{align}
where we have defined
\begin{align}
	\mathfrak{a}_m^\lambda (x)=\frac{1}{2}\int d^4 x_1 \Big( \hat{m}^{\lambda\eta} (x), \hat{m}_{\lambda\eta} (x_1) \Big)(x_1-x)^\lambda .
\end{align}
Employing,~\eqref{freq_dep_trans} from Appendix~\ref{correlators}, we can write
\begin{align}
	\mathfrak{a}_m^\lambda &= u^\lambda \mathscr{W}_{m m} = i u^\lambda \lim_{\omega \to 0}\frac{d}{d \omega}\bar{\rho}_{{}_{m m}}(\omega) . \label{freqm}
\end{align}
Here, the frequency dependent transport coefficient $\bar{\rho}_{{}_{m m}}(\omega)$ is defined by
\begin{align}
	 \bar{\rho}_{{}_{m m}}(\omega) &= \frac{1}{2}\int d^4 x_1  e^{i\omega(t-t_1)}\Big( \hat{m}^{\lambda\eta} (x), \hat{m}_{\lambda\eta} (x_1) \Big) .
\end{align}
Thus, employing~\eqref{freqm}, we simplify~\eqref{mnl3} to
\begin{align}
	\big< \hat{m}^{\mu\nu}\big>^{\text{(2),NL}}_2
	&=2\big[ {}^{\cll}D \bar{\chi}^{\mu\nu\crr}  + \bar{\mathscr{X}}^{\cll \mu}  D b^{\nu\crr}  + H \bar{\mathscr{Y}}^{\cll \mu} D u^{\nu\crr} \big] \mathscr{W}_{m m}. \label{final_m_nonlocal}
\end{align}
For brevity, explicit arguments are omitted.

\subsubsection{Local correction from the two-point correlation function using extended thermodynamic forces: $\big< \hat{m}^{\mu\nu} (x)\big>^{\text{(2),ET}}_2$}

Because the extended thermodynamic force $\hat{\mathcal{C}}_{S}$
does not contain any rank-two tensor components, Curie’s symmetry principle ensures that it cannot couple to the rank-two shear tensor operator $\hat{m}^{\mu\nu}$. Therefore, the associated two-point Kubo correlation identically vanishes:
\begin{align}
	\Big< \hat{m}^{\mu\nu} (x)\Big>^{\text{(2),ET}}_2 =	\int d^4x_1\,\big(\hat{m}^{\mu\nu}(x),\,\hat{\mathcal{C}}_{S}(x_1)\big)=0. \label{final_m_extended}
\end{align}

\subsubsection{Local correction from the three-point correlation function: $\big< \hat{m}^{\mu\nu} (x)\big>^{\text{(3)}}_2$}

Using~\eqref{tpc}, we obtain
\begin{align}
	\Big< \hat{m}^{\mu\nu} (x)\Big>^{\text{(3)}}_2 = 	\int d^4x_1 d^4x_2\,\big(\hat{m}^{\mu\nu}(x),\,\hat{\mathcal{C}}_{F}(x_1), \,\hat{\mathcal{C}}_{F}(x_2)\big) \big|_{local} , \label{m3local}
\end{align}
where $\hat{\mathcal{C}}_{F}$ is defined in~\eqref{CF}. We extract the local contributions by setting $x_1=x$ and $x_2=x$ in the thermodynamic forces appearing in $\hat{\mathcal{C}}_F(x_1)$ and $\hat{\mathcal{C}}_F(x_2)$, and following the procedure outlined in Sec.~\ref{secPi3P}, we find six nonvanishing three-point correlators that contribute to the local correction from the three-point function. Thus, we can write
\begin{align}
	\Big< \hat{m}^{\mu\nu} (x)\Big>^{\text{(3)}}_2 = \mathfrak{c}^{\mu\nu}_1 (x)+\mathfrak{c}^{\mu\nu}_2 (x)+\mathfrak{c}^{\mu\nu}_3 (x)+\mathfrak{c}^{\mu\nu}_4 (x)+\mathfrak{c}^{\mu\nu}_5 (x)+\mathfrak{c}^{\mu\nu}_6 (x)  . \label{m3local2}
\end{align}
Here,
\begin{align}
	\mathfrak{c}^{\mu\nu}_1 (x) =& \,\Gc^{\mu\nu\rho\alpha}\,\bar{\mathscr{X}}_\rho(x)\,\bar{\mathscr{Y}}_\alpha(x)\int d^4x_1 \,d^4x_2  \big( \hat m^{\lambda \eta}(x),\, \hat{\mathscr{J}}_\lambda(x_1), \, \hat{\mathscr{K}}_\eta(x_2) \big) , \\
	\mathfrak{c}^{\mu\nu}_2 (x)=& \,\Gc^{\mu\nu\rho\alpha}\,\bar{\mathscr{Y}}_\rho(x)\,\bar{\mathscr{X}}_\alpha(x) \int d^4x_1 \,d^4x_2 \big( \hat m^{\lambda \eta}(x),\, \hat{\mathscr{K}}_\lambda(x_1), \, \hat{\mathscr{J}}_\eta(x_2) \big) , \\
		\mathfrak{c}^{\mu\nu}_3 (x)=& \,- \bar{\chi}^{\mu\nu}(x) \, \bar{\theta}_{\perp}(x)\int d^4x_1 \,d^4x_2 \big( \hat m^{\lambda\eta}(x),\, \hat m_{\lambda\eta}(x_1), \, \hat{P}_\perp(x_2) \big) , \\
		\mathfrak{c}^{\mu\nu}_4 (x)=&\,- \bar{\chi}^{\mu\nu}(x)\, \bar{\theta}_{\perp}(x)\int d^4x_1 \,d^4x_2  \big( \hat m^{\lambda\eta}(x),\, \hat{P}_\perp(x_1), \, \hat m_{\lambda\eta}(x_2) \big) , \\
		\mathfrak{c}^{\mu\nu}_5 (x)=& \,- \bar{\chi}^{\mu\nu}(x) \, \bar{\theta}_{\parallel}(x)\int d^4x_1 \,d^4x_2 \big( \hat m^{\lambda\eta}(x),\, \hat m_{\lambda\eta}(x_1), \, \hat{P}_\parallel(x_2) \big) , \\
		\mathfrak{c}^{\mu\nu}_6 (x) =& \,- \bar{\chi}^{\mu\nu}(x) \, \bar{\theta}_{\parallel}(x) \int d^4x_1 \,d^4x_2  \big( \hat m^{\lambda\eta}(x),\, \hat{P}_\parallel(x_1), \, \hat m_{\lambda\eta}(x_2) \big) .
\end{align}
Inferring to~\eqref{threepointsymmetry}, we conclude that $\mathfrak{c}_1^{\mu\nu}=\mathfrak{c}_2^{\mu\nu}$, $\mathfrak{c}_3^{\mu\nu}=\mathfrak{c}_4^{\mu\nu}$ and $\mathfrak{c}_5^{\mu\nu}=\mathfrak{c}_6^{\mu\nu}$. Thus we obtain
\begin{align}
	\big< \hat{m}^{\mu\nu}\big>^{\text{(3)}}_{2} = 2\,\bar{\mathscr{X}}^{\cll\mu}\,\bar{\mathscr{Y}}^{\nu\crr}\, \bar{\rho}_{{}_{m\mathscr{J}\mathscr{K}}}
	- 2\,\bar{\chi}^{\mu\nu}\,\bar{\theta}_\perp \, \bar{\rho}_{{}_{m m P_\perp}}
	- 2\,\bar{\chi}^{\mu\nu}\,\bar{\theta}_\parallel \, \bar{\rho}_{{}_{m m P_\parallel}}  . \label{final_m_three_point}
\end{align} 
Here, we defined the transport coefficients 
\begin{align}
	\bar{\rho}_{{}_{m\mathscr{J}\mathscr{K}}}(x) &= \int d^4x_1 \,d^4x_2  \big( \hat m^{\lambda \eta}(x),\, \hat{\mathscr{J}}_\lambda(x_1), \, \hat{\mathscr{K}}_\eta(x_2) \big) , \\
	 \bar{\rho}_{{}_{m m P_\perp}}(x) &= \int d^4x_1 \,d^4x_2 \big( \hat m^{\lambda\eta}(x),\, \hat m_{\lambda\eta}(x_1), \, \hat{P}_\perp(x_2) \big) , \\
	 \bar{\rho}_{{}_{m m P_\parallel}}(x) &= \int d^4x_1 \,d^4x_2 \big( \hat m^{\lambda\eta}(x),\, \hat m_{\lambda\eta}(x_1), \, \hat{P}_\parallel(x_2) \big), 
\end{align}
and for brevity, explicit arguments are omitted in~\eqref{final_m_three_point}.

\subsubsection{Full second-order expression and the evolution equation for $m^{\mu\nu}$}

Combining the first-order correction from Eq.~\eqref{transm1} with the second-order contributions obtained from Eqs.~\eqref{final_m_nonlocal}, \eqref{final_m_extended}, and \eqref{final_m_three_point}, we can write
\begin{align}
	m^{\mu\nu} =&  \, 2 \, \bar{\rho}_\perp \bar{\chi}^{\mu\nu} + 2\big[ {}^{\cll}D \bar{\chi}^{\mu\nu\crr}  + \bar{\mathscr{X}}^{\cll \mu}  D b^{\nu\crr}  + H \bar{\mathscr{Y}}^{\cll \mu} D u^{\nu\crr} \big] \mathscr{W}_{m m}\nonumber\\
	& + 2\,\bar{\mathscr{X}}^{\cll\mu}\,\bar{\mathscr{Y}}^{\nu\crr}\, \bar{\rho}_{{}_{m\mathscr{J}\mathscr{K}}}
	- 2\,\bar{\chi}^{\mu\nu}\,\bar{\theta}_\perp \, \bar{\rho}_{{}_{m m P_\perp}}
	- 2\,\bar{\chi}^{\mu\nu}\,\bar{\theta}_\parallel \, \bar{\rho}_{{}_{m m P_\parallel}} . \label{fullm}
\end{align}

Further, we employ the first-order Navier-Srokes relation~\eqref{transm1} to approximate the term ${}^{\cll}D \bar{\chi}^{\mu\nu\crr}$, by setting $2 \bar{\chi}^{\mu\nu}\sim m^{\mu\nu} \bar{\rho}_\perp^{-1}$. This yields,
\begin{align}
	2 \mathscr{W}_{m m} {}^{\cll}D \bar{\chi}^{\mu\nu\crr}  \simeq \frac{ \mathscr{W}_{m m}}{\bar{\rho}_\perp} \big[ \dot{m}^{\mu\nu} - m^{\mu\nu} D \ln \bar{\rho}_\perp \big] . \label{nav_appro_m}
\end{align}
Here, we have defined $\dot{m}^{\mu\nu} = \Gc^{\mu\nu}_{\h\h\alpha\beta}D m^{\alpha\beta}$. At this point, a characteristic relaxation timescale becomes apparent, which we define as
\begin{align}
	\frac{ \mathscr{W}_{m m}}{\bar{\rho}_\perp} = -\tau_m . \label{timem}
\end{align}
Substituting~\eqref{nav_appro_m} and~\eqref{timem} in~\eqref{fullm}, we obtain the evolution equation for $m^{\mu\nu}$, given by
\begin{align}
	m^{\mu\nu} +\tau_m \dot{m}^{\mu\nu}  =&  \, 2 \, \bar{\rho}_\perp \bar{\chi}^{\mu\nu} + \tau_m m^{\mu\nu} D \ln \bar{\rho}_\perp + 2\big[    \bar{\mathscr{X}}^{\cll \mu}  D b^{\nu\crr}  + H \bar{\mathscr{Y}}^{\cll \mu} D u^{\nu\crr} \big] \mathscr{W}_{m m}\nonumber\\
	& + 2\,\bar{\mathscr{X}}^{\cll\mu}\,\bar{\mathscr{Y}}^{\nu\crr}\, \bar{\rho}_{{}_{m\mathscr{J}\mathscr{K}}}
	- 2\,\bar{\chi}^{\mu\nu}\,\bar{\theta}_\perp \, \bar{\rho}_{{}_{m m P_\perp}}
	- 2\,\bar{\chi}^{\mu\nu}\,\bar{\theta}_\parallel \, \bar{\rho}_{{}_{m m P_\parallel}} .  \label{evom}
\end{align}

\subsection{Second-order correction and evolution equation for Bulk Viscous Pressure}

Quoting Eq.~\eqref{fullbulkparallel} from Appendix~\ref{Bulk_viscous}, the full expression for the bulk viscous pressure parallel to the magnetic field, valid up to second order in gradients, is
\begin{align}
	\Pi_{\parallel} (x)
	= \big< \hat P_{\parallel}(x) \big>_{1}
	+ \big< \hat P_{\parallel}(x) \big>_{2}
	+ \frac{1}{2}\,\big< \hat\varepsilon (x)\big>_{1}^{2}\,
	\frac{\partial^{2}p_{\parallel}}{\partial \varepsilon^{2}}\bigg|_x
	+ \frac{1}{2}\,\big< \hat B (x)\big>_{1}^{2}\,
	\frac{\partial^{2}p_{\parallel}}{\partial B^{2}}\bigg|_x
	+ \big< \hat\varepsilon (x)\big>_{1} \big<\hat B (x)\big>_{1}\,
	\frac{\partial^{2}p_{\parallel}}{\partial \varepsilon\,\partial B}\bigg|_x. \label{fullbulk}
\end{align}
Using Eqs.~\eqref{FOC} and~\eqref{CF}, together with Curie’s principle, the first-order corrections to the energy density and magnetic flux density are
\begin{align}
\Big< \hat{\varepsilon}(x)\Big>_1&= -\bar{\theta}_\parallel (x)\,\bar{\zeta}_{{}_{\varepsilon P_{\parallel}}} (x) - \bar{\theta}_\perp (x)\,\bar{\zeta}_{{}_{\varepsilon P_{\perp}}} (x) , \label{energyfirstorder}\\
\Big< \hat{B}(x)\Big>_1&= -\bar{\theta}_\parallel (x)\,\bar{\zeta}_{{}_{B P_{\parallel}}}(x) - \bar{\theta}_\perp (x)\,\bar{\zeta}_{{}_{B P_{\perp}}} (x) , \label{magnetic_fluxfirstorder}
\end{align} 
where the associated transport coefficients are defined by
\begin{align}
	\bar{\zeta}_{{}_{\varepsilon P_{\parallel}}} (x) &=\int d^4 x_1 \Big( \hat{\varepsilon}(x), \hat{P}_\parallel (x_1) \Big),~ \bar{\zeta}_{{}_{\varepsilon P_{\perp}}} (x) =\int d^4 x_1 \Big( \hat{\varepsilon}(x), \hat{P}_\perp (x_1) \Big),\label{bulktrans1}\\
	\bar{\zeta}_{{}_{B P_{\parallel}}} (x) &=\int d^4 x_1 \Big( \hat{B}(x), \hat{P}_\parallel (x_1) \Big),~ \bar{\zeta}_{{}_{B P_{\perp}}} (x) =\int d^4 x_1 \Big( \hat{B}(x), \hat{P}_\perp (x_1) \Big). \label{bulktrans2}
\end{align}
We also introduce 
\begin{align}
	\psi_{{}_{\parallel \varepsilon\varepsilon}}(x) = \frac{1}{2}	\frac{\partial^{2}p_{\parallel}}{\partial \varepsilon^{2}}\bigg|_x,~ \psi_{{}_{\parallel B B}}(x) = \frac{1}{2}	\frac{\partial^{2}p_{\parallel}}{\partial B^{2}}\bigg|_x,~\psi_{{}_{\parallel \varepsilon B}}(x) = \frac{1}{2}	\frac{\partial^{2}p_{\parallel}}{\partial \varepsilon \partial B}\bigg|_x  . \label{bulk_partial_definition}
\end{align}
Substituting Eqs.~\eqref{energyfirstorder}, \eqref{magnetic_fluxfirstorder}, and \eqref{bulk_partial_definition} into Eq.~\eqref{fullbulk}, we find
\begin{align}
	\Pi_\parallel =& -\bar{\zeta}_\parallel \bar{\theta}_\parallel -  \bar{\zeta}_\times \bar{\theta}_\perp 
	+\bar{\theta}^2_\parallel \Big[  \bar{\zeta}^2_{{}_{\varepsilon P_{\parallel}}} \psi_{{}_{\parallel \varepsilon\varepsilon}} + \bar{\zeta}^2_{{}_{B P_{\parallel}}} \psi_{{}_{\parallel BB}} + 2 \,\bar{\zeta}_{{}_{\varepsilon P_{\parallel}}}\bar{\zeta}_{{}_{B P_{\parallel}}} \psi_{{}_{\parallel \varepsilon B}} \Big] 
	 +\bar{\theta}^2_\perp \Big[  \bar{\zeta}^2_{{}_{\varepsilon P_{\perp}}} \psi_{{}_{\parallel \varepsilon\varepsilon}} + \bar{\zeta}^2_{{}_{B P_{\perp}}} \psi_{{}_{\parallel BB}} + 2 \,\bar{\zeta}_{{}_{\varepsilon P_{\perp}}}\bar{\zeta}_{{}_{B P_{\perp}}} \psi_{{}_{\parallel \varepsilon B}} \Big] \nonumber\\
	 & + 2\, \bar{\theta}_\parallel \bar{\theta}_\perp \Big[ 
	 \bar{\zeta}_{{}_{\varepsilon P_{\parallel}}} \bar{\zeta}_{{}_{\varepsilon P_{\perp}}} \psi_{{}_{\parallel \varepsilon\varepsilon}} + \bar{\zeta}_{{}_{B P_{\parallel}}} \bar{\zeta}_{{}_{B P_{\perp}}} \psi_{{}_{\parallel BB}} +\Big( \bar{\zeta}_{{}_{\varepsilon P_{\parallel}}} \bar{\zeta}_{{}_{B P_{\perp}}}+\bar{\zeta}_{{}_{\varepsilon P_{\perp}}} \bar{\zeta}_{{}_{B P_{\parallel}}} \Big)\psi_{{}_{\parallel \varepsilon B}} 
	  \Big]+ \big< \hat P_{\parallel} \big>_{2} . \label{parallel_bulk_solve}
\end{align}
Here, for brevity, we omitted the explicit arguments and used Eq.~\eqref{pi1parZ} to substitute the first-order contribution. 
We now proceed to compute the second-order correction, $\big< \hat P_{\parallel} (x) \big>_{2}$. Using~\eqref{SOC}, we can write
\begin{align}
	\left< \hat{P}_\parallel(x)\right>_2 = \left< \hat{P}_\parallel(x)\right>_2^{(2),NL} + \left< \hat{P}_\parallel(x)\right>_2^{(2),ET} +\left< \hat{P}_\parallel(x)\right>_2^{(3)}  . \label{P_parallel_second}
\end{align}

\subsubsection{Nonlocal correction from the two-point correlation function: $\big< \hat{P}_\parallel(x)\big>^{\text{(2),NL}}_2$}

Using~\eqref{nlc}, we obtain
\begin{align}
	\Big< \hat{P}_\parallel(x)\Big>^{\text{(2),NL}}_2
	&=\int d^4 x_1 \Big( \hat{P}_\parallel (x), [\partial_\lambda \hat{\mathcal{C}}(x_1)]_{x_1=x} \Big)(x_1-x)^\lambda. \label{pnl1}
\end{align}
Setting $x_1=x$ in all hydrodynamic variables (but not in the operators) in Eq.~\eqref{delC3}, and invoking Curie’s symmetry principle, we find
\begin{align}
	\Big< \hat{P}_\parallel(x)\Big>^{\text{(2),NL}}_2
	=& -\partial_\lambda \bar{\theta}_\parallel(x)	\,\mathfrak{p}_{\parallel}^\lambda (x) 
	 -\partial_\lambda \bar{\theta}_\perp (x)\,\mathfrak{p}_{\perp}^\lambda (x) +  \bar{\theta}_\parallel(x)\partial_\lambda \gamma_\parallel (x)\,\mathfrak{p}_{\varepsilon}^\lambda (x)
	+  \bar{\theta}_\parallel (x)\partial_\lambda \phi_\parallel(x)\,\mathfrak{p}_{B}^\lambda (x) \nonumber\\
&+  \bar{\theta}_\perp(x)\partial_\lambda \gamma_\perp (x)\,\mathfrak{p}_{\varepsilon}^\lambda (x)
+  \bar{\theta}_\perp (x)\partial_\lambda \phi_\perp(x)\,\mathfrak{p}_{B}^\lambda (x) . \label{pnl2}
\end{align}
Here we have defined
\begin{align}
	\mathfrak{p}_{\parallel}^\lambda (x) &= \int d^4 x_1 \Big( \hat{P}_\parallel (x),  \hat{P}_\parallel(x_1)     \Big)(x_1-x)^\lambda, ~~ \mathfrak{p}_{\perp}^\lambda (x)= \int d^4 x_1 \Big( \hat{P}_\parallel (x),  \hat{P}_\perp(x_1)     \Big)(x_1-x)^\lambda,\\
	\mathfrak{p}_{\varepsilon}^\lambda(x) &= \int d^4 x_1 \Big( \hat{P}_\parallel (x),  \hat{\varepsilon}(x_1)     \Big)(x_1-x)^\lambda, ~~ \mathfrak{p}_{B}^\lambda(x) = \int d^4 x_1 \Big( \hat{P}_\parallel (x),  \hat{B}(x_1)     \Big)(x_1-x)^\lambda . 
\end{align}
Employing,~\eqref{freq_dep_trans} from Appendix~\ref{correlators}, we can write
\begin{align}
	\mathfrak{p}_{\parallel}^\lambda &= u^\lambda \mathscr{W}_{P_\parallel P_\parallel} = u^\lambda i \lim_{\omega\to 0} \frac{d}{d\omega} \bar{\zeta}_{{}_{P_\parallel P_\parallel}}(\omega), \label{freqp1}\\
	\mathfrak{p}_{\perp}^\lambda &= u^\lambda  \mathscr{W}_{P_\parallel P_\perp} = u^\lambda i \lim_{\omega\to 0} \frac{d}{d\omega} \bar{\zeta}_{{}_{P_\parallel P_\perp}}(\omega) , \label{freqp2}\\
	\mathfrak{p}_{\varepsilon}^\lambda &= u^\lambda \mathscr{W}_{P_\parallel \varepsilon} = u^\lambda i \lim_{\omega\to 0} \frac{d}{d\omega} \bar{\zeta}_{{}_{P_\parallel \varepsilon}}(\omega)  , \label{freqp3}\\
	\mathfrak{p}_{B}^\lambda &=u^\lambda \mathscr{W}_{P_\parallel B} = u^\lambda i \lim_{\omega\to 0} \frac{d}{d\omega} \bar{\zeta}_{{}_{P_\parallel B}}(\omega) , \label{freqp4}
\end{align}
where, the frequency dependent transport coefficients are defined as
\begin{align}
	\bar{\zeta}_{{}_{P_\parallel P_\parallel}}(\omega) &= \int d^4 x_1  e^{i\omega(t-t_1)}\Big( \hat{P}_\parallel (x),  \hat{P}_\parallel(x_1)     \Big),  \\
	\bar{\zeta}_{{}_{P_\parallel P_\perp}}(\omega) &= \int d^4 x_1  e^{i\omega(t-t_1)}\Big( \hat{P}_\parallel (x),  \hat{P}_\perp(x_1)  \Big) , \\
	\bar{\zeta}_{{}_{P_\parallel \varepsilon}}(\omega) &= \int d^4 x_1  e^{i\omega(t-t_1)}\Big( \hat{P}_\parallel (x),  \hat{\varepsilon}(x_1)     \Big), \\
	\bar{\zeta}_{{}_{P_\parallel B}}(\omega) &= \int d^4 x_1  e^{i\omega(t-t_1)}\Big( \hat{P}_\parallel (x),  \hat{B}(x_1)     \Big).
\end{align}
Thus, using~\eqref{freqp1}-\eqref{freqp4}, Eq.~\eqref{pnl2} simplifies to
\begin{align}
	\big< \hat{P}_\parallel \big>^{\text{(2),NL}}_2
	=& -D \bar{\theta}_\parallel	\mathscr{W}_{P_\parallel P_\parallel} 
	-D \bar{\theta}_\perp \mathscr{W}_{P_\parallel P_\perp}  +  \bar{\theta}_\parallel D \gamma_\parallel \mathscr{W}_{P_\parallel \varepsilon} 
	+  \bar{\theta}_\parallel D \phi_\parallel \mathscr{W}_{P_\parallel B} \nonumber\\
	&+  \bar{\theta}_\perp D \gamma_\perp \mathscr{W}_{P_\parallel \varepsilon}
	+  \bar{\theta}_\perp D \phi_\perp \mathscr{W}_{P_\parallel B}  . \label{final_P_parallel_nonlocal}
\end{align}

\subsubsection{Local correction from the two-point correlation function using extended thermodynamic forces: $\big< \hat{P}_\parallel(x)\big>^{\text{(2),ET}}_2$}

With the help of Eqs.~\eqref{etc} and~\eqref{CS}, and employing Curie’s symmetry principle, the local correction arising from two-point correlation function using extended thermodynamic forces is obtained as
\begin{align}
		\Big< \hat{P}_\parallel (x)\Big>^{\text{(2),ET}}_2 =\Theta_T (x)\, \zeta_{{}_{P_\parallel \beta}} (x)	- \Theta_F (x)\,	\bar{\zeta}_{{}_{P_\parallel \mathcal{H}}}(x) .  \label{final_P_parallel_extended}
\end{align}
Here, we have defined
\begin{align}
	\Theta_T &=-u_\nu \partial_\mu T_1^{\mu\nu},~~\bar{\zeta}_{{}_{P_\parallel \beta}} (x)= \int d^4x_1\,\big(\hat{P}_\parallel(x),\,\hat{\beta}^*(x_1)\big),  \\
	 \Theta_F &= -b_\nu \partial_\mu \tilde F_1^{\mu\nu},~~ 	\bar{\zeta}_{{}_{P_\parallel \mathcal{H}}}(x) = \int d^4x_1\,\big(\hat{P}_\parallel(x),\,\hat{\mathcal{H}}^*(x_1)\big)  .
\end{align}

\subsubsection{Local correction from the three-point correlation function: $\big< \hat{P}_\parallel(x)\big>^{\text{(3)}}_2$}

Using~\eqref{tpc}, we obtain
\begin{align}
	\Big< \hat{P}_\parallel (x)\Big>^{\text{(3)}}_2 = 	\int d^4x_1 d^4x_2\,\big(\hat{P}_\parallel(x),\,\hat{\mathcal{C}}_{F}(x_1), \,\hat{\mathcal{C}}_{F}(x_2)\big) \big|_{local} . \label{p3local}
\end{align}
Substituting Eq.~\eqref{CF} into Eq.~\eqref{p3local} and following the procedure outlined in Sec.~\ref{secPi3P}, we find four nonvanishing three-point correlators: 
\begin{align}
		\big( \hat P_\parallel(x),\, \hat{\mathscr{J}}^\rho(x_1), \, \hat{\mathscr{J}}^\alpha(x_2) \big) &= \frac{1}{2}\G^{\rho\alpha}	\big( \hat P_\parallel(x),\, \hat{\mathscr{J}}^\lambda(x_1), \, \hat{\mathscr{J}}_\lambda(x_2) \big) , \label{corrP1}\\
	 \big( \hat P_\parallel(x),\, \hat{\mathscr{K}}^\rho(x_1), \, \hat{\mathscr{K}}^\alpha(x_2) \big) &=\frac{1}{2}\G^{\rho\alpha}\big( \hat P_\parallel(x),\, \hat{\mathscr{K}}^\lambda(x_1), \, \hat{\mathscr{K}}_\lambda(x_2) \big) , \label{corrP2}\\
		\big( \hat P_\parallel(x),\, \hat m^{\rho\sigma}(x_1), \, \hat m^{\alpha\beta}(x_2) \big) &=\Gc^{\rho\sigma\alpha\beta}\big( \hat P_\parallel(x),\, \hat m^{\lambda\eta}(x_1), \, \hat m_{\lambda\eta}(x_2) \big) , \label{corrP3}\\
	 \big( \hat P_\parallel(x),\, \hat \pi_\perp^{\rho\sigma}(x_1), \, \hat \pi_\perp^{\alpha\beta}(x_2) \big) &=\frac{1}{2}\G^{\rho\sigma\alpha\beta}\big( \hat P_\parallel(x),\, \hat \pi_\perp^{\lambda\eta}(x_1), \, \hat \pi_{\perp\lambda\eta}(x_2) \big). \label{corrP4}
\end{align}
Upon substituting the correlators \eqref{corrP1}-\eqref{corrP4} into Eq.~\eqref{p3local}, we arrive at
\begin{align}
	\big< \hat{P}_\parallel \big>^{\text{(3)}}_2 =  \bar{\zeta}_{{}_{P_\parallel  \mathscr{J}\mathscr{J}}} \bar{\mathscr{X}}_\mu  \bar{\mathscr{X}}^\mu  
	+ \bar{\zeta}_{{}_{P_\parallel  \mathscr{K}\mathscr{K}}} \bar{\mathscr{Y}}_\mu  \bar{\mathscr{Y}}^\mu
	+ 	\bar{\zeta}_{{}_{P_\parallel  m m }} \bar{\chi}_{\mu\nu} \bar{\chi}^{\mu\nu} 
	+ 	\bar{\zeta}_{{}_{P_\parallel  \pi \pi }} \bar{\sigma}_{\perp\mu\nu} \bar{\sigma}_{\perp}^{\mu\nu}, \label{final_P_parallel_three}
\end{align}
where the transport coefficients are defined as
\begin{align}
	\bar{\zeta}_{{}_{P_\parallel  \mathscr{J}\mathscr{J}}}(x) &= \frac{1}{2}\int d^4x_1 d^4x_2 	\big( \hat P_\parallel(x),\, \hat{\mathscr{J}}^\lambda(x_1), \, \hat{\mathscr{J}}_\lambda(x_2) \big) , \\
	\bar{\zeta}_{{}_{P_\parallel  \mathscr{K}\mathscr{K}}}(x) &= \frac{1}{2}\int d^4x_1 d^4x_2 	\big( \hat P_\parallel(x),\, \hat{\mathscr{K}}^\lambda(x_1), \, \hat{\mathscr{K}}_\lambda(x_2) \big) , \\
	\bar{\zeta}_{{}_{P_\parallel  m m }}(x) &= \int d^4x_1 d^4x_2 \big( \hat P_\parallel(x),\, \hat m^{\lambda\eta}(x_1), \, \hat m_{\lambda\eta}(x_2) \big) , \\
		\bar{\zeta}_{{}_{P_\parallel  \pi \pi }} (x) &=\frac{1}{2}\int d^4x_1 d^4x_2  \big( \hat P_\parallel(x),\, \hat \pi_\perp^{\lambda\eta}(x_1), \, \hat \pi_{\perp\lambda\eta}(x_2) \big).
\end{align}
Also, we have omitted the explicit arguments for brevity in~\eqref{final_P_parallel_three}.

\subsubsection{Full second order expression and the evolution equation for the bulk viscous pressure}

Substituting~\eqref{final_P_parallel_nonlocal},~\eqref{final_P_parallel_extended}, and~\eqref{final_P_parallel_three} in~\eqref{P_parallel_second}, we get
\begin{align}
	\big< \hat{P}_\parallel\big>_2 =& -D \bar{\theta}_\parallel	\mathscr{W}_{P_\parallel P_\parallel} 
	-D \bar{\theta}_\perp \mathscr{W}_{P_\parallel P_\perp}  +  \bar{\theta}_\parallel D \gamma_\parallel \mathscr{W}_{P_\parallel \varepsilon} 
	+  \bar{\theta}_\parallel D \phi_\parallel \mathscr{W}_{P_\parallel B} \nonumber\\
	&+  \bar{\theta}_\perp D \gamma_\perp \mathscr{W}_{P_\parallel \varepsilon}
	+  \bar{\theta}_\perp D \phi_\perp \mathscr{W}_{P_\parallel B} 
	+ \Theta_T \, \zeta_{{}_{P_\parallel \beta}}	- \Theta_F \,	\bar{\zeta}_{{}_{P_\parallel \mathcal{H}}} \nonumber\\
	& + \bar{\zeta}_{{}_{P_\parallel  \mathscr{J}\mathscr{J}}} \bar{\mathscr{X}}_\mu  \bar{\mathscr{X}}^\mu  
	+ \bar{\zeta}_{{}_{P_\parallel  \mathscr{K}\mathscr{K}}} \bar{\mathscr{Y}}_\mu  \bar{\mathscr{Y}}^\mu
	+ 	\bar{\zeta}_{{}_{P_\parallel  m m }} \bar{\chi}_{\mu\nu} \bar{\chi}^{\mu\nu} 
	+ 	\bar{\zeta}_{{}_{P_\parallel  \pi \pi }} \bar{\sigma}_{\perp\mu\nu} \bar{\sigma}_{\perp}^{\mu\nu}, 
\end{align}
which is further substituted in~\eqref{parallel_bulk_solve}, to get the full expression of bulk viscous pressure parallel to the magnetic field. Thus, we obtain
\begin{align}
		\Pi_\parallel =& -\bar{\zeta}_\parallel \bar{\theta}_\parallel -  \bar{\zeta}_\times \bar{\theta}_\perp 
	+\bar{\theta}^2_\parallel \Big[  \bar{\zeta}^2_{{}_{\varepsilon P_{\parallel}}} \psi_{{}_{\parallel \varepsilon\varepsilon}} + \bar{\zeta}^2_{{}_{B P_{\parallel}}} \psi_{{}_{\parallel BB}} + 2 \,\bar{\zeta}_{{}_{\varepsilon P_{\parallel}}}\bar{\zeta}_{{}_{B P_{\parallel}}} \psi_{{}_{\parallel \varepsilon B}} \Big] 
	+\bar{\theta}^2_\perp \Big[  \bar{\zeta}^2_{{}_{\varepsilon P_{\perp}}} \psi_{{}_{\parallel \varepsilon\varepsilon}} + \bar{\zeta}^2_{{}_{B P_{\perp}}} \psi_{{}_{\parallel BB}} + 2 \,\bar{\zeta}_{{}_{\varepsilon P_{\perp}}}\bar{\zeta}_{{}_{B P_{\perp}}} \psi_{{}_{\parallel \varepsilon B}} \Big] \nonumber\\
	& + 2\, \bar{\theta}_\parallel \bar{\theta}_\perp \Big[ 
	\bar{\zeta}_{{}_{\varepsilon P_{\parallel}}} \bar{\zeta}_{{}_{\varepsilon P_{\perp}}} \psi_{{}_{\parallel \varepsilon\varepsilon}} + \bar{\zeta}_{{}_{B P_{\parallel}}} \bar{\zeta}_{{}_{B P_{\perp}}} \psi_{{}_{\parallel BB}} +\Big( \bar{\zeta}_{{}_{\varepsilon P_{\parallel}}} \bar{\zeta}_{{}_{B P_{\perp}}}+\bar{\zeta}_{{}_{\varepsilon P_{\perp}}} \bar{\zeta}_{{}_{B P_{\parallel}}} \Big)\psi_{{}_{\parallel \varepsilon B}} 
	\Big]\nonumber\\
	& -D \bar{\theta}_\parallel	\mathscr{W}_{P_\parallel P_\parallel} 
	-D \bar{\theta}_\perp \mathscr{W}_{P_\parallel P_\perp}  +  \bar{\theta}_\parallel D \gamma_\parallel \mathscr{W}_{P_\parallel \varepsilon} 
	+  \bar{\theta}_\parallel D \phi_\parallel \mathscr{W}_{P_\parallel B} \nonumber\\
	&+  \bar{\theta}_\perp D \gamma_\perp \mathscr{W}_{P_\parallel \varepsilon}
	+  \bar{\theta}_\perp D \phi_\perp \mathscr{W}_{P_\parallel B} 
	+ \Theta_T \, \zeta_{{}_{P_\parallel \beta}} 	- \Theta_F \,	\bar{\zeta}_{{}_{P_\parallel \mathcal{H}}} \nonumber\\
	& + \bar{\zeta}_{{}_{P_\parallel  \mathscr{J}\mathscr{J}}} \bar{\mathscr{X}}_\mu  \bar{\mathscr{X}}^\mu  
	+ \bar{\zeta}_{{}_{P_\parallel  \mathscr{K}\mathscr{K}}} \bar{\mathscr{Y}}_\mu  \bar{\mathscr{Y}}^\mu
	+ 	\bar{\zeta}_{{}_{P_\parallel  m m }} \bar{\chi}_{\mu\nu} \bar{\chi}^{\mu\nu} 
	+ 	\bar{\zeta}_{{}_{P_\parallel  \pi \pi }} \bar{\sigma}_{\perp\mu\nu} \bar{\sigma}_{\perp}^{\mu\nu} . \label{final_bulk_paralel}
\end{align}

From Eq.~\eqref{fullbulkperp}, we observe that the structure of the bulk viscous pressure perpendicular to the magnetic field mirrors that of the parallel bulk viscous pressure. Accordingly, we follow the same procedure as given in the parallel case to derive the second-order expression for the perpendicular bulk viscous pressure. This yields
\begin{align}
	\Pi_\perp =& -\bar{\zeta}_\perp \bar{\theta}_\perp -  \bar{\zeta}'_\times \bar{\theta}_\parallel 
	+\bar{\theta}^2_\perp \Big[  \bar{\zeta}^2_{{}_{\varepsilon P_{\perp}}} \psi_{{}_{\perp \varepsilon\varepsilon}} + \bar{\zeta}^2_{{}_{B P_{\perp}}} \psi_{{}_{\perp BB}} + 2 \,\bar{\zeta}_{{}_{\varepsilon P_{\perp}}}\bar{\zeta}_{{}_{B P_{\perp}}} \psi_{{}_{\perp \varepsilon B}} \Big]
	+\bar{\theta}^2_\parallel \Big[  \bar{\zeta}^2_{{}_{\varepsilon P_{\parallel}}} \psi_{{}_{\perp \varepsilon\varepsilon}} + \bar{\zeta}^2_{{}_{B P_{\parallel}}} \psi_{{}_{\perp BB}} + 2 \,\bar{\zeta}_{{}_{\varepsilon P_{\parallel}}}\bar{\zeta}_{{}_{B P_{\parallel}}} \psi_{{}_{\perp \varepsilon B}} \Big] 
	 \nonumber\\
	& + 2\, \bar{\theta}_\parallel \bar{\theta}_\perp \Big[ 
	\bar{\zeta}_{{}_{\varepsilon P_{\parallel}}} \bar{\zeta}_{{}_{\varepsilon P_{\perp}}} \psi_{{}_{\perp \varepsilon\varepsilon}} + \bar{\zeta}_{{}_{B P_{\parallel}}} \bar{\zeta}_{{}_{B P_{\perp}}} \psi_{{}_{\perp BB}} +\Big( \bar{\zeta}_{{}_{\varepsilon P_{\parallel}}} \bar{\zeta}_{{}_{B P_{\perp}}}+\bar{\zeta}_{{}_{\varepsilon P_{\perp}}} \bar{\zeta}_{{}_{B P_{\parallel}}} \Big)\psi_{{}_{\perp \varepsilon B}} 
	\Big] \nonumber\\
	& -D \bar{\theta}_\parallel	\mathscr{W}_{P_\perp P_\parallel} 
	-D \bar{\theta}_\perp \mathscr{W}_{P_\perp P_\perp}  +  \bar{\theta}_\parallel D \gamma_\parallel \mathscr{W}_{P_\perp \varepsilon} 
	+  \bar{\theta}_\parallel D \phi_\parallel \mathscr{W}_{P_\perp B} \nonumber\\
	&+  \bar{\theta}_\perp D \gamma_\perp \mathscr{W}_{P_\perp \varepsilon}
	+  \bar{\theta}_\perp D \phi_\perp \mathscr{W}_{P_\perp B}
	+	\Theta_T\, \zeta_{{}_{P_\perp \beta}} - \Theta_F \,	\bar{\zeta}_{{}_{P_\perp \mathcal{H}}}  \nonumber\\
	& + \bar{\zeta}_{{}_{P_\perp  \mathscr{J}\mathscr{J}}} \bar{\mathscr{X}}_\mu  \bar{\mathscr{X}}^\mu  
	+ \bar{\zeta}_{{}_{P_\perp  \mathscr{K}\mathscr{K}}} \bar{\mathscr{Y}}_\mu  \bar{\mathscr{Y}}^\mu
	+ 	\bar{\zeta}_{{}_{P_\perp  m m }} \bar{\chi}_{\mu\nu} \bar{\chi}^{\mu\nu} 
	+ 	\bar{\zeta}_{{}_{P_\perp  \pi \pi }} \bar{\sigma}_{\perp\mu\nu} \bar{\sigma}_{\perp}^{\mu\nu} . \label{final_bulk_perp}
\end{align}
Here, the newly introduced parameters ($\psi$’s and $\mathscr{W}$’s), as well as the transport coefficients, are defined in direct analogy with their parallel counterparts.

Further, we employ the first-order Navier-Srokes relation~\eqref{pi1parZ} to approximate the term $D\bar{\theta}_\parallel$ in~\eqref{final_bulk_paralel}, by setting $\bar{\theta}_\parallel\sim - (\Pi_\parallel+\bar{\theta}_\perp \bar{\zeta}_\times)\bar{\zeta}_\parallel^{-1}$. This yields,
\begin{align}
	 -D \bar{\theta}_\parallel	\mathscr{W}_{P_\parallel P_\parallel}  \simeq \frac{ \mathscr{W}_{P_\parallel P_\parallel}}{\bar{\zeta}_\parallel} \big[ \dot{\Pi}_\parallel + \bar{\zeta}_\times D\bar{\theta}_\perp + \bar{\theta}_\perp D\bar{\zeta}_\times- (\Pi_\parallel +\bar{\theta}_\perp \bar{\zeta}_\times)D \ln \bar{\zeta}_\parallel \big] , \label{nav_appro_pipar}
\end{align}
Here, we have defined $\dot{\Pi}_\parallel=D\Pi_\parallel$. The characteristic relaxation timescale is define as
\begin{align}
	\frac{ \mathscr{W}_{P_\parallel P_\parallel}}{\bar{\zeta}_\parallel} = -\tau_{\Pi_\parallel} . \label{timepipar}
\end{align}
Substituting~\eqref{nav_appro_pipar} and~\eqref{timepipar} in~\eqref{final_bulk_paralel}, we obtain the evolution equation for $\Pi_\parallel$, given by
\begin{align}
		\Pi_\parallel +\tau_{\Pi_\parallel} \dot{\Pi}_\parallel =& -\bar{\zeta}_\parallel \bar{\theta}_\parallel -  \bar{\zeta}_\times \bar{\theta}_\perp -\bar{\zeta}_\times D \bar{\theta}_\perp (\tau_{\Pi_\parallel}-\tau_{\Pi_\times}) + \tau_{\Pi_\parallel} (\Pi_\parallel +\bar{\theta}_\perp \bar{\zeta}_\times) D\ln \bar{\zeta}_\parallel  - \tau_{\Pi_\parallel}\bar{\theta}_\perp D\bar{\zeta}_\times\nonumber\\
	&+\bar{\theta}^2_\parallel \, \mathfrak{z}_{{}_{\parallel}} +\bar{\theta}^2_\perp \, \mathfrak{z}_{{}_{\perp}} + 2\, \bar{\theta}_\parallel \bar{\theta}_\perp \, \mathfrak{z}_{{}_{\times}}  +  \bar{\theta}_\parallel D \gamma_\parallel \mathscr{W}_{P_\parallel \varepsilon} 
	+  \bar{\theta}_\parallel D \phi_\parallel \mathscr{W}_{P_\parallel B} \nonumber\\
	&+  \bar{\theta}_\perp D \gamma_\perp \mathscr{W}_{P_\parallel \varepsilon}
	+  \bar{\theta}_\perp D \phi_\perp \mathscr{W}_{P_\parallel B} 
	+ \Theta_T \, \zeta_{{}_{P_\parallel \beta}} 	- \Theta_F \,	\bar{\zeta}_{{}_{P_\parallel \mathcal{H}}} \nonumber\\
	& + \bar{\zeta}_{{}_{P_\parallel  \mathscr{J}\mathscr{J}}} \bar{\mathscr{X}}_\mu  \bar{\mathscr{X}}^\mu  
	+ \bar{\zeta}_{{}_{P_\parallel  \mathscr{K}\mathscr{K}}} \bar{\mathscr{Y}}_\mu  \bar{\mathscr{Y}}^\mu
	+ 	\bar{\zeta}_{{}_{P_\parallel  m m }} \bar{\chi}_{\mu\nu} \bar{\chi}^{\mu\nu} 
	+ 	\bar{\zeta}_{{}_{P_\parallel  \pi \pi }} \bar{\sigma}_{\perp\mu\nu} \bar{\sigma}_{\perp}^{\mu\nu}.  \label{evopipar}
\end{align}
Here, we have defined
\begin{align}
	\frac{ \mathscr{W}_{P_\parallel P_\perp}}{\bar{\zeta}_\times} = -\tau_{\Pi_\times} , \label{timepipar_cross}
\end{align}
and
\begin{align}
	\mathfrak{z}_{{}_{\parallel}} &= \Big[  \bar{\zeta}^2_{{}_{\varepsilon P_{\parallel}}} \psi_{{}_{\parallel \varepsilon\varepsilon}} + \bar{\zeta}^2_{{}_{B P_{\parallel}}} \psi_{{}_{\parallel BB}} + 2 \,\bar{\zeta}_{{}_{\varepsilon P_{\parallel}}}\bar{\zeta}_{{}_{B P_{\parallel}}} \psi_{{}_{\parallel \varepsilon B}} \Big] , \\
	\mathfrak{z}_{{}_{\perp}} &= \Big[  \bar{\zeta}^2_{{}_{\varepsilon P_{\perp}}} \psi_{{}_{\parallel \varepsilon\varepsilon}} + \bar{\zeta}^2_{{}_{B P_{\perp}}} \psi_{{}_{\parallel BB}} + 2 \,\bar{\zeta}_{{}_{\varepsilon P_{\perp}}}\bar{\zeta}_{{}_{B P_{\perp}}} \psi_{{}_{\parallel \varepsilon B}} \Big] , \\
	\mathfrak{z}_{{}_{\times}} &= \Big[ 
	\bar{\zeta}_{{}_{\varepsilon P_{\parallel}}} \bar{\zeta}_{{}_{\varepsilon P_{\perp}}} \psi_{{}_{\parallel \varepsilon\varepsilon}} + \bar{\zeta}_{{}_{B P_{\parallel}}} \bar{\zeta}_{{}_{B P_{\perp}}} \psi_{{}_{\parallel BB}} +\Big( \bar{\zeta}_{{}_{\varepsilon P_{\parallel}}} \bar{\zeta}_{{}_{B P_{\perp}}}+\bar{\zeta}_{{}_{\varepsilon P_{\perp}}} \bar{\zeta}_{{}_{B P_{\parallel}}} \Big)\psi_{{}_{\parallel \varepsilon B}} 
	\Big].
\end{align} 

In an analogous manner, the evolution equation for $\Pi_\perp$ is obtained by approximating $D\bar{\theta}_\perp$ in~\eqref{final_bulk_perp}, using $\bar{\theta}_\perp \sim - (\Pi_\perp+\bar{\theta}_\parallel \bar{\zeta}'_\times)\bar{\zeta}_\perp^{-1}$. This leads to the evolution equation for $\Pi_\perp$, given by
\begin{align}
	\Pi_\perp +\tau_{\Pi_\perp} \dot{\Pi}_\perp =& -\bar{\zeta}_\perp \bar{\theta}_\perp -  \bar{\zeta}'_\times \bar{\theta}_\parallel  
	-\bar{\zeta}'_{\times}D \bar{\theta}_\parallel	(\tau_{\Pi_\perp}-\tau'_{\Pi_\times}) + \tau_{\Pi_\perp} (\Pi_\perp+\bar{\theta}_\parallel\bar{\zeta}'_{\times}) D\ln \bar{\zeta}_\perp -\tau_{\Pi_\perp}\bar{\theta}_\parallel D\bar{\zeta}'_{\times}
	\nonumber\\
	& +\bar{\theta}^2_\perp \, \mathfrak{z}'_{{}_{\perp}}
	+\bar{\theta}^2_\parallel \,\mathfrak{z}'_{{}_{\parallel}}  + 2\, \bar{\theta}_\parallel \bar{\theta}_\perp \, \mathfrak{z}'_{{}_{\times}}  +  \bar{\theta}_\parallel D \gamma_\parallel \mathscr{W}_{P_\perp \varepsilon} 
	+  \bar{\theta}_\parallel D \phi_\parallel \mathscr{W}_{P_\perp B} \nonumber\\
	&+  \bar{\theta}_\perp D \gamma_\perp \mathscr{W}_{P_\perp \varepsilon}
	+  \bar{\theta}_\perp D \phi_\perp \mathscr{W}_{P_\perp B}
	+	\Theta_T\, \zeta_{{}_{P_\perp \beta}} - \Theta_F \,	\bar{\zeta}_{{}_{P_\perp \mathcal{H}}}  \nonumber\\
	& + \bar{\zeta}_{{}_{P_\perp  \mathscr{J}\mathscr{J}}} \bar{\mathscr{X}}_\mu  \bar{\mathscr{X}}^\mu  
	+ \bar{\zeta}_{{}_{P_\perp  \mathscr{K}\mathscr{K}}} \bar{\mathscr{Y}}_\mu  \bar{\mathscr{Y}}^\mu
	+ 	\bar{\zeta}_{{}_{P_\perp  m m }} \bar{\chi}_{\mu\nu} \bar{\chi}^{\mu\nu} 
	+ 	\bar{\zeta}_{{}_{P_\perp  \pi \pi }} \bar{\sigma}_{\perp\mu\nu} \bar{\sigma}_{\perp}^{\mu\nu} . \label{evopiperp}
\end{align} 
Here, we have defined
\begin{align}
		\tau_{\Pi_\perp}=-\frac{ \mathscr{W}_{P_\perp P_\perp}}{\bar{\zeta}_\perp}, ~~~ \tau'_{\Pi_\times}= -\frac{ \mathscr{W}_{P_\perp P_\parallel}}{\bar{\zeta}'_\times} , \label{timepiperp_cross}
\end{align}
and
\begin{align}
 	\mathfrak{z}'_{{}_{\perp}} &=  \Big[  \bar{\zeta}^2_{{}_{\varepsilon P_{\perp}}} \psi_{{}_{\perp \varepsilon\varepsilon}} + \bar{\zeta}^2_{{}_{B P_{\perp}}} \psi_{{}_{\perp BB}} + 2 \,\bar{\zeta}_{{}_{\varepsilon P_{\perp}}}\bar{\zeta}_{{}_{B P_{\perp}}} \psi_{{}_{\perp \varepsilon B}} \Big] , \\
 	\mathfrak{z}'_{{}_{\parallel}} &= \Big[  \bar{\zeta}^2_{{}_{\varepsilon P_{\parallel}}} \psi_{{}_{\perp \varepsilon\varepsilon}} + \bar{\zeta}^2_{{}_{B P_{\parallel}}} \psi_{{}_{\perp BB}} + 2 \,\bar{\zeta}_{{}_{\varepsilon P_{\parallel}}}\bar{\zeta}_{{}_{B P_{\parallel}}} \psi_{{}_{\perp \varepsilon B}} \Big] , \\
	\mathfrak{z}'_{{}_{\times}} &= \Big[ 
	\bar{\zeta}_{{}_{\varepsilon P_{\parallel}}} \bar{\zeta}_{{}_{\varepsilon P_{\perp}}} \psi_{{}_{\perp \varepsilon\varepsilon}} + \bar{\zeta}_{{}_{B P_{\parallel}}} \bar{\zeta}_{{}_{B P_{\perp}}} \psi_{{}_{\perp BB}} +\Big( \bar{\zeta}_{{}_{\varepsilon P_{\parallel}}} \bar{\zeta}_{{}_{B P_{\perp}}}+\bar{\zeta}_{{}_{\varepsilon P_{\perp}}} \bar{\zeta}_{{}_{B P_{\parallel}}} \Big)\psi_{{}_{\perp \varepsilon B}} 
	\Big] . 
\end{align}

\subsection{Second-order correction and evolution equation for  $\hat{\mathscr{J}}^{\mu}$}

\subsubsection{Nonlocal correction from the two-point correlation function: $\big< \hat{\mathscr{J}}^\mu(x)\big>^{\text{(2),NL}}_2$}

We use Eq.~\eqref{nlc} and evaluate $[\partial_\lambda \hat{\mathcal{C}}(x_1)]_{x_1=x}$ using~\eqref{delC3}. Employing Curie’s symmetry principle thereafter, we obtain
\begin{align}
	\Big< \hat{\mathscr{J}}^\mu(x)\Big>^{\text{(2),NL}}_2
	=&\frac{1}{2}\G^{\mu\rho}\partial_\lambda \bar{\mathscr{X}}_\rho (x)\int d^4 x_1 \Big( \hat{\mathscr{J}}^\eta(x), \hat{\mathscr{J}}_\eta (x_1) \Big)(x_1-x)^\lambda \nonumber\\
	&+ \frac{1}{2}\G^{\mu\rho}\Big[2DB(x) \partial_\lambda u_\rho(x) -\bar{\mathscr{X}}_\rho(x) \partial_\lambda B'(x)\Big]\int d^4 x_1 \Big( \hat{\mathscr{J}}^\eta(x),\hat{h}_\eta (x_1) \Big)(x_1-x)^\lambda \nonumber\\
	& -\frac{1}{2}\G^{\mu\rho}\Big[D\mathcal{H}(x)+H(x)\bar{\theta}_\parallel(x)\Big]\partial_\lambda u_\rho(x) \int d^4 x_1 \Big( \hat{\mathscr{J}}^\eta(x),\hat{l}_\eta (x_1) \Big)(x_1-x)^\lambda . 
\end{align}
Further, employing~\eqref{freq_dep_trans} from Appendix~\ref{correlators}, we get
\begin{align}
	\big< \hat{\mathscr{J}}^\mu\big>^{\text{(2),NL}}_2
	=&\G^{\mu\rho} D\bar{\mathscr{X}}_\rho \mathscr{W}_{\mathscr{J}\mathscr{J}} + \Big[2DB \,\G^{\mu\rho}D u_\rho -\bar{\mathscr{X}}^\mu DB'\Big] \mathscr{W}_{\mathscr{J} h} - \G^{\mu\rho}\Big[D\mathcal{H}+H\bar{\theta}_\parallel\Big]D u_\rho \mathscr{W}_{\mathscr{J} l}   . \label{final_J_nonlocal}
\end{align}
Here, for brevity, we omitted the explicit arguments and defined
\begin{align}
	\mathscr{W}_{\mathscr{J}\mathscr{J}} &= i \lim_{\omega\to 0} \frac{d}{d\omega} \bar{\rho}_{{}_{\mathscr{J}\mathscr{J}}}(\omega) , \\
	\mathscr{W}_{\mathscr{J} h} &= i \lim_{\omega\to 0} \frac{d}{d\omega} \bar{\rho}_{{}_{\mathscr{J} h}}(\omega) , \\
	\mathscr{W}_{\mathscr{J} l} &= i \lim_{\omega\to 0} \frac{d}{d\omega} \bar{\rho}_{{}_{\mathscr{J} l}}(\omega) ,
\end{align}
where the frequency dependent transport coefficients are defined by
\begin{align}
	\bar{\rho}_{{}_{\mathscr{J}\mathscr{J}}} (\omega)&=\frac{1}{2}\int d^4 x_1 e^{i\omega(t-t_1)}\Big( \hat{\mathscr{J}}^\eta(x), \hat{\mathscr{J}}_\eta (x_1) \Big) , \\
	\bar{\rho}_{{}_{\mathscr{J} h}} (\omega)&= \frac{1}{2} \int d^4 x_1 e^{i\omega(t-t_1)}\Big( \hat{\mathscr{J}}^\eta(x),\hat{h}_\eta (x_1) \Big) , \\
	\bar{\rho}_{{}_{\mathscr{J} l}} (\omega)&= \frac{1}{2} \int d^4 x_1 e^{i\omega(t-t_1)}\Big( \hat{\mathscr{J}}^\eta(x),\hat{l}_\eta (x_1) \Big) .
\end{align}

\subsubsection{Local correction from the two-point correlation function using extended thermodynamic forces: $\big< \hat{\mathscr{J}}^\mu(x)\big>^{\text{(2),ET}}_2$}

Employing Eqs.~\eqref{etc} and~\eqref{CS}, and invoking Curie’s symmetry principle, we get
\begin{align}
	\Big< \hat{\mathscr{J}}^\mu(x)\Big>^{\text{(2),ET}}_2 = \bar{\mathfrak{U}}^\mu(x)\,\bar{\rho}_{{}_{\mathscr{J}h}} (x) ,  \label{final_J_extended}
\end{align}
where, we have defined
\begin{align}
	\bar{\mathfrak{U}}^\mu = \beta\, \G^{\mu\rho} \,(D u_\rho)_2,~~\bar{\rho}_{{}_{\mathscr{J}h}} (x) =\frac{1}{2}\int d^4 x_1 \Big( \hat{\mathscr{J}}^\lambda(x), \hat{h}_\lambda (x_1) \Big) .
\end{align}
Because $\hat{\mathscr{J}}^\mu$ and $\hat{g}^\mu$ have different parity, Curie’s principle forbids their coupling; hence no $\hat{g}^\mu$ term appears in the result.

\subsubsection{Local correction from the three point correlation function: $\big< \hat{\mathscr{J}}^\mu(x)\big>^{\text{(3)}}_2$}

Employing Eqs.~\eqref{tpc} and~\eqref{CF}, and following the procedure described in Sec.~\ref{secPi3P}, we identify eight nonvanishing three-point correlators that contribute to the local correction arising from the three-point correlation function in $\hat{\mathscr{J}}^\mu$. Consequently, we obtain
\begin{align}
	\Big< \hat{\mathscr{J}}^\mu (x)\Big>^{\text{(3)}}_2= \mathfrak{e}_1^\mu (x)+ \mathfrak{e}_2^\mu (x)+\mathfrak{e}_3^\mu (x)+\mathfrak{e}_4^\mu (x)+\mathfrak{e}_5^\mu (x)+\mathfrak{e}_6^\mu (x)+\mathfrak{e}_7^\mu (x)+\mathfrak{e}_8^\mu (x) . \label{J3point}
\end{align}
Here, we have defined
\begin{align}
	\mathfrak{e}_1^\mu (x)&=	-\frac{1}{2}\,\G^{\mu\rho}\,\bar{\mathscr{X}}_\rho (x) \, \bar{\theta}_\parallel(x)\int d^4x_1 d^4x_2\,\big( \hat{\mathscr{J}}^\lambda(x),\, \hat{\mathscr{J}}_\lambda(x_1), \, \hat{P}_\parallel(x_2) \big) ,  \\
	 \mathfrak{e}_2^\mu (x)&=	-\frac{1}{2}\,\G^{\mu\alpha}\,\bar{\theta}_\parallel(x) \,\bar{\mathscr{X}}_\alpha (x)  \int d^4x_1 d^4x_2\,\big( \hat{\mathscr{J}}^\lambda(x),\, \hat{P}_\parallel(x_1), \, \hat{\mathscr{J}}_\lambda(x_2) \big) , \\
	\mathfrak{e}_3^\mu (x) &=		-\frac{1}{2}\,\G^{\mu\rho}\,\bar{\mathscr{X}}_\rho (x) \, \bar{\theta}_\perp(x)\int d^4x_1 d^4x_2\,\big( \hat{\mathscr{J}}^\lambda(x),\, \hat{\mathscr{J}}_\lambda(x_1), \, \hat{P}_\perp(x_2) \big) , \\
	 \mathfrak{e}_4^\mu (x)&=	-\frac{1}{2}\,\G^{\mu\alpha}	\,\bar{\theta}_\perp(x) \,\bar{\mathscr{X}}_\alpha (x) \int d^4x_1 d^4x_2\,\big( \hat{\mathscr{J}}^\lambda(x),\, \hat{P}_\perp(x_1), \, \hat{\mathscr{J}}_\lambda(x_2) \big) , \\
	 \mathfrak{e}_5^\mu (x)&=	\frac{1}{2}\,\G^{\mu\rho\alpha\beta}\,\bar{\mathscr{X}}_\rho (x)\,\bar{\sigma}_{\perp\alpha\beta}(x)\int d^4x_1 d^4x_2\,\big( \hat{\mathscr{J}}^\lambda(x),\, \hat{\mathscr{J}}^\eta(x_1), \, \hat{\pi}_{\perp\lambda\eta}(x_2) \big) , \\
	 \mathfrak{e}_6^\mu (x)&=	\frac{1}{2}\,\G^{\mu\rho\sigma\alpha}	\,\bar{\sigma}_{\perp\rho\sigma}(x)\bar{\mathscr{X}}_\alpha (x) \int d^4x_1 d^4x_2\,\big( \hat{\mathscr{J}}^\lambda(x),\, \hat{\pi}_{\perp\lambda\eta}(x_1), \, \hat{\mathscr{J}}^\eta(x_2) \big) , \\
	 \mathfrak{e}_7^\mu (x)&=	\Gc^{\mu\rho\alpha\beta}\,\bar{\mathscr{Y}}_\rho (x)\,\bar{\chi}_{\alpha\beta}(x)\int d^4x_1 d^4x_2\,\big( \hat{\mathscr{J}}^\lambda(x),\, \hat{\mathscr{K}}^\eta(x_1), \, \hat{m}_{\lambda\eta}(x_2) \big) , \\
	 \mathfrak{e}_8^\mu (x)&=	\Gc^{\mu\alpha\rho\sigma}\,\bar{\chi}_{\rho\sigma}(x)\bar{\mathscr{Y}}_\alpha (x)\int d^4x_1 d^4x_2\,\big( \hat{\mathscr{J}}^\lambda(x),\, \hat{m}_{\lambda\eta}(x_1), \, \hat{\mathscr{K}}^\eta(x_2) \big) .
\end{align}
Using the symmetry~\eqref{threepointsymmetry}, we can infer that $\mathfrak{e}_1^\mu = \mathfrak{e}_2^\mu$, $\mathfrak{e}_3^\mu = \mathfrak{e}_4^\mu$, $\mathfrak{e}_5^\mu = \mathfrak{e}_6^\mu$, and $\mathfrak{e}_7^\mu = \mathfrak{e}_8^\mu$. Further, we can rewrite~\eqref{J3point} as
\begin{align}
	\big< \hat{\mathscr{J}}^\mu\big>^{\text{(3)}}_2 = -2\,\bar{\rho}_{{}_{\mathscr{J}\mathscr{J}P_\parallel}} \bar{\mathscr{X}}^\mu  \, \bar{\theta}_\parallel
	- 2\,\bar{\rho}_{{}_{\mathscr{J}\mathscr{J}P_\perp}} \bar{\mathscr{X}}^\mu  \, \bar{\theta}_\perp
	+ 2\,\bar{\rho}_{{}_{\mathscr{J}\mathscr{J} \pi}}\bar{\mathscr{X}}_\rho \,\bar{\sigma}_{\perp}^{\mu\rho}
	+ 2\, 	\bar{\rho}_{{}_{\mathscr{J}\mathscr{K} m}}\bar{\mathscr{Y}}_\rho \,\bar{\chi}^{\mu\rho} . \label{final_J_three_point}
\end{align}
Here, we defined the transport coefficients as
\begin{align}
	\bar{\rho}_{{}_{\mathscr{J}\mathscr{J}P_\parallel}}(x) &= \frac{1}{2}\int d^4x_1 d^4x_2\, \big( \hat{\mathscr{J}}^\lambda(x),\, \hat{\mathscr{J}}_\lambda(x_1), \, \hat{P}_\parallel(x_2) \big) , \\
	\bar{\rho}_{{}_{\mathscr{J}\mathscr{J}P_\perp}}(x) &= \frac{1}{2}\int d^4x_1 d^4x_2\, \big( \hat{\mathscr{J}}^\lambda(x),\, \hat{\mathscr{J}}_\lambda(x_1), \, \hat{P}_\perp(x_2) \big) , \\
	\bar{\rho}_{{}_{\mathscr{J}\mathscr{J} \pi}}(x) &=\frac{1}{2}\int d^4x_1 d^4x_2\,\big( \hat{\mathscr{J}}^\lambda(x),\, \hat{\mathscr{J}}^\eta(x_1), \, \hat{\pi}_{\perp\lambda\eta}(x_2) \big) , \\
	\bar{\rho}_{{}_{\mathscr{J}\mathscr{K} m}}(x) &=\int d^4x_1 d^4x_2\,\big( \hat{\mathscr{J}}^\lambda(x),\, \hat{\mathscr{K}}^\eta(x_1), \, \hat{m}_{\lambda\eta}(x_2) \big).
\end{align}
Also, we have omitted the explicit arguments for the sake of brevity in~\eqref{final_J_three_point}.

\subsubsection{Full second order expression and the evolution equation for $\hat{\mathscr{J}}^{\mu}$}

In order to get the full expression of $\hat{\mathscr{J}}^{\mu}$ up to the second order, we collect the first-order term from~\eqref{transJ1} and second-order terms from~\eqref{final_J_nonlocal},~\eqref{final_J_extended}, and~\eqref{final_J_three_point}, we obtain
\begin{align}
	\mathscr{J}^{\mu} =& \, \bar{\rho}_\parallel \,\bar{\mathscr{X}}^\mu 
	+ \G^{\mu\rho} D\bar{\mathscr{X}}_\rho \mathscr{W}_{\mathscr{J}\mathscr{J}} + \Big[2DB \,\G^{\mu\rho}D u_\rho -\bar{\mathscr{X}}^\mu DB'\Big] \mathscr{W}_{\mathscr{J} h} - \G^{\mu\rho}\Big[D\mathcal{H}+H\bar{\theta}_\parallel\Big]D u_\rho \mathscr{W}_{\mathscr{J} l} \nonumber\\
	& + \bar{\mathfrak{U}}^\mu\,\bar{\rho}_{{}_{\mathscr{J}h}} 
	 -2\,\bar{\rho}_{{}_{\mathscr{J}\mathscr{J}P_\parallel}} \bar{\mathscr{X}}^\mu  \, \bar{\theta}_\parallel
	- 2\,\bar{\rho}_{{}_{\mathscr{J}\mathscr{J}P_\perp}} \bar{\mathscr{X}}^\mu  \, \bar{\theta}_\perp
	+ 2\,\bar{\rho}_{{}_{\mathscr{J}\mathscr{J} \pi}}\bar{\mathscr{X}}_\rho \,\bar{\sigma}_{\perp}^{\mu\rho}
	+ 2\, 	\bar{\rho}_{{}_{\mathscr{J}\mathscr{K} m}}\bar{\mathscr{Y}}_\rho \,\bar{\chi}^{\mu\rho} . \label{fullJ}
\end{align} 

Further, we use first-order Navier-Srokes equation~\eqref{transJ1} and approximate $\bar{\mathscr{X}}_\rho\sim \mathscr{J}_\rho \bar{\rho}_\parallel^{-1}$ in the term $\G^{\mu\rho} D\bar{\mathscr{X}}_\rho \mathscr{W}_{\mathscr{J}\mathscr{J}}$. This yields the evolution equation for $\mathscr{J}^\mu$ given by
\begin{align}
	\mathscr{J}^{\mu} +\tau_{\mathscr{J}}\dot{\mathscr{J}}^\mu=& \, \bar{\rho}_\parallel \,\bar{\mathscr{X}}^\mu + \tau_{\mathscr{J}} \mathscr{J}^\mu D \ln\bar{\rho}_\parallel
	 + \Big[2DB \,\G^{\mu\rho}D u_\rho -\bar{\mathscr{X}}^\mu DB'\Big] \mathscr{W}_{\mathscr{J} h} - \G^{\mu\rho}\Big[D\mathcal{H}+H\bar{\theta}_\parallel\Big]D u_\rho \mathscr{W}_{\mathscr{J} l} \nonumber\\
	& + \bar{\mathfrak{U}}^\mu\,\bar{\rho}_{{}_{\mathscr{J}h}} 
	-2\,\bar{\rho}_{{}_{\mathscr{J}\mathscr{J}P_\parallel}} \bar{\mathscr{X}}^\mu  \, \bar{\theta}_\parallel
	- 2\,\bar{\rho}_{{}_{\mathscr{J}\mathscr{J}P_\perp}} \bar{\mathscr{X}}^\mu  \, \bar{\theta}_\perp
	+ 2\,\bar{\rho}_{{}_{\mathscr{J}\mathscr{J} \pi}}\bar{\mathscr{X}}_\rho \,\bar{\sigma}_{\perp}^{\mu\rho}
	+ 2\, 	\bar{\rho}_{{}_{\mathscr{J}\mathscr{K} m}}\bar{\mathscr{Y}}_\rho \,\bar{\chi}^{\mu\rho} . \label{evoJ}
\end{align}
Here we used $\dot{\mathscr{J}}^\mu=\G^\mu_{\h\nu}D\mathscr{J}^\nu$ and defined
\begin{eqnarray}
	\tau_{\mathscr{J}}=-\frac{ \mathscr{W}_{\mathscr{J} \mathscr{J}}}{\bar{\rho}_\parallel} .
\end{eqnarray}

\subsection{Second-order correction and evolution equation for $\hat{\mathscr{K}}^{\mu}$}

\subsubsection{Nonlocal correction from the two-point correlation function: $\big< \hat{\mathscr{K}}^\mu(x)\big>^{\text{(2),NL}}_2$}

Applying Eq.~\eqref{nlc} and evaluating $[\partial_\lambda \hat{\mathcal{C}}(x_1)]_{x_1=x}$ via Eq.~\eqref{delC3}, and subsequently invoking Curie’s symmetry principle, we arrive at
\begin{align}
	\Big< \hat{\mathscr{K}}^\mu(x)\Big>^{\text{(2),NL}}_2
	=&\frac{1}{2}\G^{\mu\rho}\,\partial_\lambda \bar{\mathscr{Y}}_\rho (x)\int d^4 x_1 \Big( \hat{\mathscr{K}}^\eta(x), \hat{\mathscr{K}}_\eta (x_1) \Big)(x_1-x)^\lambda \nonumber\\
	&+\G^{\mu\rho}\,\bar{\theta}_\parallel (x)\partial_\lambda b_\rho(x)\int d^4 x_1 \Big( \hat{\mathscr{K}}^\eta(x),\hat{f}_\eta (x_1) \Big)(x_1-x)^\lambda \nonumber\\
	& -\frac{1}{2}\G^{\mu\rho}\Big[(D\mathcal{H}(x)+H(x)\bar{\theta}_\parallel(x)) \partial_\lambda b_\rho(x)-\bar{\mathscr{Y}}_\rho(x) \partial_\lambda H (x) \Big]  \int d^4 x_1 \Big( \hat{\mathscr{K}}^\eta(x),\hat{g}_\eta (x_1) \Big)(x_1-x)^\lambda .
\end{align}
Further, employing~\eqref{freq_dep_trans} from Appendix~\ref{correlators}, we get
\begin{align}
	\big< \hat{\mathscr{K}}^\mu\big>^{\text{(2),NL}}_2
	=&\, \G^{\mu\rho} D\bar{\mathscr{Y}}_\rho \mathscr{W}_{\mathscr{K}\mathscr{K}} +  2 \G^{\mu\rho}Db_\rho \bar{\theta}_\parallel\mathscr{W}_{\mathscr{K} f} - \Big[(D\mathcal{H}+H\bar{\theta}_\parallel)\G^{\mu\rho}Db_\rho - \bar{\mathscr{Y}}^\mu D H \Big] \mathscr{W}_{\mathscr{K} g}  .  \label{final_K_nonlocal}
\end{align}
Here, for brevity, we omitted the explicit arguments and defined
\begin{align}
	\mathscr{W}_{\mathscr{K}\mathscr{K}} &= i \lim_{\omega\to 0} \frac{d}{d\omega} \bar{\eta}_{{}_{\mathscr{K}\mathscr{K}}}(\omega) , \\
	\mathscr{W}_{\mathscr{K} f} &= i \lim_{\omega\to 0} \frac{d}{d\omega} \bar{\eta}_{{}_{\mathscr{K} f}}(\omega) , \\
	\mathscr{W}_{\mathscr{K} g} &= i \lim_{\omega\to 0} \frac{d}{d\omega} \bar{\eta}_{{}_{\mathscr{J} g}}(\omega) ,
\end{align}
where the frequency dependent transport coefficients are defined by
\begin{align}
	\bar{\eta}_{{}_{\mathscr{K}\mathscr{K}}} (\omega)&=\frac{1}{2}\int d^4 x_1 e^{i\omega(t-t_1)}\Big( \hat{\mathscr{K}}^\eta(x), \hat{\mathscr{K}}_\eta (x_1) \Big) , \\
	\bar{\eta}_{{}_{\mathscr{K} f}} (\omega)&= \frac{1}{2} \int d^4 x_1 e^{i\omega(t-t_1)}\Big( \hat{\mathscr{K}}^\eta(x),\hat{f}_\eta (x_1) \Big) , \\
	\bar{\eta}_{{}_{\mathscr{K} g}} (\omega)&= \frac{1}{2} \int d^4 x_1 e^{i\omega(t-t_1)}\Big( \hat{\mathscr{K}}^\eta(x),\hat{g}_\eta (x_1) \Big).
\end{align}

\subsubsection{Local correction from the two-point correlation function using extended thermodynamic forces: $\big< \hat{\mathscr{K}}^\mu(x)\big>^{\text{(2),ET}}_2$}

Employing Eqs.~\eqref{etc} and~\eqref{CS}, and invoking Curie’s symmetry principle, we get
\begin{align}
	\Big< \hat{\mathscr{K}}^\mu(x)\Big>^{\text{(2),ET}}_2 = H(x)\bar{\mathfrak{B}}^\mu(x)\,\bar{\eta}_{{}_{\mathscr{K}g}} (x) , \label{final_K_extended}
\end{align}
where, we have defined
\begin{align}
	\bar{\mathfrak{B}}^\mu = \beta\, \G^{\mu\rho} \,(D b_\rho)_2,~~\bar{\eta}_{{}_{\mathscr{K}g}} (x) =\frac{1}{2}\int d^4 x_1 \Big( \hat{\mathscr{K}}^\lambda(x), \hat{g}_\lambda (x_1) \Big) .
\end{align}
Since $\hat{\mathscr{K}}^\mu$ and $\hat{h}^\mu$ possess opposite parity, Curie’s symmetry principle forbids their coupling.

\subsubsection{Local correction from the three point correlation function: $\big< \hat{\mathscr{K}}^\mu(x)\big>^{\text{(3)}}_2$}

Using Eqs.~\eqref{tpc} and~\eqref{CF}, and applying the method detailed in Sec.~\ref{secPi3P}, we determine that eight three-point correlators contribute nonvanishing terms to the local correction associated with the three-point function in $\hat{\mathscr{K}}^\mu$. Accordingly, we find
\begin{align}
	\Big< \hat{\mathscr{K}}^\mu (x)\Big>^{\text{(3)}}_2= \mathfrak{K}_1^\mu (x)+ \mathfrak{K}_2^\mu (x)+\mathfrak{K}_3^\mu (x)+\mathfrak{K}_4^\mu (x)+\mathfrak{K}_5^\mu (x)+\mathfrak{K}_6^\mu (x)+\mathfrak{K}_7^\mu (x)+\mathfrak{K}_8^\mu (x) . \label{K3point}
\end{align}
Here, we defined
\begin{align}
	\mathfrak{K}^\mu_1 (x)&=	-\frac{1}{2}\,\G^{\mu\rho}\,\bar{\mathscr{Y}}_\rho (x) \, \bar{\theta}_\parallel(x)\int d^4x_1 d^4x_2\,\big( \hat{\mathscr{K}}^\lambda(x),\, \hat{\mathscr{K}}_\lambda(x_1), \, \hat{P}_\parallel(x_2) \big) , \\
	\mathfrak{K}^\mu_2 (x)&=-	\frac{1}{2}\,\G^{\mu\alpha}\,\bar{\theta}_\parallel(x) \,\bar{\mathscr{Y}}_\alpha (x)  \int d^4x_1 d^4x_2\,\big( \hat{\mathscr{K}}^\lambda(x),\, \hat{P}_\parallel(x_1), \, \hat{\mathscr{K}}_\lambda(x_2) \big) ,  \\
	\mathfrak{K}^\mu_3 (x)&=-	\frac{1}{2}\,\G^{\mu\rho}\,\bar{\mathscr{Y}}_\rho (x) \, \bar{\theta}_\perp(x)\int d^4x_1 d^4x_2\,\big( \hat{\mathscr{K}}^\lambda(x),\, \hat{\mathscr{K}}_\lambda(x_1), \, \hat{P}_\perp(x_2) \big) , \\
	\mathfrak{K}^\mu_4 (x)&=-	\frac{1}{2}\,\G^{\mu\alpha}	\,\bar{\theta}_\perp(x) \,\bar{\mathscr{Y}}_\alpha (x) \int d^4x_1 d^4x_2\,\big( \hat{\mathscr{K}}^\lambda(x),\, \hat{P}_\perp(x_1), \, \hat{\mathscr{K}}_\lambda(x_2) \big) , \\
	\mathfrak{K}^\mu_5 (x)&=	\frac{1}{2}\,\G^{\mu\rho\alpha\beta}\,\bar{\mathscr{Y}}_\rho (x)\,\bar{\sigma}_{\perp\alpha\beta}(x)\int d^4x_1 d^4x_2\,\big( \hat{\mathscr{K}}^\lambda(x),\, \hat{\mathscr{K}}^\eta(x_1), \, \hat{\pi}_{\perp\lambda\eta}(x_2) \big) , \\
	\mathfrak{K}^\mu_6 (x)&=	\frac{1}{2}\,\G^{\mu\rho\sigma\alpha}	\,\bar{\sigma}_{\perp\rho\sigma}(x)\bar{\mathscr{Y}}_\alpha (x) \int d^4x_1 d^4x_2\,\big( \hat{\mathscr{K}}^\lambda(x),\, \hat{\pi}_{\perp\lambda\eta}(x_1), \, \hat{\mathscr{K}}^\eta(x_2) \big) , \\
	\mathfrak{K}^\mu_7 (x)&=	\Gc^{\mu\rho\alpha\beta}\,\bar{\mathscr{X}}_\rho (x)\,\bar{\chi}_{\alpha\beta}(x)\int d^4x_1 d^4x_2\,\big( \hat{\mathscr{K}}^\lambda(x),\, \hat{\mathscr{J}}^\eta(x_1), \, \hat{m}_{\lambda\eta}(x_2) \big) , \\
	\mathfrak{K}^\mu_8 (x)&=	\Gc^{\mu\alpha\rho\sigma}\,\bar{\chi}_{\rho\sigma}(x)\bar{\mathscr{X}}_\alpha (x)\int d^4x_1 d^4x_2\,\big( \hat{\mathscr{K}}^\lambda(x),\, \hat{m}_{\lambda\eta}(x_1), \, \hat{\mathscr{J}}^\eta(x_2) \big) .
\end{align}
Using the symmetry~\eqref{threepointsymmetry}, we can infer that $\mathfrak{K}_1^\mu = \mathfrak{K}_2^\mu$, $\mathfrak{K}_3^\mu = \mathfrak{K}_4^\mu$, $\mathfrak{K}_5^\mu = \mathfrak{K}_6^\mu$, and $\mathfrak{K}_7^\mu = \mathfrak{K}_8^\mu$. Hence, we can rewrite~\eqref{K3point} as
\begin{align}
	\big< \hat{\mathscr{K}}^\mu\big>^{\text{(3)}}_2 = -2\,\bar{\eta}_{{}_{\mathscr{K}\mathscr{K}P_\parallel}} \bar{\mathscr{Y}}^\mu  \, \bar{\theta}_\parallel
	- 2\,\bar{\eta}_{{}_{\mathscr{K}\mathscr{K}P_\perp}} \bar{\mathscr{Y}}^\mu  \, \bar{\theta}_\perp
	+ 2\,\bar{\eta}_{{}_{\mathscr{K}\mathscr{K} \pi}}\bar{\mathscr{Y}}_\rho \,\bar{\sigma}_{\perp}^{\mu\rho}
	+ 2\, 	\bar{\eta}_{{}_{\mathscr{K}\mathscr{J} m}}\bar{\mathscr{X}}_\rho \,\bar{\chi}^{\mu\rho} . \label{final_K_three_point}
\end{align}
Here, for brevity, we omitted the explicit arguments and defined the transport coefficients as
\begin{align}
	\bar{\eta}_{{}_{\mathscr{K}\mathscr{K}P_\parallel}}(x) &= \frac{1}{2}\int d^4x_1 d^4x_2\, \big( \hat{\mathscr{K}}^\lambda(x),\, \hat{\mathscr{K}}_\lambda(x_1), \, \hat{P}_\parallel(x_2) \big) , \\
	\bar{\eta}_{{}_{\mathscr{K}\mathscr{K}P_\perp}}(x) &= \frac{1}{2}\int d^4x_1 d^4x_2\, \big( \hat{\mathscr{K}}^\lambda(x),\, \hat{\mathscr{K}}_\lambda(x_1), \, \hat{P}_\perp(x_2) \big) ,  \\
	\bar{\eta}_{{}_{\mathscr{K}\mathscr{K} \pi}}(x) &=\frac{1}{2}\int d^4x_1 d^4x_2\,\big( \hat{\mathscr{K}}^\lambda(x),\, \hat{\mathscr{K}}^\eta(x_1), \, \hat{\pi}_{\perp\lambda\eta}(x_2) \big) , \\
	\bar{\eta}_{{}_{\mathscr{K}\mathscr{J} m}}(x) &=\int d^4x_1 d^4x_2\,\big( \hat{\mathscr{K}}^\lambda(x),\, \hat{\mathscr{J}}^\eta(x_1), \, \hat{m}_{\lambda\eta}(x_2) \big) .
\end{align}

\subsubsection{Full second order expression and the evolution equation for $\hat{\mathscr{K}}^{\mu}$}

In order to get the full expression of $\hat{\mathscr{K}}^{\mu}$ up to the second order, we collect the first-order term from~\eqref{transK1} and second-order terms from~\eqref{final_K_nonlocal},~\eqref{final_K_extended}, and~\eqref{final_K_three_point}, we obtain
\begin{align}
	\mathscr{K}^{\mu} =&  \, \bar{\eta}_\parallel\,\bar{\mathscr{Y}}^\mu  
	+ \G^{\mu\rho} D\bar{\mathscr{Y}}_\rho \mathscr{W}_{\mathscr{K}\mathscr{K}} +  2 \G^{\mu\rho}Db_\rho \bar{\theta}_\parallel\mathscr{W}_{\mathscr{K} f} - \Big[(D\mathcal{H}+H\bar{\theta}_\parallel)\G^{\mu\rho}Db_\rho - \bar{\mathscr{Y}}^\mu D H \Big] \mathscr{W}_{\mathscr{K} g}   \nonumber\\
	& + H\,\bar{\mathfrak{B}}^\mu\,\bar{\eta}_{{}_{\mathscr{K}g}} 
	-2\,\bar{\eta}_{{}_{\mathscr{K}\mathscr{K}P_\parallel}} \bar{\mathscr{Y}}^\mu  \, \bar{\theta}_\parallel
	- 2\,\bar{\eta}_{{}_{\mathscr{K}\mathscr{K}P_\perp}} \bar{\mathscr{Y}}^\mu  \, \bar{\theta}_\perp
	+ 2\,\bar{\eta}_{{}_{\mathscr{K}\mathscr{K} \pi}}\bar{\mathscr{Y}}_\rho \,\bar{\sigma}_{\perp}^{\mu\rho}
	+ 2\, 	\bar{\eta}_{{}_{\mathscr{K}\mathscr{J} m}}\bar{\mathscr{X}}_\rho \,\bar{\chi}^{\mu\rho} . \label{fullK}
\end{align}

Further, we use first-order Navier-Srokes equation~\eqref{transK1} and approximate $\bar{\mathscr{Y}}_\rho\sim \mathscr{K}_\rho \bar{\eta}_\parallel^{-1}$ in the term $\G^{\mu\rho} D\bar{\mathscr{Y}}_\rho \mathscr{W}_{\mathscr{K}\mathscr{K}}$. This yields the evolution equation for $\mathscr{K}^\mu$ given by
\begin{align}
	\mathscr{K}^{\mu} + \tau_{\mathscr{K}} \dot{\mathscr{K}}^\mu =&  \, \bar{\eta}_\parallel\,\bar{\mathscr{Y}}^\mu  
	+ \tau_{\mathscr{K}}\mathscr{K}^\mu D\ln \bar{\eta}_\parallel +  2 \G^{\mu\rho}Db_\rho \bar{\theta}_\parallel\mathscr{W}_{\mathscr{K} f} - \Big[(D\mathcal{H}+H\bar{\theta}_\parallel)\G^{\mu\rho}Db_\rho - \bar{\mathscr{Y}}^\mu D H \Big] \mathscr{W}_{\mathscr{K} g}   \nonumber\\
	& + H\,\bar{\mathfrak{B}}^\mu\,\bar{\eta}_{{}_{\mathscr{K}g}} 
	-2\,\bar{\eta}_{{}_{\mathscr{K}\mathscr{K}P_\parallel}} \bar{\mathscr{Y}}^\mu  \, \bar{\theta}_\parallel
	- 2\,\bar{\eta}_{{}_{\mathscr{K}\mathscr{K}P_\perp}} \bar{\mathscr{Y}}^\mu  \, \bar{\theta}_\perp
	+ 2\,\bar{\eta}_{{}_{\mathscr{K}\mathscr{K} \pi}}\bar{\mathscr{Y}}_\rho \,\bar{\sigma}_{\perp}^{\mu\rho}
	+ 2\, 	\bar{\eta}_{{}_{\mathscr{K}\mathscr{J} m}}\bar{\mathscr{X}}_\rho \,\bar{\chi}^{\mu\rho} .  \label{evoK}
\end{align}
Here we used $\dot{\mathscr{K}}^\mu=\G^\mu_{\h\nu}D\mathscr{K}^\nu$ and defined
\begin{eqnarray}
	\tau_{\mathscr{K}}=-\frac{ \mathscr{W}_{\mathscr{K} \mathscr{K}}}{\bar{\eta}_\parallel}.
\end{eqnarray}

	\section{Summary and discussion}
	
	For completeness and convenience, we collect the evolution equations for all dissipative tensors below:
	\begin{align}
		\pi_\perp^{\mu\nu} +\tau_\pi \dot{\pi}^{\mu\nu}_\perp =& \, 2 \,\bar{\eta}_\perp \bar\sigma_{\perp}^{\mu\nu} + \tau_\pi \pi^{\mu\nu}_\perp D \ln \bar{\eta}_\perp - 2\big[ B'\bar{\mathscr{X}}^{<\mu} D u^{\nu>} +  \bar{\mathscr{Y}}^{<\mu} D b^{\nu>}\big]\mathscr{W}_{\pi\pi} \nonumber\\
		& +  \bar{\mathscr{X}}^{<\mu}\,\bar{\mathscr{X}}^{\nu>}\, \bar{\eta}_{{}_{\pi\mathscr{J}\mathscr{J}}}
		+ \bar{\mathscr{Y}}^{<\mu}\,\bar{\mathscr{Y}}^{\nu>}\, \bar{\eta}_{{}_{\pi\mathscr{K}\mathscr{K}}}
		- 2\,\bar{\sigma}^{\mu\nu}_{\perp}\,\bar{\theta}_\perp \, \bar{\eta}_{{}_{\pi\pi P_\perp}}
		- 2\,\bar{\sigma}^{\mu\nu}_{\perp}\,\bar{\theta}_\parallel \, \bar{\eta}_{{}_{\pi\pi P_\parallel}}  .\\
		\	\mathscr{K}^{\mu} + \tau_{\mathscr{K}} \dot{\mathscr{K}}^\mu =&  \, \bar{\eta}_\parallel\,\bar{\mathscr{Y}}^\mu  
		+ \tau_{\mathscr{K}}\mathscr{K}^\mu D\ln \bar{\eta}_\parallel +  2 \G^{\mu\rho}Db_\rho \bar{\theta}_\parallel\mathscr{W}_{\mathscr{K} f} - \Big[(D\mathcal{H}+H\bar{\theta}_\parallel)\G^{\mu\rho}Db_\rho - \bar{\mathscr{Y}}^\mu D H \Big] \mathscr{W}_{\mathscr{K} g}   \nonumber\\
		& + H\,\bar{\mathfrak{B}}^\mu\,\bar{\eta}_{{}_{\mathscr{K}g}} 
		-2\,\bar{\eta}_{{}_{\mathscr{K}\mathscr{K}P_\parallel}} \bar{\mathscr{Y}}^\mu  \, \bar{\theta}_\parallel
		- 2\,\bar{\eta}_{{}_{\mathscr{K}\mathscr{K}P_\perp}} \bar{\mathscr{Y}}^\mu  \, \bar{\theta}_\perp
		+ 2\,\bar{\eta}_{{}_{\mathscr{K}\mathscr{K} \pi}}\bar{\mathscr{Y}}_\rho \,\bar{\sigma}_{\perp}^{\mu\rho}
		+ 2\, 	\bar{\eta}_{{}_{\mathscr{K}\mathscr{J} m}}\bar{\mathscr{X}}_\rho \,\bar{\chi}^{\mu\rho} .
	\end{align}
	\begin{align}
		\Pi_\parallel +\tau_{\Pi_\parallel} \dot{\Pi}_\parallel =& -\bar{\zeta}_\parallel \bar{\theta}_\parallel -  \bar{\zeta}_\times \bar{\theta}_\perp -\bar{\zeta}_\times D \bar{\theta}_\perp (\tau_{\Pi_\parallel}-\tau_{\Pi_\times}) + \tau_{\Pi_\parallel} (\Pi_\parallel +\bar{\theta}_\perp \bar{\zeta}_\times) D\ln \bar{\zeta}_\parallel  - \tau_{\Pi_\parallel}\bar{\theta}_\perp D\bar{\zeta}_\times\nonumber\\
		&+\bar{\theta}^2_\parallel \, \mathfrak{z}_{{}_{\parallel}} +\bar{\theta}^2_\perp \, \mathfrak{z}_{{}_{\perp}} + 2\, \bar{\theta}_\parallel \bar{\theta}_\perp \, \mathfrak{z}_{{}_{\times}}  +  \bar{\theta}_\parallel D \gamma_\parallel \mathscr{W}_{P_\parallel \varepsilon} 
		+  \bar{\theta}_\parallel D \phi_\parallel \mathscr{W}_{P_\parallel B} \nonumber\\
		&+  \bar{\theta}_\perp D \gamma_\perp \mathscr{W}_{P_\parallel \varepsilon}
		+  \bar{\theta}_\perp D \phi_\perp \mathscr{W}_{P_\parallel B} 
		+ \Theta_T \, \zeta_{{}_{P_\parallel \beta}} 	- \Theta_F \,	\bar{\zeta}_{{}_{P_\parallel \mathcal{H}}} \nonumber\\
		& + \bar{\zeta}_{{}_{P_\parallel  \mathscr{J}\mathscr{J}}} \bar{\mathscr{X}}_\mu  \bar{\mathscr{X}}^\mu  
		+ \bar{\zeta}_{{}_{P_\parallel  \mathscr{K}\mathscr{K}}} \bar{\mathscr{Y}}_\mu  \bar{\mathscr{Y}}^\mu
		+ 	\bar{\zeta}_{{}_{P_\parallel  m m }} \bar{\chi}_{\mu\nu} \bar{\chi}^{\mu\nu} 
		+ 	\bar{\zeta}_{{}_{P_\parallel  \pi \pi }} \bar{\sigma}_{\perp\mu\nu} \bar{\sigma}_{\perp}^{\mu\nu}.\\
		\Pi_\perp +\tau_{\Pi_\perp} \dot{\Pi}_\perp =& -\bar{\zeta}_\perp \bar{\theta}_\perp -  \bar{\zeta}'_\times \bar{\theta}_\parallel  
		-\bar{\zeta}'_{\times}D \bar{\theta}_\parallel	(\tau_{\Pi_\perp}-\tau'_{\Pi_\times}) + \tau_{\Pi_\perp} (\Pi_\perp+\bar{\theta}_\parallel\bar{\zeta}'_{\times}) D\ln \bar{\zeta}_\perp -\tau_{\Pi_\perp}\bar{\theta}_\parallel D\bar{\zeta}'_{\times}
		\nonumber\\
		& +\bar{\theta}^2_\perp \, \mathfrak{z}'_{{}_{\perp}}
		+\bar{\theta}^2_\parallel \,\mathfrak{z}'_{{}_{\parallel}}  + 2\, \bar{\theta}_\parallel \bar{\theta}_\perp \, \mathfrak{z}'_{{}_{\times}}  +  \bar{\theta}_\parallel D \gamma_\parallel \mathscr{W}_{P_\perp \varepsilon} 
		+  \bar{\theta}_\parallel D \phi_\parallel \mathscr{W}_{P_\perp B} \nonumber\\
		&+  \bar{\theta}_\perp D \gamma_\perp \mathscr{W}_{P_\perp \varepsilon}
		+  \bar{\theta}_\perp D \phi_\perp \mathscr{W}_{P_\perp B}
		+	\Theta_T\, \zeta_{{}_{P_\perp \beta}} - \Theta_F \,	\bar{\zeta}_{{}_{P_\perp \mathcal{H}}}  \nonumber\\
		& + \bar{\zeta}_{{}_{P_\perp  \mathscr{J}\mathscr{J}}} \bar{\mathscr{X}}_\mu  \bar{\mathscr{X}}^\mu  
		+ \bar{\zeta}_{{}_{P_\perp  \mathscr{K}\mathscr{K}}} \bar{\mathscr{Y}}_\mu  \bar{\mathscr{Y}}^\mu
		+ 	\bar{\zeta}_{{}_{P_\perp  m m }} \bar{\chi}_{\mu\nu} \bar{\chi}^{\mu\nu} 
		+ 	\bar{\zeta}_{{}_{P_\perp  \pi \pi }} \bar{\sigma}_{\perp\mu\nu} \bar{\sigma}_{\perp}^{\mu\nu} .
	\end{align} 
	\begin{align}
			m^{\mu\nu} +\tau_m \dot{m}^{\mu\nu}  =&  \, 2 \, \bar{\rho}_\perp \bar{\chi}^{\mu\nu} + \tau_m m^{\mu\nu} D \ln \bar{\rho}_\perp + 2\big[    \bar{\mathscr{X}}^{\cll \mu}  D b^{\nu\crr}  + H \bar{\mathscr{Y}}^{\cll \mu} D u^{\nu\crr} \big] \mathscr{W}_{m m}\nonumber\\
		& + 2\,\bar{\mathscr{X}}^{\cll\mu}\,\bar{\mathscr{Y}}^{\nu\crr}\, \bar{\rho}_{{}_{m\mathscr{J}\mathscr{K}}}
		- 2\,\bar{\chi}^{\mu\nu}\,\bar{\theta}_\perp \, \bar{\rho}_{{}_{m m P_\perp}}
		- 2\,\bar{\chi}^{\mu\nu}\,\bar{\theta}_\parallel \, \bar{\rho}_{{}_{m m P_\parallel}} .\\
		\mathscr{J}^{\mu} +\tau_{\mathscr{J}}\dot{\mathscr{J}}^\mu=& \, \bar{\rho}_\parallel \,\bar{\mathscr{X}}^\mu + \tau_{\mathscr{J}} \mathscr{J}^\mu D \ln\bar{\rho}_\parallel
		+ \Big[2DB \,\G^{\mu\rho}D u_\rho -\bar{\mathscr{X}}^\mu DB'\Big] \mathscr{W}_{\mathscr{J} h} - \G^{\mu\rho}\Big[D\mathcal{H}+H\bar{\theta}_\parallel\Big]D u_\rho \mathscr{W}_{\mathscr{J} l} \nonumber\\
		& + \bar{\mathfrak{U}}^\mu\,\bar{\rho}_{{}_{\mathscr{J}h}} 
		-2\,\bar{\rho}_{{}_{\mathscr{J}\mathscr{J}P_\parallel}} \bar{\mathscr{X}}^\mu  \, \bar{\theta}_\parallel
		- 2\,\bar{\rho}_{{}_{\mathscr{J}\mathscr{J}P_\perp}} \bar{\mathscr{X}}^\mu  \, \bar{\theta}_\perp
		+ 2\,\bar{\rho}_{{}_{\mathscr{J}\mathscr{J} \pi}}\bar{\mathscr{X}}_\rho \,\bar{\sigma}_{\perp}^{\mu\rho}
		+ 2\, 	\bar{\rho}_{{}_{\mathscr{J}\mathscr{K} m}}\bar{\mathscr{Y}}_\rho \,\bar{\chi}^{\mu\rho} . 
	\end{align}

This work develops a complete second-order formulation of relativistic magnetohydrodynamics (RMHD) within the framework of Zubarev’s nonequilibrium statistical operator (NESO), based on the modern, symmetry-driven foundation in which the total energy-momentum conservation and the Bianchi identity (magnetic-flux conservation) serve as the fundamental dynamical equations. This approach avoids the conceptual limitations of the conventional RMHD formulation, which separates the fluid and electromagnetic energy-momentum tensors and relies on the assumption of infinite conductivity to remove the electric field from the hydrodynamic sector. By contrast, in the present formalism the magnetic field is treated as a true hydrodynamic variable at $\mathcal{O}(\partial^0)$, whereas the electric field appears only as a higher-order induced effect encoded in the dissipative corrections. 

We include a full extraction of nonlocal contributions by expanding not only the thermodynamic forces coupled to $\hat{T}^{\mu\nu}$  and $\hat{\tilde{F}}^{\mu\nu}$, but also the hydrodynamic fields entering the tensor structures of these operators themselves. This improves upon previous treatments, which expanded only selected thermodynamic forces and omitted the hydrodynamic fields inside the operators. This refinement leads to the elimination of several terms from the evolution equations for both the bulk viscous pressure and the dissipative vectors. By fully expanding all hydrodynamic quantities contained in $\hat{\mathcal{C}}$, we find that the evolution equation for the bulk viscous pressure no longer contains terms of the form $\dot{u}_\mu \bar{\mathscr{X}}^\mu$, $\dot{u}_\mu \bar{\mathscr{Y}}^\mu$. Likewise, the evolution equations for the dissipative vectors $\mathscr{J}^\mu$ and $\mathscr{K}^\mu$ no longer include contributions of the type $\dot{u}_\nu \bar{\sigma}_\perp^{\mu\nu}$ and $\dot{u}_\nu \bar{\chi}^{\mu\nu}$.

Furthermore, we find that the nonlinear terms involving three-point correlations between the second-rank tensors operators $\hat{\pi}_\perp^{\mu\nu}$ and $\hat{m}^{\mu\nu}$ are absent in the evolution equations for $\pi_\perp^{\mu\nu}$ and $m^{\mu\nu}$. This can be traced to two reasons.
First, certain three-point correlators—such as $\big( \hat\pi_{\perp}^{\mu\nu}(x),\, \hat m^{\rho\sigma}(x_1), \, \hat\pi_{\perp}^{\alpha\beta}(x_2) \big)$—do not preserve the parity of the equilibrium state. By Curie’s principle, such terms must vanish and are therefore discarded. Second, even among the correlators that do preserve parity—such as $\big( \hat\pi_{\perp}^{\mu\nu}(x),\, \hat\pi_{\perp}^{\rho\sigma}(x_1), \, \hat\pi_{\perp}^{\alpha\beta}(x_2) \big)$ and $\big( \hat\pi_{\perp}^{\mu\nu}(x),\, \hat m^{\rho\sigma}(x_1), \, \hat m^{\alpha\beta}(x_2) \big)$—the corresponding transport coefficients are found to vanish due to the specific tensor structures involved.

In this work, we focus on a parity-conserving, charge-conjugation-symmetric plasma and therefore do not include Hall transport. An important direction for future research is to extend the present formulation to systems exhibiting Hall responses within the fully conserved and symmetry-based framework developed here, and to derive the corresponding evolution equations for Hall transport. It would also be of considerable interest to incorporate spin degrees of freedom together with magnetic fields. Although first-order theories combining spin and magnetohydrodynamics have been formulated—both within kinetic theory~\cite{Bhadury:2022ulr} and nonequilibrium statistical approaches~\cite{Fang:2024hxa}—a complete second-order relativistic hydrodynamic theory that consistently includes both spin and magnetic fields has not yet been established.

%\section*{Acknowledgment}

	 \begin{appendix}
	 	\section{Three-point correlators}\label{three_point_correlatos}

		We begin by expressing the correlator 
		$\big( \hat\pi_{\perp}^{\mu\nu}(x),\, \hat\pi_{\perp}^{\rho\sigma}(x_1),\, \hat\pi_{\perp}^{\alpha\beta}(x_2) \big)$ 
		in the tensor basis provided by the decomposition:
		\begin{align}
			\Big( \hat\pi_{\perp}^{\mu\nu}(x),\, \hat\pi_{\perp}^{\rho\sigma}(x_1), \, \hat\pi_{\perp}^{\alpha\beta}(x_2) \Big) &= a_1\,\G^{\mu\nu}\G^{\rho\alpha}\G^{\sigma\beta} +a_2 \, \G^{\mu\nu}\G^{\rho\beta}\G^{\sigma\alpha} +b_1 \, \G^{\rho\sigma}\G^{\mu\alpha}\G^{\nu\beta} +b_2\, \G^{\rho\sigma}\G^{\mu\beta}\G^{\nu\alpha}\nonumber\\
			&~~~ +c_1 \,\G^{\alpha\beta}\G^{\mu\rho}\G^{\nu\sigma} +c_2 \, \G^{\alpha\beta}\G^{\mu\sigma}\G^{\nu\rho} +d\, \G^{\mu\nu}\G^{\rho\sigma}\G^{\alpha\beta} \nonumber\\
			&~~~ +e_1 \,\G^{\mu\rho}\G^{\nu\alpha}\G^{\sigma\beta} +e_2\, \G^{\mu\rho}\G^{\nu\beta}\G^{\sigma\alpha} +e_3\, \G^{\mu\sigma}\G^{\nu\alpha}\G^{\rho\beta} +e_4\, \G^{\mu\sigma}\G^{\nu\beta}\G^{\rho\alpha} \nonumber\\
			&~~~ +e_5\, \G^{\mu\alpha}\G^{\nu\rho}\G^{\sigma\beta} +e_6\, \G^{\mu\alpha}\G^{\nu\sigma}\G^{\rho\beta} +e_7\, \G^{\mu\beta}\G^{\nu\rho}\G^{\sigma\alpha} +e_8\, \G^{\mu\beta}\G^{\nu\sigma}\G^{\rho\alpha}, \label{tenst1}
		\end{align}

		Since $\hat\pi_{\perp}^{\mu\nu}$ is a symmetric, traceless tensor, the correlator must be symmetric under the exchanges 
		$\mu \leftrightarrow \nu$, $\rho \leftrightarrow \sigma$, and $\alpha \leftrightarrow \beta$, 
		and it must vanish upon contraction $\mu=\nu$, $\rho=\sigma$, and $\alpha=\beta$ in accordance with tracelessness. 
		Imposing, in particular, the symmetry under $\alpha \leftrightarrow \beta$ yields
		\begin{align}
			a_1 = a_2,~ b_1 = b_2,~ e_1 = e_2,~ e_3 = e_4,~ e_5 = e_7,~ e_6 = e_8 . \label{value1}
		\end{align}
		Similarly, enforcing symmetry under the exchange $\rho \leftrightarrow \sigma$ yields
		\begin{align}
			c_1 = c_2,~ e_2 = e_4,~ e_7 = e_8 . \label{value2}
		\end{align}
		Finally, imposing symmetry under the exchange $\mu \leftrightarrow \nu$, we get
		\begin{align}
			e_4 = e_8 . \label{value3}
		\end{align}
		Substituting~\eqref{value1}, \eqref{value2}, and \eqref{value3} in~\eqref{tenst1}, we obtain
		\begin{align}
			\Big( \hat\pi_{\perp}^{\mu\nu}(x),\, \hat\pi_{\perp}^{\rho\sigma}(x_1), \, \hat\pi_{\perp}^{\alpha\beta}(x_2) \Big) &= a_2\,(\G^{\mu\nu}\G^{\rho\alpha}\G^{\sigma\beta} + \, \G^{\mu\nu}\G^{\rho\beta}\G^{\sigma\alpha}) +b_2 (\, \G^{\rho\sigma}\G^{\mu\alpha}\G^{\nu\beta} +\, \G^{\rho\sigma}\G^{\mu\beta}\G^{\nu\alpha})\nonumber\\
			&~~~ +c_2 (\,\G^{\alpha\beta}\G^{\mu\rho}\G^{\nu\sigma} + \, \G^{\alpha\beta}\G^{\mu\sigma}\G^{\nu\rho}) +d\, \G^{\mu\nu}\G^{\rho\sigma}\G^{\alpha\beta} \nonumber\\
			&~~~ +e_8 \Big(\,\G^{\mu\rho}\G^{\nu\alpha}\G^{\sigma\beta} +\, \G^{\mu\rho}\G^{\nu\beta}\G^{\sigma\alpha} +\, \G^{\mu\sigma}\G^{\nu\alpha}\G^{\rho\beta} +\, \G^{\mu\sigma}\G^{\nu\beta}\G^{\rho\alpha} \nonumber\\
			&~~~ +\, \G^{\mu\alpha}\G^{\nu\rho}\G^{\sigma\beta} +\, \G^{\mu\alpha}\G^{\nu\sigma}\G^{\rho\beta} +\, \G^{\mu\beta}\G^{\nu\rho}\G^{\sigma\alpha} +\, \G^{\mu\beta}\G^{\nu\sigma}\G^{\rho\alpha}\Big), \label{tenst2}
		\end{align}
		Next, we impose tracelessness by contracting $\mu=\nu$, $\rho=\sigma$, and $\alpha=\beta$ and equate the corresponding tensor coefficients to zero, we obtain the following six linear constraints:
		\begin{align}
			2 a_2 + 4 e_8 &= 0, \label{valeq1}\\
			 2 b_2 + 2 c_2 + 2 d &= 0,\\
			  2 b_2 + 4 e_8 &= 0, \\
			2 c_2 + 2 d + 2 a_2 &= 0,\\
			 2 c_2 + 4 e_8 &= 0, \\
			 2 a_2 + 2 b_2 + 2 d &= 0 . \label{valeq2} 
		\end{align} 
		Solving this linear equations for $a_{2},\, b_{2},\, c_{2},\, d,$ and $e_{8}$ yields		
		\begin{align}
			b_2 = a_2,~ c_2= a_2,~ d = -2 a_2,~ e_8 = -(a_2/2) . \label{value}
		\end{align}
		Finally, substituting Eq.~\eqref{value} into Eq.~\eqref{tenst2}, we find that the trace of the correlator 
		$\big( \hat\pi_{\perp}^{\mu\nu}(x),\, \hat\pi_{\perp}^{\rho\sigma}(x_1),\, \hat\pi_{\perp}^{\alpha\beta}(x_2) \big)$ 
		vanishes, $\Big( \hat\pi_{\perp\eta}^{\lambda}(x),\, \hat\pi_{\perp\delta}^{\eta}(x_1), \, \hat\pi_{\perp\lambda}^{\delta}(x_2) \Big)=0$
		where we used $\G^{\mu\nu}_{\h\h\mu\nu}=2$. 
		Consequently, the transport coefficient associated with 
		$\big( \hat\pi_{\perp}^{\mu\nu}(x),\, \hat\pi_{\perp}^{\rho\sigma}(x_1),\, \hat\pi_{\perp}^{\alpha\beta}(x_2) \big)$ 
		is zero. 
		By the same reasoning, one also obtains 
		$\Big( \hat\pi_{\perp\eta}^{\lambda}(x),\, \hat m_{\h\delta}^{\eta}(x_1), \, \hat m_{\h\lambda}^{\delta}(x_2) \Big)=0$.

		\section{Important relations}\label{important_relation}

	Let us simplify the term $\Uc^{\alpha\beta}_{\h\h\mu}\partial_\alpha \mathcal{H}_\beta$. 
	Substituting $\Uc^{\alpha\beta}_{\h\h\mu}$ from Eq.~\eqref{cutprojectorU} and using $\mathcal{H}_\beta=-\mathcal{H}\,b_\beta$, together with the transversality properties of $\G^{\alpha}_{\h \mu}$ with respect to $u^\mu$ and $b^\mu$, we obtain
		\begin{align}
			\Uc^{\alpha\beta}_{\h\h\mu}\partial_\alpha \mathcal{H}_\beta =  \frac{1}{2} \mathcal{H} \left[
			b_\mu b^\alpha u^\beta \partial_\alpha b_\beta 
			- D b_\mu + u^\beta \partial_\mu b_\beta 	\right] . \label{UcdelH1}
		\end{align}
		From Eq.~\eqref{Db}, we retain the terms up to first order in the gradient expansion. Thus we can write
		\begin{align}
			D b_{\mu}
			= \beta^{-1} u_{\mu}  b^\nu\,\partial_{\nu}\beta
			+  b^{\nu}\partial_{\nu}u_{\mu}
			-  b_{\mu}\theta_{\parallel} . \label{Db1}
		\end{align}
		Substituting the value of $Db_\mu$ from~\eqref{Db1} in~\eqref{UcdelH1}, we obtain 
		\begin{align}
		2\mathcal{H}^{-1} \Uc^{\alpha\beta}_{\h\h\mu}\partial_\alpha \mathcal{H}_\beta
		=
		\Big[
		2 b_\mu \theta_{\parallel}
		+ u^\beta \partial_\mu b_\beta 
		- \beta^{-1} u_\mu b^\nu \partial_\nu \beta
		- b^\nu \partial_\nu u_\mu
		\Big].  \label{UcdelH2}
	\end{align}

		Next, we simplify the term $\B^{\alpha\beta}_{\h\h\mu}\partial_\alpha \beta_\beta$. 
		Substituting $\B^{\alpha\beta}_{\h\h\mu}$ from Eq.~\eqref{noncutprojectorB} and using $\beta_\beta=\beta\,u_\beta$, together with the transversality properties of $\G^{\alpha}_{\h \mu}$ with respect to $u^\mu$ and $b^\mu$, we obtain	
		\begin{align}
		\B^{\alpha\beta}_{\h\h\mu}\partial_\alpha \beta_\beta
		&=
		\frac{1}{2} \beta \Big[
		2 b^\alpha b^\beta b_\mu \partial_\alpha u_\beta
		- u_\mu b^\beta D u_\beta
		+ b^\alpha \partial_\alpha u_\mu
		+ b^\beta \partial_\mu u_\beta
		\Big]\nonumber\\
	&	= -\frac{1}{2}\beta
		\Big[
		2 b_\mu \theta_{\parallel}
		+ u^\beta \partial_\mu b_\beta
		- \beta^{-1} u_\mu b^\nu \partial_\nu \beta
		- b^\nu \partial_\nu u_\mu
		\Big],  \label{Bdelbeta1}
		\end{align}
		where we have used the identities 
		$b^\alpha b^\beta \partial_\alpha u_\beta = -\theta_\parallel$ 
		and 
		$b^\beta D u_\beta = -\,u^\beta D b_\beta$, 
		together with Eq.~\eqref{Db1}.

		Inferring to~\eqref{UcdelH2} and~\eqref{Bdelbeta1}, we obtain the relation  valid up to first order in gradient expansion, given by
		\begin{align}
			\Uc^{\alpha\beta}_{\h\h\mu}\partial_\alpha \mathcal{H}_\beta
			=
			- H \, \B^{\alpha\beta}_{\h\h\mu}\partial_\alpha \beta_\beta = -\frac{1}{2} H \, \bar{\mathscr{Y}}_\mu , \label{UctoBrelation}
		\end{align}
		where we have used the definition, $\bar{\mathscr{Y}}_\mu=2\,\B^{\alpha\beta}_{\h\h\mu}\partial_\alpha \beta_\beta$.

		Furthermore, we simplify the structure $\U^{\alpha\beta}_{\h\h\mu}\,\partial_\alpha \beta_\beta$. 
		Inserting $\U^{\alpha\beta}_{\h\h\mu}$ from Eq.~\eqref{noncutprojectorU} and using $\beta_\beta=\beta\,u_\beta$, together with the transversality of $\G^{\alpha}_{\h \mu}$ with respect to $u^\mu$ and $b^\mu$, we obtain
	\begin{align}
			\U^{\alpha\beta}_{\h\h\mu}\partial_\alpha \beta_\beta = \frac{1}{2}\G^\alpha_{\h\mu}\partial_\alpha\beta + \frac{1}{2} \beta (b_\mu b^\beta Du_\beta + Du_\mu) .
		\end{align}
		Using the identity $b^\beta D u_\beta = -\,u^\beta D b_\beta$ and substituting the first-order expressions for $D b_\beta$ and $D u_\mu$ from Eqs.~\eqref{Db} and~\eqref{Du}, respectively, we obtain
		\begin{align}
			\U^{\alpha\beta}_{\h\h\mu}\partial_\alpha \beta_\beta &= \frac{1}{2}\G^\alpha_{\h\mu}\partial_\alpha\beta -\frac{1}{2}\nabla_\mu\beta -\frac{1}{2}b_\mu b^\nu \partial_\nu \beta\nonumber\\
			&~~ - \frac{B}{\varepsilon + p_\perp} \Big( -b^\nu \nabla_{[\mu}\mathcal{H}_{\nu]}+\frac{1}{2}u_\mu \beta H \theta_\parallel \Big).
		\end{align}
		Next, employing $\nabla_\mu\equiv\Delta^\alpha_{\h\mu}\partial_\alpha$ and $\G^\alpha_{\mu}-b^\alpha b_\mu=\Delta^\alpha_{\h\mu}$, we can write
		\begin{align}
			\U^{\alpha\beta}_{\h\h\mu}\partial_\alpha \beta_\beta &= - \frac{B}{\varepsilon + p_\perp} \Big( b^\nu \nabla_{[\nu}\mathcal{H}_{\mu]}+\frac{1}{2}u_\mu \beta H \theta_\parallel \Big) . \label{Udelbeta1}
		\end{align}

		We now analyze the term $\Bc^{\alpha\beta}_{\h\h\mu}\,\partial_\alpha \mathcal{H}_\beta$. 
		Substituting $\Bc^{\alpha\beta}_{\h\h\mu}$ from Eq.~\eqref{cutprojectorB} and using $\mathcal{H}_\beta = -\,\mathcal{H}\, b_\beta$, it follows that
		\begin{align}
			\Bc^{\alpha\beta}_{\h\h\mu}\partial_\alpha \mathcal{H}_\beta = \frac{1}{2} u_\mu D\mathcal{H}+\frac{1}{2} u_\mu \mathcal{H}\theta_\parallel + b^\alpha \partial_{[\alpha}\mathcal{H}_{\mu]} . 
		\end{align}
		Using $\partial_\alpha\equiv u_\alpha D +\nabla_\alpha$, we obtain
		\begin{align}
			\Bc^{\alpha\beta}_{\h\h\mu}\partial_\alpha \mathcal{H}_\beta = \frac{1}{2}u_\mu \mathcal{H}\theta_\parallel -b^\alpha \nabla_{[\mu}\mathcal{H}_{\alpha]} . \label{BcdelH1}
		\end{align}
		Inferring to~\eqref{Udelbeta1} and~\eqref{BcdelH1}, up to first order in gradient expansion, we obtain
		\begin{align}
				\U^{\alpha\beta}_{\h\h\mu}\partial_\alpha \beta_\beta &= -B'\,\Bc^{\alpha\beta}_{\h\h\mu}\partial_\alpha \mathcal{H}_\beta=-\frac{1}{2}B' \,\bar{\mathscr{X}}_\mu , \label{UtoBcrelation}
		\end{align}
		where we defined $B'=\frac{B}{(\varepsilon + p_\perp)}$ and used the definition $\bar{\mathscr{X}}_\mu=2\,\Bc^{\alpha\beta}_{\h\h\mu}\partial_\alpha \mathcal{H}_\beta$.
		%%%%%%%%%%%%%%%%%%%%%%%%%%%%%%%%%%%%%%%%%%%%%%%%%%%%%%%%%%%%%%%%%%%%%%%%%
		%%%%%%%%%%%%%%%%%%%%%%%%%%%%%%%%%%%%%%%%%%%%%%%%%%%%%%%%%%%%%%%%%%%%%%%%

		\section{Complete Derivation of $\hat{\mathcal{C}}$}\label{cal_C}
	
	Substituting~\eqref{To_decom} and~\eqref{Fo_decom} in~\eqref{C}, we get
	\begin{align}
		\hat{\mathcal{C}} &= \hat\varepsilon\,u^{\mu}u^{\nu}\partial_{\mu}\beta_{\nu}
			+ \hat p_{\parallel}\, b^{\mu}b^{\nu}\partial_{\mu}\beta_{\nu}
			- \hat p_{\perp}\, \G^{\mu\nu} \partial_{\mu}\beta_{\nu}
			+ (\hat h^{\mu}u^{\nu} + \hat h^{\nu}u^{\mu})\partial_{\mu}\beta_{\nu}
			+ (\hat f^{\mu}b^{\nu} + \hat f^{\nu}b^{\mu})\partial_{\mu}\beta_{\nu}
			+ \hat \pi^{\mu\nu}_{\perp}\partial_{\mu}\beta_{\nu}  \nonumber\\
			&~~~+  \hat B\,(b^{\mu}u^{\nu}-b^{\nu}u^{\mu}) \partial_{\mu}\mathcal{H}_{\nu}
			- (\hat l^{\mu}b^{\nu}-\hat l^{\nu}b^{\mu})\partial_{\mu}\mathcal{H}_{\nu}
			+(\hat g^{\mu}u^{\nu} - \hat g^{\nu}u^{\mu})\partial_{\mu}\mathcal{H}_{\nu} + \hat m^{\mu\nu}\partial_{\mu}\mathcal{H}_{\nu} \nonumber\\
			&= \hat\varepsilon\,D\beta
			- \hat p_{\parallel}\, \bar{\theta}_\parallel
			- \hat p_{\perp}\, \bar{\theta}_\perp -\hat{B}\,D\mathcal{H} - \hat B\, \mathcal{H}\, \theta_{\parallel}
			+ (\hat h^{\alpha}u^{\beta} + \hat h^{\beta}u^{\alpha})\partial_{\alpha}\beta_{\beta}
			+ (\hat f^{\alpha}b^{\beta} + \hat f^{\beta}b^{\alpha})\partial_{\alpha}\beta_{\beta}
			+ \hat \pi^{\alpha\beta}_{\perp}\partial_{\alpha}\beta_{\beta}  \nonumber\\
			&~~~
			- (\hat l^{\alpha}b^{\beta}-\hat l^{\beta}b^{\alpha})\partial_{\alpha}\mathcal{H}_{\beta}
			+(\hat g^{\alpha}u^{\beta} - \hat g^{\beta}u^{\alpha})\partial_{\alpha}\mathcal{H}_{\beta} + \hat m^{\alpha\beta}\partial_{\alpha}\mathcal{H}_{\beta} .
		\end{align}
		Here, we have used $\beta_\nu = \beta u_\nu$ and $\mathcal{H}_\nu = -\mathcal{H} b_\nu$ in the terms involving $\hat{\varepsilon}$ and $\hat{B}$. Also we have used the relations $b^\mu b^\nu \partial_{\mu}\beta_{\nu} = -\bar{\theta}_{\parallel}$ and $\mathcal{G}^{\mu\nu} \partial_{\mu}\beta_{\nu} = \bar{\theta}_{\perp}$ to simplify. Further,
		with the help of projectors given in~\eqref{Gprojectos}-\eqref{cutprojectorB}, we can write
	\begin{align}
		\hat h^{\alpha}u^{\beta} + \hat h^{\beta}u^{\alpha}&=2 \U^{\alpha\beta}_{\h\h \mu} \hat{h}^\mu , \label{proonh}\\
		\hat f^{\alpha}b^{\beta} + \hat f^{\beta}b^{\alpha}&=2 \B^{\alpha\beta}_{\h\h \mu} \hat{f}^\mu , \label{proonf} \\
		\hat l^{\alpha}b^{\beta} - \hat l^{\beta}b^{\alpha}&=-2 \Bc^{\alpha\beta}_{\h\h \mu} \hat{l}^\mu, \label{proonl}\\
		\hat g^{\alpha}u^{\beta} - \hat g^{\beta}u^{\alpha}&=-2 \Uc^{\alpha\beta}_{\h\h \mu} \hat{g}^\mu , \label{proong} \\
		\hat{\pi}_\perp^{\alpha\beta}&= \hat{\pi}_\perp^{\mu\nu} \G^{\alpha\beta}_{\h\h\mu\nu}, \label{proonpi}\\
		\hat{m}^{\alpha\beta}&= \hat{m}^{\mu\nu} \Gc^{\alpha\beta}_{\h\h\mu\nu}. \label{proonm}
	\end{align} 
	Thus, we obtain
		\begin{align}
		\hat{\mathcal{C}} &= \hat\varepsilon\, D\beta - \hat B\, D \mathcal{H}
		- \hat p_{\parallel}\,\bar{\theta}_{\parallel}
		- \hat p_{\perp}\,\bar{\theta}_{\perp}
		- \hat B\, \mathcal{H}\, \theta_{\parallel}\nonumber\\
		&\quad + 2 \hat h^\mu \, \U^{\alpha\beta}_{\h\h \mu} \partial_\alpha \beta_\beta + 2 \hat f^\mu \, \B^{\alpha\beta}_{\h\h \mu} \partial_\alpha \beta_\beta - 2 \hat g^\mu \, \Uc^{\alpha\beta}_{\h\h \mu} \partial_\alpha \mathcal{H}_\beta + 2 \hat l^\mu \, \Bc^{\alpha\beta}_{\h\h \mu} \partial_\alpha \mathcal{H}_\beta \nonumber \\
		&~~~+ \hat \pi_{\perp}^{\mu\nu}\,\G^{\alpha\beta}_{\h\h \mu\nu} \partial_\alpha \beta_\beta 
		+ \hat m^{\mu \nu} \Gc^{\alpha\beta}_{\h\h \mu\nu} \partial_\alpha \mathcal{H}_\beta  . \label{cal_cc2}
	\end{align}
	Next, we define the thermodynamic forces: $ \bar{\sigma}_{\perp \mu\nu}=\G^{\alpha\beta}_{\h\h \mu\nu}\,\partial_{\alpha}\beta_{\beta}$, $\bar{\chi}_{ \mu\nu}=\Gc^{\alpha\beta}_{\h\h \mu\nu}\,\partial_{\alpha} \mathcal{H}_{\beta}$, $\bar{\mathscr{X}}_\mu=2 \, \Bc^{\alpha\beta}_{\h\h \mu} \partial_\alpha \mathcal{H}_\beta$, and $\bar{\mathscr{Y}}_\mu=2 \,\B^{\alpha\beta}_{\h\h \mu} \partial_\alpha \beta_\beta$. Thus, we can rewrite~\eqref{cal_cc2} as
	\begin{align}
		\hat{\mathcal{C}} &= \hat\varepsilon\, D\beta - \hat B\, D \mathcal{H}
		- \hat p_{\parallel}\,\bar{\theta}_{\parallel}
		- \hat p_{\perp}\,\bar{\theta}_{\perp}
		- \hat B\, \mathcal{H}\, \theta_{\parallel}\nonumber\\
		&\quad + 2 \hat h^\mu \, \U^{\alpha\beta}_{\h\h \mu} \partial_\alpha \beta_\beta +  \hat f^\mu \, \bar{\mathscr{Y}}_\mu - 2 \hat g^\mu \, \Uc^{\alpha\beta}_{\h\h \mu} \partial_\alpha \mathcal{H}_\beta +  \hat l^\mu \, \bar{\mathscr{X}}_\mu \nonumber \\
		&~~~+ \hat \pi_{\perp}^{\mu\nu}\bar{\sigma}_{\perp \mu\nu}
		+ \hat m^{\mu \nu} \bar{\chi}_{ \mu\nu}  .
	\end{align}
Further, we separate the first- and second-order contributions from $\U^{\alpha\beta}_{\h\h \mu}\,\partial_\alpha \beta_\beta$ and $\Uc^{\alpha\beta}_{\h\h \mu}\,\partial_\alpha \mathcal{H}_\beta$, and use~\eqref{UtoBcrelation} and~\eqref{UctoBrelation} to rewrite the first-order contributions. Thus we get
	\begin{align}
		\hat{\mathcal C}
		&= \hat\varepsilon\, D\beta - \hat B\, D \mathcal{H}
		- \hat p_{\parallel}\,\bar{\theta}_{\parallel}
		- \hat p_{\perp}\,\bar{\theta}_{\perp}
		- \hat B\, \mathcal{H}\, \theta_{\parallel}\nonumber\\
		&~~~+\hat{\mathscr{J}}^{\mu}
	\bar{\mathscr{X}}_\mu + \hat{\mathscr{K}}^{\mu}
		\bar{\mathscr{Y}}_\mu
		+ \hat \pi_{\perp}^{\mu\nu}\bar{\sigma}_{\perp \mu\nu} + \hat m^{\mu \nu} \bar{\chi}_{ \mu\nu} \nonumber\\
		&~~~+ \hat h^{\mu}\,\beta\,(D u_\mu)_2
		+ \hat g^{\mu}\,\mathcal{H}\,(D b_\mu)_2 , \label{cal_c2}
	\end{align}
	where we have defined
	\begin{equation}
		\hat{\mathscr{J}}^{\mu} \equiv\hat l^{\mu}-\frac{B}{\varepsilon+p_{\perp}}\;\hat h^{\mu},
		~~ \hat{\mathscr{K}}^{\mu}\equiv\hat f^{\mu}+H\,\hat g^{\mu}.
	\end{equation}
	To proceed with the evaluation of $\hat{\mathcal C}$, we must first establish several useful relations. We begin by considering the thermodynamic identities
		\begin{align}
			ds &= \beta\, d\varepsilon - \mathcal{H}\, dB , \label{dsther}\\
			s &= \beta\,\varepsilon + \beta\, p_{\perp} - B \mathcal{H}, \label{sther}\\
			dp_{\perp} &= -\beta^{-1}(\varepsilon+p_{\perp})\, d\beta + \beta^{-1} B\, d\mathcal{H} . \label{dpther}
		\end{align}
		From~\eqref{dsther}, we get
		\begin{align}
			\frac{\partial s}{\partial \varepsilon}\bigg|_{B} &= \beta, ~~
			\frac{\partial s}{\partial B}\bigg|_{\beta} = -\mathcal{H},~\text{and}~~\frac{\partial \beta}{\partial B}\bigg|_{\varepsilon}
			= - \frac{\partial \mathcal{H}}{\partial \varepsilon}\bigg|_{B} , \label{therrel1}
		\end{align}
		and from~\eqref{dpther}, we get
		\begin{align}
			-\beta^{-1}(\varepsilon+p_{\perp}) &= \frac{\partial p_{\perp}}{\partial \beta}\bigg|_{\mathcal{H}},~
			\beta^{-1} B = \frac{\partial p_{\perp}}{\partial \mathcal{H}}\bigg|_{\beta} . \label{therrel2}
		\end{align}
		With the help of the two independent thermodynamic variables, the total energy density $\varepsilon$ and the magnetic flux density $B$, all other thermodynamic quantities may be regarded as functions of $(\varepsilon,B)$. 
		Accordingly, their comoving derivatives follow from the chain rule. 
		Thus, for $\beta=\beta(\varepsilon,B)$ one has
		\begin{align}
			D\beta &= \frac{\partial \beta}{\partial \varepsilon}\, D\varepsilon
			+ \frac{\partial \beta}{\partial B}\, DB, \\[0.25em]
		\implies	D\beta &= \frac{\partial \beta}{\partial \varepsilon}
			\Big[-(\varepsilon+p_{\perp})\theta + H B\,\theta_{\parallel}\Big]
			+ \frac{\partial \beta}{\partial B}\,(-B\theta_{\perp}) \\
			&\quad
			+ \frac{\partial \beta}{\partial \varepsilon}\big[ -u_{\nu}\partial_{\mu}T_{1}^{\mu\nu} \big] + \frac{\partial \beta}{\partial B}
			\big[ -b_{\nu}\partial_{\mu}\tilde F_1^{\mu\nu} \big].
		\end{align}
	Here we have used Eqs.~\eqref{udelT} and~\eqref{bdelF} to substitute the values of $D\varepsilon$ and $D B$, respectively. Also, we have used the relations $\theta=\theta_\parallel+\theta_\perp$ and $p_\perp=p_\parallel+BH$. 
	We now decompose $D\beta$ into its first- and second-order gradient contributions, writing
		\begin{align}
			D\beta &= (D\beta)_1 + (D\beta)_2, \label{dbeta}
		\end{align}
		where
		\begin{align}
			(D\beta)_1 &= \frac{\partial\beta}{\partial\varepsilon}\Big[ -(\varepsilon+p_{\perp})\,\theta + H B\,\theta_{\parallel} \Big]
			+ \frac{\partial\beta}{\partial B}\,(-B\theta_{\perp}),  \\
			(D\beta)_2 &= \frac{\partial \beta}{\partial \varepsilon}\big[ -u_{\nu}\partial_{\mu}T_{1}^{\mu\nu} \big] + \frac{\partial \beta}{\partial B}
			\big[ -b_{\nu}\partial_{\mu}\tilde F_1^{\mu\nu} \big] . \label{dbeta2}
		\end{align}
		 Using~\eqref{therrel1} and~\eqref{therrel2}, $(D\beta)_1$ can be further simplified as
		\begin{align*}
			(D\beta)_1 &= \frac{\partial\beta}{\partial\varepsilon}\Big[-(\varepsilon+p_{\perp})\theta + H B\,\theta_{\parallel}\Big]
			+ \frac{\partial \mathcal{H}}{\partial \varepsilon} \, B\, \theta_{\perp} \\
			&= \frac{\partial\beta}{\partial\varepsilon}\Big[\beta \frac{\partial p_{\perp}}{\partial\beta}\,\theta + H B\,\theta_{\parallel}\Big]
			+ \beta\,\theta_{\perp}\,\frac{\partial \mathcal{H}}{\partial\varepsilon} \frac{\partial p_\perp}{\partial \mathcal{H}}\\
			&= \beta\,\frac{\partial\beta}{\partial\varepsilon}\frac{\partial p_{\perp}}{\partial\beta}\,\theta_\parallel
			+ \frac{\partial\beta}{\partial\varepsilon} H B\,\theta_{\parallel}
			+ \beta\,\theta_{\perp}\Big[\frac{\partial p_\perp}{\partial \beta}\frac{\partial \beta}{\partial \varepsilon}
			+ \frac{\partial \beta}{\partial \mathcal{H}}\frac{\partial \mathcal{H}}{\partial \varepsilon}\Big] \\
			&= \beta\,\theta_{\parallel}\Big[\frac{\partial \beta}{\partial \varepsilon}\frac{\partial p_{\perp}}{\partial \beta}
			+ \frac{\partial p_\perp}{\partial \mathcal{H}} \frac{\partial \mathcal{H}}{\partial \varepsilon}\Big]
			- \beta\,\theta_{\parallel} \frac{\partial p_\perp}{\partial \mathcal{H}} \frac{\partial \mathcal{H}}{\partial \varepsilon}
			+ \frac{\partial \beta}{\partial \varepsilon} H B\,\theta_{\parallel}  + \beta\,\theta_{\perp}\,\frac{\partial p_\perp}{\partial \varepsilon} \\
			&= \beta\,\theta_{\parallel}\,\frac{\partial p_{\perp}}{\partial \varepsilon}
			+ \beta\,\theta_{\perp}\,\frac{\partial p_{\perp}}{\partial \varepsilon}
			- \beta  \theta_{\parallel} \frac{\partial \mathcal{H}}{\partial \varepsilon}(\beta^{-1}B)
			+ \frac{\partial \beta}{\partial \varepsilon} H B\,\theta_{\parallel} \\
			&= \beta\,\theta_{\parallel}\,\frac{\partial p_{\perp}}{\partial \varepsilon}
			+ \beta\,\theta_{\perp}\,\frac{\partial p_{\perp}}{\partial \varepsilon}
			- B\,\theta_\parallel \left[\frac{\partial \mathcal{H}}{\partial \varepsilon}-H\frac{\partial \beta}{\partial \varepsilon}\right].
		\end{align*}
		Using the relation $\mathcal{H}=\beta\,H$, we obtain
		\begin{align}
			(D\beta)_1 = \bar{\theta}_{\parallel}\,\frac{\partial p_{\perp}}{\partial \varepsilon} + \bar{\theta}_{\perp}\,\frac{\partial p_{\perp}}{\partial \varepsilon}- B\,\frac{\partial H}{\partial \varepsilon}\,\bar{\theta}_{\parallel} . \label{dbeta1}
		\end{align}
		Similarly, we can also decompose the comoving derivative of the reduced magnetic field ($\mathcal{H}$) into first- and second-order contributions:
		\begin{align}
			D \mathcal{H} = (D \mathcal{H})_1 + (D \mathcal{H})_2 \label{dh} , 
		\end{align}
		where
		\begin{align}
			(D \mathcal{H})_1 &= \frac{\partial \mathcal{H}}{\partial \varepsilon}\Big[-(\varepsilon+p_{\perp})\,\theta + H B\,\theta_{\parallel}\Big]
			+ \frac{\partial \mathcal{H}}{\partial B}\,(-B\theta_{\perp}),  \\[0.25em]
			(D \mathcal{H})_2 &= \frac{\partial \mathcal{H}}{\partial \varepsilon}\big[ -u_{\nu}\partial_{\mu}T_{1}^{\mu\nu} \big] + \frac{\partial \mathcal{H}}{\partial B}
			\big[ -b_{\nu}\partial_{\mu}\tilde F_1^{\mu\nu} \big]. \label{dH2}
		\end{align}
		Using~\eqref{therrel1} and~\eqref{therrel2}, $(D \mathcal{H})_1$ simplifies to
		\begin{align}
			(D \mathcal{H})_1 = -\bar{\theta}_{\parallel}\,\frac{\partial p_{\perp}}{\partial B}
			- \bar{\theta}_{\perp}\,\frac{\partial p_{\perp}}{\partial B}
			+ B\,\bar{\theta}_{\parallel}\,\frac{\partial H}{\partial B}. \label{dH1}
		\end{align}

		%%%%%%%%%%%%%%%%%%%%%%%%%%%%%%%%%%%%%%%%%%%%%%%%%%%%%%%%%%%%%%%%%%
		
		Next we collect the terms involving scalar operators in Eq.~\eqref{cal_c2} and substitute $D\beta$ and $D\mathcal{H}$ from Eqs.~\eqref{dbeta} and~\eqref{dh}, with the first-order pieces $(D\beta)_1$ and $(D\mathcal{H})_1$ given by Eqs.~\eqref{dbeta1} and~\eqref{dH1}. We obtain	
		\begin{align}
			\hat{\mathcal{C}}_{scalar}&=  \hat\varepsilon\, D\beta - \hat B\, D \mathcal{H}
			- \hat p_{\parallel}\,\bar{\theta}_{\parallel}
			- \hat p_{\perp}\,\bar{\theta}_{\perp}
			- \hat B\, \mathcal{H}\, \theta_{\parallel}\nonumber\\
			&= -\bar{\theta}_{\perp}\Big[\,\hat p_{\perp}
			- \hat\varepsilon\,\frac{\partial p_{\perp}}{\partial \varepsilon}
			- \hat B\,\frac{\partial p_{\perp}}{\partial B}\,\Big] \\
			&\quad
			-\bar{\theta}_{\parallel}\Big[\,\hat p_{\parallel}
			- \hat\varepsilon\,\frac{\partial p_{\perp}}{\partial \varepsilon}
			- \hat B\,\frac{\partial p_{\perp}}{\partial B}
			+ \hat\varepsilon\, B\,\frac{\partial H}{\partial \varepsilon}
			+ \hat B\, B\,\frac{\partial H}{\partial B}
			+ \hat B\, H \,\Big] \\
			&\qquad + \big[\hat\varepsilon\,(D\beta)_2 - \hat B\, (D \mathcal{H})_2\big].
		\end{align}
		For the terms multiplied to $\bar{\theta}_\parallel$, we use $p_{\perp}=p_{\parallel}+BH$ to express all occurrences of $p_\perp$ in terms of $p_\parallel$. We then obtain
		\begin{align}
			\hat{\mathcal{C}}_{scalar}&=-\bar{\theta}_{\perp}\Big[\,\hat p_{\perp}
			- \hat\varepsilon\,\frac{\partial p_{\perp}}{\partial \varepsilon}
			- \hat B\,\frac{\partial p_{\perp}}{\partial B}\Big]  -\bar{\theta}_{\parallel}\Big[\,\hat p_{\parallel}
			- \hat\varepsilon\,\frac{\partial p_{\parallel}}{\partial \varepsilon}
			- \hat B\,\frac{\partial p_{\parallel}}{\partial B}\Big]  \nonumber\\
			&~~~ + \big(\hat\varepsilon\,D\beta_2 - \hat B\,D H_2\big). \label{cscalarsolve}
		\end{align}
		Since the $\varepsilon$ and $B$ are the independent variables, hence their partial differentiating with respect to each other vanishes.
		We now use~\eqref{therrel1} and rewrite \((D\beta)_2\) and \((D \mathcal{H})_2\) given in~\eqref{dbeta2} and~\eqref{dH2}. We obtain
		\begin{align}
			\hat\varepsilon\,D\beta_2 - \hat B\, D \mathcal{H}_2
			&= \hat\beta^{*}\big[ -u_{\nu}\partial_{\mu}T_{1}^{\mu\nu} \big]
			  - \hat{\mathcal{H}}^{*}
			  \big[ -b_{\nu}\partial_{\mu}\tilde F_1^{\mu\nu} \big],
		\end{align}
		where we defined
		\begin{align}
			\hat\beta^{*} &= \hat\varepsilon\,\frac{\partial \beta}{\partial \varepsilon}
			+ \hat B\,\frac{\partial \beta}{\partial B}, \qquad
			\hat{\mathcal{H}}^{*} = \hat\varepsilon\,\frac{\partial \mathcal{H}}{\partial \varepsilon}
			+ \hat B\,\frac{\partial \mathcal{H}}{\partial B}.
		\end{align}
		Thus~\eqref{cscalarsolve} reduce to
		\begin{align}
			\hat{\mathcal{C}}_{scalar}
			= -\bar{\theta}_{\perp} \hat P_{\perp}
			- \bar{\theta}_{\parallel} \hat P_{\parallel}
			+ \hat\beta^{*}\big[ -u_{\nu}\partial_{\mu}T_{1}^{\mu\nu} \big]
			- \hat{\mathcal{H}}^{*}\big[ -b_{\nu}\partial_{\mu}\tilde F_1^{\mu\nu} \big].\label{Cscalar}
		\end{align}
	Here, 
		\begin{align}
			\hat P_{\perp} &= \hat p_{\perp}
			- \hat\varepsilon\,\frac{\partial p_{\perp}}{\partial \varepsilon}
			- \hat B\,\frac{\partial p_{\perp}}{\partial B}, \\
			\hat P_{\parallel} &= \hat p_{\parallel}
			- \hat\varepsilon\,\frac{\partial p_{\parallel}}{\partial \varepsilon}
			- \hat B\,\frac{\partial p_{\parallel}}{\partial B}.
		\end{align}	
		Substituting Eq.~\eqref{Cscalar} back into Eq.~\eqref{cal_c2}, we decompose $\hat{\mathcal{C}}$ into its first-order part $\hat{\mathcal{C}}_{F}$ and second-order part $\hat{\mathcal{C}}_{S}$, given by
		\begin{align}
			\hat{\mathcal C}_F
			&= -\bar{\theta}_{\perp}\,\hat{P}_{\perp}
			-\bar{\theta}_{\parallel}\,\hat{P}_{\parallel}
			+ \hat{\mathscr{K}}^{\mu}\,\bar{\mathscr{Y}}_\mu
			+ \hat{\mathscr{J}}^{\mu}\, \bar{\mathscr{X}}_\mu
			+ \hat \pi_{\perp}^{\mu\nu} \bar{\sigma}_{\perp \mu\nu}
			+ \hat m^{\mu\nu} \bar{\chi}_{ \mu\nu},\label{CF_appen}\\
			\hat{\mathcal C}_S
			&= \hat\beta^{*}\big[-u_\nu \partial_\mu T_1^{\mu\nu}\big]
			- \hat{\mathcal{H}}^{*}\big[-b_\nu \partial_\mu \tilde F_1^{\mu\nu}\big]
			+ \hat h^{\mu}\,\beta\,(D u_\mu)_2
			+ \hat g^{\mu}\,\mathcal{H}\,(D b_\mu)_2 .\label{CS_appen}
		\end{align}

		%%%%%%%%%%%%%%%%%%%%%%%%%%%%%%%%%%%%%%%%%%%%%%%%%%%

	\section{Complete Derivation of $\partial_\lambda \hat{\mathcal{C}}$}\label{delC_cal}

	Starting from~\eqref{delC1}, we write
	\begin{align}
		\partial_\lambda \hat{\mathcal{C}}
		&= \hat{\varepsilon}\, \partial_\lambda (u^\mu u^\nu \partial_\mu \beta_\nu)
		+ \hat{p}_\parallel\, \partial_\lambda (b^\mu b^\nu \partial_\mu \beta_\nu)
		- \hat{p}_\perp\,  \partial_\lambda(\G^{\mu\nu}  \partial_\mu \beta_\nu)
		\nonumber\\
		&\quad
		+ \hat{h}^\mu \partial_\lambda (u^\nu \partial_\mu \beta_\nu)
		+ \hat{h}^\nu \partial_\lambda (u^\mu \partial_\mu \beta_\nu)
		+ \hat{f}^\mu \partial_\lambda (b^\nu \partial_\mu \beta_\nu)
		+ \hat{f}^\nu \partial_\lambda (b^\mu \partial_\mu \beta_\nu)
		\nonumber\\
		&\quad
		+ \hat{\pi}_{\perp}^{\mu\nu}\, \partial_\lambda \partial_\mu \beta_\nu
		+ \hat{B}\, \partial_\lambda 
		\big[(b^\mu u^\nu - b^\nu u^\mu)\partial_\mu \mathcal{H}_\nu \big]
		\nonumber\\
		&\quad
		- \hat{l}^\mu \partial_\lambda (b^\nu \partial_\mu \mathcal{H}_\nu)
		+ \hat{l}^\nu \partial_\lambda (b^\mu \partial_\mu \mathcal{H}_\nu)
		+ \hat{g}^\mu \partial_\lambda (u^\nu \partial_\mu \mathcal{H}_\nu)
		- \hat{g}^\nu \partial_\lambda (u^\mu \partial_\mu \mathcal{H}_\nu)
		\nonumber\\
		&\quad
		+ \hat{m}^{\mu\nu}\, \partial_\lambda \partial_\mu \mathcal{H}_\nu  . \label{delcappen1}
	\end{align}
	We first evaluate the contributions involving scalar operators, 
	\begin{align}
		(\partial_\lambda \hat{\mathcal{C}})_{scalar}&=\hat{\varepsilon}\,\partial_\lambda(u^\mu u^\nu \partial_\mu \beta_\nu)
		+ \hat{p}_\parallel \partial_\lambda(b^\mu b^\nu \partial_\mu \beta_\nu)
		- \hat{p}_\perp\partial_\lambda(\G^{\mu\nu} \partial_\mu \beta_\nu)
		+ \hat{B}\,\partial_\lambda[(b^\mu u^\nu - b^\nu u^\mu)\partial_\mu \mathcal{H}_\nu]\nonumber\\
		&=
		\hat{\varepsilon}\,\partial_\lambda D\beta
		- \hat{p}_\parallel \partial_\lambda \bar{\theta}_\parallel
		- \hat{p}_\perp \partial_\lambda \bar{\theta}_\perp
		- \hat{B}\, \partial_\lambda D\mathcal{H}
		- \hat{B}\, \partial_\lambda (H\bar{\theta}_\parallel) . \label{delcappen2}
	\end{align}
	Here we have used the relations
	$b^\mu b^\nu \partial_\mu \beta_\nu = -\,\bar{\theta}_\parallel$, 
	$\G^{\mu\nu} \partial_\mu \beta_\nu = \bar{\theta}_\perp$, 
	$\bar{\theta} = \beta\,\theta$, and $\mathcal{H} = \beta\,H$.
	
Acting with $\partial_\lambda$ on Eqs.~\eqref{dbeta1} and~\eqref{dH1}, we obtain
	\begin{align}
		\partial_\lambda (D \beta)_1
		&=
		(\partial_\lambda \bar{\theta}_{\parallel})\,\frac{\partial p_{\perp}}{\partial \varepsilon}
		+ \bar{\theta}_{\parallel}\,\partial_\lambda\!\left(\frac{\partial p_{\perp}}{\partial \varepsilon}\right)
		+ (\partial_\lambda \bar{\theta}_{\perp})\,\frac{\partial p_{\perp}}{\partial \varepsilon}
		+ \bar{\theta}_{\perp}\,\partial_\lambda\!\left(\frac{\partial p_{\perp}}{\partial \varepsilon}\right)
		-\; (\partial_\lambda \bar{\theta}_{\parallel})\,B\,\frac{\partial H}{\partial \varepsilon}
		-\; \bar{\theta}_{\parallel}\,\partial_\lambda\left(B\,\frac{\partial H}{\partial \varepsilon}\right) , \label{delDbeta}\\
		\partial_\lambda (D \mathcal{H})_1
		&=
		-(\partial_\lambda \bar{\theta}_{\parallel})\,\frac{\partial p_{\perp}}{\partial B}
		- \bar{\theta}_{\parallel}\,\partial_\lambda\!\left(\frac{\partial p_{\perp}}{\partial B}\right)
		- (\partial_\lambda \bar{\theta}_{\perp})\,\frac{\partial p_{\perp}}{\partial B}
		- \bar{\theta}_{\perp}\,\partial_\lambda\!\left(\frac{\partial p_{\perp}}{\partial B}\right)
		+\; (\partial_\lambda \bar{\theta}_{\parallel})\,B\,\frac{\partial H}{\partial B}
		+\; \bar{\theta}_{\parallel}\,\partial_\lambda\left(B\,\frac{\partial H}{\partial B}\right) . \label{delDH}
	\end{align}
Here we do not differentiate $(D\beta)_{2}$ or $(D\mathcal{H})_{2}$, since doing so would generate third-order terms, which are beyond the scope of the present work. Further, using~\eqref{delDbeta} and~\eqref{delDH}, and using the relation $p_\perp =p_\parallel +BH$, we can simplify~\eqref{delcappen2} as
	\begin{align}
		(\partial_\lambda \hat{\mathcal{C}})_{scalar}&=\hat{\varepsilon}\,\partial_\lambda(D\beta)
		- \hat{B}\,\partial_\lambda(D\mathcal{H})
		- \hat{p}_{\parallel}\,\partial_\lambda \bar{\theta}_{\parallel}
		- \hat{p}_{\perp}\,\partial_\lambda \bar{\theta}_{\perp}
		- \hat{B} H\,\partial_\lambda \bar{\theta}_{\parallel}
		- \bar{\theta}_{\parallel}\,\hat{B}\,\partial_\lambda H \nonumber\\
		&=
		-(\partial_\lambda \bar{\theta}_{\parallel})
		\left[
		\hat{p}_{\parallel}
		- \hat{\varepsilon}\frac{\partial p_{\parallel}}{\partial \varepsilon}
		- \hat{B}\frac{\partial p_{\parallel}}{\partial B}
		\right]
		- (\partial_\lambda \bar{\theta}_{\perp})
		\left[
		\hat{p}_{\perp}
		- \hat{\varepsilon}\frac{\partial p_{\perp}}{\partial \varepsilon}
		- \hat{B}\frac{\partial p_{\perp}}{\partial B}
		\right] \nonumber\\
		&\qquad
		+ \bar{\theta}_{\parallel}
		\left[
		\hat{\varepsilon}\,\partial_\lambda\left(\frac{\partial p_{\parallel}}{\partial \varepsilon}\right)
		+ \hat{B}\,\partial_\lambda\left(\frac{\partial p_{\parallel}}{\partial B}\right)
		\right] 
		+ \bar{\theta}_{\perp}
		\left[
		\hat{\varepsilon}\,\partial_\lambda\left(\frac{\partial p_{\perp}}{\partial \varepsilon}\right)
		+ \hat{B}\,\partial_\lambda\left(\frac{\partial p_{\perp}}{\partial B}\right)
		\right] \nonumber\\
		&=
		-\hat{P}_\parallel \partial_\lambda \bar{\theta}_\parallel  - \hat{P}_\perp \partial_\lambda \bar{\theta}_\perp +  \bar{\theta}_\parallel [\hat \varepsilon \,\partial_\lambda \gamma_\parallel + \hat B \,\partial_\lambda \phi_\parallel]  + \bar{\theta}_\perp [\hat \varepsilon \,\partial_\lambda \gamma_\perp + \hat B \,\partial_\lambda \phi_\perp] .  \label{delcscalarappen1}
	\end{align}
	Here we defined, 
	\begin{align}
		\gamma_{\parallel} &= \frac{\partial p_\parallel}{\partial \varepsilon},~ \gamma_{\perp} = \frac{\partial p_\perp}{\partial \varepsilon},\\
		\phi_{\parallel} &= \frac{\partial p_\parallel}{\partial B},~ \phi_{\perp} = \frac{\partial p_\perp}{\partial B} .
	\end{align}
	Recall that $\varepsilon$ and $B$ are independent variables; hence their partial derivatives with respect to each other  vanish.

Next, we simplify the contributions involving vector operators in Eq.~\eqref{delcappen1}. We can write
	\begin{align}
		(\partial_\lambda \hat{\mathcal{C}})_{vector}&=\hat{h}^\mu \partial_\lambda (u^\nu \partial_\mu \beta_\nu)
		+
		\hat{h}^\nu \partial_\lambda (u^\mu \partial_\mu \beta_\nu) +\hat{f}^\mu \partial_\lambda (b^\nu \partial_\mu \beta_\nu)
		+
		\hat{f}^\nu \partial_\lambda (b^\mu \partial_\mu \beta_\nu)\nonumber\\
		&~~~ -\hat{l}^\mu \partial_\lambda (b^\nu \partial_\mu \mathcal{H}_\nu)
		+
		\hat{l}^\nu \partial_\lambda (b^\mu \partial_\mu \mathcal{H}_\nu) 
		+\hat{g}^\mu \partial_\lambda (u^\nu \partial_\mu \mathcal{H}_\nu)
		+
		\hat{g}^\nu \partial_\lambda (u^\mu \partial_\mu \mathcal{H}_\nu) \nonumber\\
		&=
		(\hat{h}^\mu u^\nu + \hat{h}^\nu u^\mu)\,\partial_\lambda \partial_\mu \beta_\nu
		+ (\hat{h}^\mu \partial_\lambda u^\nu + \hat{h}^\nu \partial_\lambda u^\mu)\,\partial_\mu \beta_\nu  \nonumber\\
		&~~~+ 	(\hat{f}^\mu b^\nu + \hat{f}^\nu b^\mu)\,\partial_\lambda \partial_\mu \beta_\nu
		+ (\hat{f}^\mu \partial_\lambda b^\nu + \hat{f}^\nu \partial_\lambda b^\mu)\,\partial_\mu \beta_\nu \nonumber\\
		&~~~-(\hat{l}^\mu b^\nu - \hat{l}^\nu b^\mu)\,\partial_\lambda \partial_\mu \mathcal{H}_\nu
		- (\hat{l}^\mu \partial_\lambda b^\nu - \hat{l}^\nu \partial_\lambda b^\mu)\,\partial_\mu \mathcal{H}_\nu\nonumber\\ 
		&~~~ + (\hat{g}^\mu u^\nu - \hat{g}^\nu u^\mu)\,\partial_\lambda \partial_\mu \mathcal{H}_\nu
		+ (\hat{g}^\mu \partial_\lambda u^\nu - \hat{g}^\nu \partial_\lambda u^\mu)\,\partial_\mu \mathcal{H}_\nu \nonumber\\
		&=
		2 \hat{h}^{\mu}\frac{1}{2}
		\left( u^\alpha \G_{\h\mu}^{\beta} + u^\beta \G_{\h\mu}^{\alpha} \right)
		\partial_\lambda \partial_\alpha \beta_\beta
		+\,2\hat{h}^{\mu}\frac{1}{2}
		\left[
		(\partial_\lambda u^\alpha)\G_{\h\mu}^{\beta}
		+ (\partial_\lambda u^\beta)\G_{\h\mu}^{\alpha}
		\right]
		\partial_\alpha \beta_\beta \nonumber\\
		&~~~ +2 \hat{f}^{\mu}\frac{1}{2}
		\left( b^\alpha \G_{\h\mu}^{\beta} + b^\beta \G_{\h\mu}^{\alpha} \right)
		\partial_\lambda \partial_\alpha \beta_\beta
		+\,2\hat{f}^{\mu}\frac{1}{2}
		\left[
		(\partial_\lambda b^\alpha)\G_{\h\mu}^{\beta}
		+ (\partial_\lambda b^\beta)\G_{\h\mu}^{\alpha}
		\right]
		\partial_\alpha \beta_\beta \nonumber\\
		&~~~ +2 \hat{l}^{\mu}\frac{1}{2}
		\left( b^\alpha \G_{\h\mu}^{\beta} - b^\beta \G_{\h\mu}^{\alpha} \right)
		\partial_\lambda \partial_\alpha \mathcal{H}_\beta
		+\,2\hat{l}^{\mu}\frac{1}{2}
		\left[
		(\partial_\lambda b^\alpha)\G_{\h\mu}^{\beta}
		- (\partial_\lambda b^\beta)\G_{\h\mu}^{\alpha}
		\right]
		\partial_\alpha \mathcal{H}_\beta \nonumber\\
		&~~~ -2 \hat{g}^{\mu}\frac{1}{2}
		\left( u^\alpha \G_{\h\mu}^{\beta} - u^\beta \G_{\h\mu}^{\alpha} \right)
		\partial_\lambda \partial_\alpha \mathcal{H}_\beta
		-\,2\hat{g}^{\mu}\frac{1}{2}
		\left[
		(\partial_\lambda u^\alpha)\G_{\h\mu}^{\beta}
		- (\partial_\lambda u^\beta)\G_{\h\mu}^{\alpha}
		\right]
		\partial_\alpha \mathcal{H}_\beta \nonumber\\
		&= 2\hat{h}^{\mu} \U^{\alpha\beta}_{\h\h\mu}\,\partial_\lambda \partial_\alpha \beta_\beta
		+ 2\hat{h}^\mu\, \U^{\alpha\beta}_{\lambda; \ \mu}
		(\partial_\alpha \beta_\beta)
		+2\hat{f}^{\mu} \B^{\alpha\beta}_{\h\h\mu}\,\partial_\lambda \partial_\alpha \beta_\beta
		+ 2\hat{f}^\mu\, \B^{\alpha\beta}_{\lambda; \ \mu}
		(\partial_\alpha \beta_\beta)\nonumber\\ 
		&~~~ +2\hat{l}^{\mu} \Bc^{\alpha\beta}_{\h\h\mu}\,\partial_\lambda \partial_\alpha \mathcal{H}_\beta
		+ 2\hat{l}^\mu\, \Bc^{\alpha\beta}_{\lambda; \ \mu}
		(\partial_\alpha \mathcal{H}_\beta) 
		-2\hat{g}^{\mu} \Uc^{\alpha\beta}_{\h\h\mu}\,\partial_\lambda \partial_\alpha \mathcal{H}_\beta
		- 2\hat{g}^\mu\, \Uc^{\alpha\beta}_{\lambda; \ \mu}
		(\partial_\alpha \mathcal{H}_\beta),
	\end{align}
	where we have defined
	\begin{align}
		\U^{\alpha\beta}_{\h\h\mu} &= \frac{1}{2}\left( u^\alpha \G_{\h\mu}^{\beta}+u^\beta \G_{\h\mu}^{\alpha} \right)
		,~~\U^{\alpha\beta}_{\lambda; \ \mu} = \frac{1}{2}
		\left[
		(\partial_\lambda u^\alpha)\G_{\h\mu}^{\beta}
		+ (\partial_\lambda u^\beta)\G_{\h\mu}^{\alpha}
		\right] , 
		\nonumber\\
		\B^{\alpha\beta}_{\h\h\mu}&=\frac{1}{2}\left( b^\alpha \G_{\h\mu}^{\beta} + b^\beta \G_{\h\mu}^{\alpha} \right)
		,~~\B^{\alpha\beta}_{\lambda; \ \mu}=\frac{1}{2}
		\left[
		(\partial_\lambda b^\alpha)\G_{\h\mu}^{\beta}
		+ (\partial_\lambda b^\beta)\G_{\h\mu}^{\alpha}
		\right] , 
		\nonumber\\
		\Bc^{\alpha\beta}_{\h\h\mu}&=\frac{1}{2}\left( b^\alpha \G_{\h\mu}^{\beta} - b^\beta \G_{\h\mu}^{\alpha} \right)
		,~~\Bc^{\alpha\beta}_{\lambda; \ \mu} = \frac{1}{2}
		\left[
		(\partial_\lambda b^\alpha)\G_{\h\mu}^{\beta}
		- (\partial_\lambda b^\beta)\G_{\h\mu}^{\alpha}
		\right] , \nonumber \\
		\Uc^{\alpha\beta}_{\h\h\mu}&=\frac{1}{2}\left( u^\alpha \G_{\h\mu}^{\beta} - u^\beta \G_{\h\mu}^{\alpha} \right)
		,~~\Uc^{\alpha\beta}_{\lambda; \ \mu}=\frac{1}{2}
		\left[
		(\partial_\lambda u^\alpha)\G_{\h\mu}^{\beta}
		- (\partial_\lambda u^\beta)\G_{\h\mu}^{\alpha}
		\right] . \nonumber
	\end{align}
	Thus,
	\begin{align}
		(\partial_\lambda \hat{\mathcal{C}})_{vector}&=
		2\hat{h}^\mu
		\partial_\lambda ( \U^{\alpha\beta}_{\h \h \mu} \partial_\alpha \beta_\beta )
		+
		2\hat{h}^{\mu}
		\left[ \U^{\alpha\beta}_{\lambda; \ \mu}
		-\partial_\lambda \U^{\alpha\beta}_{\h\h \mu}
		\right]
		\partial_\alpha \beta_\beta \nonumber\\
		&~~~ +2\hat{f}^\mu
		\partial_\lambda ( \B^{\alpha\beta}_{\h \h \mu} \partial_\alpha \beta_\beta )
		+
		2\hat{f}^{\mu}
		\left[ \B^{\alpha\beta}_{\lambda; \ \mu}
		-\partial_\lambda \B^{\alpha\beta}_{\h\h \mu}
		\right]
		\partial_\alpha \beta_\beta \nonumber\\
		&~~~ +2\hat{l}^\mu
		\partial_\lambda ( \Bc^{\alpha\beta}_{\h \h \mu} \partial_\alpha \mathcal{H}_\beta )
		+
		2\hat{l}^{\mu}
		\left[ \Bc^{\alpha\beta}_{\lambda; \ \mu}
		-\partial_\lambda \Bc^{\alpha\beta}_{\h\h \mu}
		\right]
		\partial_\alpha \mathcal{H}_\beta \nonumber\\
		&~~~ -2\hat{g}^\mu
		\partial_\lambda ( \Uc^{\alpha\beta}_{\h \h \mu} \partial_\alpha \mathcal{H}_\beta )
		-
		2\hat{g}^{\mu}
		\left[ \Uc^{\alpha\beta}_{\lambda; \ \mu}
		-\partial_\lambda \Bc^{\alpha\beta}_{\h\h \mu}
		\right]
		\partial_\alpha \mathcal{H}_\beta.
	\end{align} 
	Evaluating the term in bracket and substituting the first-order expression for $D b^\beta$ from~\eqref{Db}, together with 
	$b^\alpha b^\beta \partial_\alpha \beta_\beta = -\,\bar{\theta}_\parallel$, we finally obtain
	\begin{align}
		(\partial_\lambda \hat{\mathcal{C}})_{vector}=&\,2\hat{h}^{\mu}\Big[\partial_\lambda ( \U^{\alpha\beta}_{\h \h \mu} \partial_\alpha \beta_\beta )
		+ D\beta\,\partial_\lambda u_\mu \Big] 
		+2\hat{f}^{\mu}\Big[\partial_\lambda ( \B^{\alpha\beta}_{\h \h \mu} \partial_\alpha \beta_\beta )
		+ \bar{\theta}_\parallel\,\partial_\lambda b_\mu \Big]	 \nonumber\\
		&+ 2\hat{l}^{\mu}\partial_\lambda ( \Bc^{\alpha\beta}_{\h \h \mu} \partial_\alpha \mathcal{H}_\beta )
		- \hat{l}^\mu\,(D\mathcal{H}+H\bar{\theta}_\parallel)\,\partial_\lambda u_\mu
		- 2\hat{g}^{\mu}\partial_\lambda ( \Uc^{\alpha\beta}_{\h \h \mu} \partial_\alpha \mathcal{H}_\beta )
		+ \hat{g}^\mu\,(D\mathcal{H}+H\bar{\theta}_\parallel)\,\partial_\lambda b_\mu . \label{delcvectorappen1}
	\end{align}
	
	Finally, we simplify the contributions involving tensor operators in~\eqref{delcappen1}, given by
	\begin{align}
		(\partial_\lambda \hat{\mathcal{C}})_{tensor} &=  \hat{\pi}_{\perp}^{\mu\nu}\,\partial_\lambda(\partial_\mu \beta_\nu) +\hat{m}^{\mu\nu}\partial_\lambda(\partial_\mu \mathcal{H}_\nu)\nonumber\\
		&= \hat{\pi}_{\perp}^{\mu\nu}\,\G^{\alpha\beta}_{\h\h\mu\nu}\,\partial_\lambda(\partial_\alpha \beta_\beta) 
		+\hat{m}^{\mu\nu}\,\Gc^{\alpha\beta}_{\h\h\mu\nu}\,\partial_\lambda(\partial_\alpha \mathcal{H}_\beta)\nonumber\\
		&= \hat{\pi}_{\perp}^{\mu\nu}\left[
		\partial_\lambda(\G^{\alpha\beta}_{\h\h\mu\nu}\,\partial_\alpha \beta_\beta)
		- \partial_\alpha \beta_\beta\,\partial_\lambda(\G^{\alpha\beta}_{\h\h\mu\nu})
		\right] 
		+ \hat{m}^{\mu\nu}\left[
		\partial_\lambda(\Gc^{\alpha\beta}_{\h\h\mu\nu}\,\partial_\alpha \mathcal{H}_\beta)
		- \partial_\alpha \mathcal{H}_\beta\,\partial_\lambda(\Gc^{\alpha\beta}_{\h\h\mu\nu})
		\right]\nonumber\\
		&=  \hat{\pi}_{\perp}^{\mu\nu}
		\partial_\lambda \bar{\sigma}_{\perp\mu\nu}
		-  \hat{\pi}_{\perp}^{\mu\nu}\partial_\alpha \beta_\beta\,\partial_\lambda(\G^{\alpha\beta}_{\h\h\mu\nu}) 
		+ \hat{m}^{\mu\nu} \partial_\lambda\bar{\chi}_{\mu\nu} - \hat{m}^{\mu\nu}\partial_\alpha \mathcal{H}_\beta\,\partial_\lambda(\Gc^{\alpha\beta}_{\h\h\mu\nu}) \nonumber\\
		&=  \hat{\pi}_{\perp}^{\mu\nu}
		\partial_\lambda \bar{\sigma}_{\perp\mu\nu} 
		-\hat{\pi}^{\mu\nu}_{\perp}\beta \big( b^\alpha \partial_\alpha u_\nu + b^\beta \partial_\nu u_\beta \big)\partial_\lambda b_\mu + \hat{\pi}^{\mu\nu}_{\perp} \big( \beta Du_\nu +\partial_\nu \beta \big)\partial_\lambda u_\mu \nonumber\\
		&~~~ +\hat{m}^{\mu\nu} \partial_\lambda\bar{\chi}_{\mu\nu} + \hat{m}^{\mu\nu} \big( \mathcal{H}b^\alpha\partial_\alpha b_\mu +\partial_\mu \mathcal{H} \big)\partial_\lambda b_\nu  + \hat{m}^{\mu\nu} \big( \mathcal{H}u^\beta\partial_\mu b_\beta- \mathcal{H} D b_\mu \big)\partial_\lambda u_\nu . \label{delctenappen1}
	\end{align}
	Since, we can write
	\begin{align}
		\beta \hat{\pi}^{\mu\nu}_\perp \big(  b^\alpha \partial_\alpha u_\nu + b^\beta \partial_\nu u_\beta \big)&=\hat{\pi}^{\mu\nu}_\perp(2 \B^{\alpha\beta}_{\h\h\mu}\partial_\alpha\beta_\beta) , \nonumber\\
		\hat{\pi}^{\mu\nu}_\perp \big( \beta Du_\nu +\partial_\nu \beta \big)&=\hat{\pi}^{\mu\nu}_\perp(2 \U^{\alpha\beta}_{\h\h\mu}\partial_\alpha\beta_\beta)  , \nonumber\\
		\hat{m}^{\mu\nu} \big( \mathcal{H}b^\alpha\partial_\alpha b_\mu +\partial_\mu \mathcal{H} \big) &= - \hat{m}^{\mu\nu}(2 \Bc^{\alpha\beta}_{\h\h\mu}\partial_\alpha\mathcal{H}_\beta) , \nonumber\\
		 \hat{m}^{\mu\nu} \big( \mathcal{H}u^\beta\partial_\mu b_\beta- \mathcal{H} D b_\mu \big) &= \hat{m}^{\mu\nu}(2 \Uc^{\alpha\beta}_{\h\h\mu}\partial_\alpha\mathcal{H}_\beta), 
	\end{align}
	which simplify~\eqref{delctenappen1} to
	\begin{align}
		(\partial_\lambda \hat{\mathcal{C}})_{tensor} &= \hat{\pi}_{\perp}^{\mu\nu}
		\Big[\partial_\lambda \bar{\sigma}_{\perp\mu\nu} 
		- \big(2 \B^{\alpha\beta}_{\h\h\mu}\partial_\alpha\beta_\beta  \big)\partial_\lambda b_\nu +  \big( 2 \U^{\alpha\beta}_{\h\h\mu}\partial_\alpha\beta_\beta  \big)\partial_\lambda u_\nu \Big]\nonumber\\
		&~~~ +  \hat{m}^{\mu\nu} \Big[
		\partial_\lambda \bar{\chi}_{\mu\nu} 
		+ \big(2 \Bc^{\alpha\beta}_{\h\h\mu}\partial_\alpha\mathcal{H}_\beta  \big)\partial_\lambda b_\nu -  \big( 2 \Uc^{\alpha\beta}_{\h\h\mu}\partial_\alpha\mathcal{H}_\beta  \big)\partial_\lambda u_\nu \Big] . \label{delctenappen2}
	\end{align}
Finally, we use~\eqref{UctoBrelation} and~\eqref{UtoBcrelation} to write $\Uc^{\alpha\beta}_{\h\h\mu}\partial_\alpha\mathcal{H}_\beta$ and $\U^{\alpha\beta}_{\h\h\mu}\partial_\alpha\beta_\beta$ in terms of $\B^{\alpha\beta}_{\h\h\mu}\partial_\alpha\beta_\beta$ and $\Bc^{\alpha\beta}_{\h\h\mu}\partial_\alpha\mathcal{H}_\beta$, respectively in eqs.~\eqref{delcscalarappen1},~\eqref{delcvectorappen1}, and~\eqref{delctenappen2}. We obtain 
\begin{align}
	\partial_\lambda \hat{\mathcal{C}}
	=&-\hat{P}_\parallel \partial_\lambda \bar{\theta}_\parallel  - \hat{P}_\perp \partial_\lambda \bar{\theta}_\perp +  \bar{\theta}_\parallel [\hat \varepsilon \,\partial_\lambda \gamma_\parallel + \hat B \,\partial_\lambda \phi_\parallel]  + \bar{\theta}_\perp [\hat \varepsilon \,\partial_\lambda \gamma_\perp + \hat B \,\partial_\lambda \phi_\perp]  \nonumber\\
	&+ \hat{\mathscr{J}}^\mu \partial_\lambda \bar{\mathscr{X}}_\mu +\hat{h}^\mu [-\bar{\mathscr{X}}_\mu \partial_\lambda B' +2DB \partial_\lambda u_\mu] - \hat{l}^\mu (D\mathcal{H}+H\bar{\theta}_\parallel)\partial_\lambda u_\mu \nonumber\\
	&+ \hat{\pi}_\perp^{\mu\nu} [\partial_\lambda \bar{\sigma}_{\perp\mu\nu} -B'\bar{\mathscr{X}}_\mu \partial_\lambda u_\nu -  \bar{\mathscr{Y}}_\mu \partial_\lambda b_\nu  ] \nonumber \\
	&+ \hat{\mathscr{K}}^\mu \partial_\lambda \bar{\mathscr{Y}}_\mu +2\hat{f}^\mu \bar{\theta}_\parallel \partial_\lambda b_\mu -\hat{g}^\mu [-\bar{\mathscr{Y}}_\mu \partial_\lambda H + (D\mathcal{H}+H\bar{\theta}_\parallel) \partial_\lambda b_\mu] \nonumber \\
	& + \hat{m}^{\mu\nu} [ \partial_\lambda \bar{\chi}_{\mu\nu} + \bar{\mathscr{X}}_\mu \partial_\lambda b_\nu  + H \bar{\mathscr{Y}}_\mu  \partial_\lambda u_\nu  ] . \label{delc3appen}
\end{align}
Here, we defined $B'=\frac{B}{(\varepsilon + p_\perp)}$ and used $\bar{\mathscr{X}}_\mu=2 \, \Bc^{\alpha\beta}_{\h\h \mu} \partial_\alpha \mathcal{H}_\beta$, and $\bar{\mathscr{Y}}_\mu=2 \,\B^{\alpha\beta}_{\h\h \mu} \partial_\alpha \beta_\beta$.

	%%%%%%%%%%%%%%%%%%%%%%%%%%%%%%%%%%%%%%%%%%%%%%%%%%%%%%%%%%%%%%%%%%%%%%%%%%%%%%%%%%%%%%%%%%%%%%%%%%%%%%%%%%%%%%%%%%%%%%%%%%%%%%%%%%%%

		\section{Bulk viscous pressure}\label{Bulk_viscous}
		
		The bulk viscous pressure is defined as the difference between the statistical average of the total pressure operator and the equilibrium pressure (Eq.~\eqref{bulk_to_eq_relation}). 
		For the bulk viscous pressure, parallel to the magnetic field, this yields
			\begin{align}
			\Pi_{\parallel} &= \langle \hat p_{\parallel} \rangle - p_{\parallel}(\varepsilon_{eq},B_{eq}). \label{bulk_appen1}
		\end{align}
		Here, $p_\parallel$ denotes the equilibrium pressure parallel to the magnetic field. 
		It is a state function of the two independent conserved quantities—the energy density $\varepsilon$ and the magnetic flux density $B$—whose dynamics are governed by the conservation equations \eqref{con_T} and \eqref{con_F}.
		
		Using~\eqref{gradient_expansion}, we can write
		\begin{align}
			\big< \hat p_{\parallel} \big>
			= \big< \hat p_{\parallel} \big>_{l}
			+ \big< \hat p_{\parallel} \big>_{1}
			+ \big< \hat p_{\parallel} \big>_{2}
			+ \cdots  . \label{bulk_expan_appen}
		\end{align}
		The local-equilibrium average of the total pressure operator is a function of the equilibrium energy density and equilibrium magnetic flux density. 
		Accordingly, using 
		$\big< \hat \varepsilon \big>_{l}=\big< \hat \varepsilon \big>-\Delta \varepsilon$ and $\big< \hat B \big>_{l}=\big< \hat B \big>-\Delta B$,
		we obtain
		\begin{align}
			\big< \hat p_{\parallel} \big>_{l}
			&= \left\langle \hat p_{\parallel}\big(\big< \hat \varepsilon \big>_{l},\ \big< \hat B \big>_{l}\big) \right\rangle_{l} \\
			&= p_{\parallel}(\varepsilon_{eq},B_{eq})
			- \frac{\partial p_{\parallel}}{\partial \varepsilon}\,\Delta\varepsilon
			- \frac{\partial p_{\parallel}}{\partial B}\,\Delta B
			+ \frac{1}{2}\,(\Delta\varepsilon)^{2}\,\frac{\partial^{2} p_{\parallel}}{\partial \varepsilon^{2}}
			+ \frac{1}{2}\,(\Delta B)^{2}\,\frac{\partial^{2} p_{\parallel}}{\partial B^{2}}
			+ (\Delta\varepsilon)(\Delta B)\,\frac{\partial^{2} p_{\parallel}}{\partial \varepsilon\,\partial B},
		\end{align}		
		where $\Delta \varepsilon$ and $\Delta B$ denote the higher-order corrections to the statistical averages of the energy density and the magnetic flux density, respectively, given by	
		\begin{align}
			\Delta\varepsilon &= \big< \hat \varepsilon \big>_{1} + \big< \hat \varepsilon \big>_{2}, \qquad
			\Delta B = \big< \hat B \big>_{1} + \big< \hat B \big>_{2}.
		\end{align}
		Hence, retaining terms up to second order in the gradient expansion, we can write
		\begin{align}
			\big< \hat p_{\parallel} \big>_{l}
			&= p_{\parallel}(\varepsilon_{eq},B_{eq})
			- \frac{\partial p_{\parallel}}{\partial \varepsilon}\,\big< \hat \varepsilon \big>_{1}
			- \frac{\partial p_{\parallel}}{\partial B}\,\big< \hat B \big>_{1}
			- \frac{\partial p_{\parallel}}{\partial \varepsilon}\,\big< \hat \varepsilon \big>_{2}
			- \frac{\partial p_{\parallel}}{\partial B}\,\big< \hat B \big>_{2} \nonumber\\
			&\quad
			+ \frac{1}{2}\,\big< \hat \varepsilon \big>_{1}^{2}\,\frac{\partial^{2} p_{\parallel}}{\partial \varepsilon^{2}}
			+ \frac{1}{2}\,\big< \hat B \big>_{1}^{2}\,\frac{\partial^{2} p_{\parallel}}{\partial B^{2}}
			+ \big< \hat \varepsilon \big>_{1}\,\big< \hat B \big>_{1}\,\frac{\partial^{2} p_{\parallel}}{\partial \varepsilon\,\partial B}. \label{bulk_appen_solve}
		\end{align}
		Using~\eqref{bulk_expan_appen} in~\eqref{bulk_appen1}, we obtain	
		\begin{align}
			\Pi_{\parallel}
			&= \big< \hat p_{\parallel} \big> - p_{\parallel}(\varepsilon_{eq},B_{eq}) \\
			&= \big< \hat p_{\parallel} \big>_{l}
			+ \big< \hat p_{\parallel} \big>_{1}
			+ \big< \hat p_{\parallel} \big>_{2}
			- p_{\parallel}(\varepsilon_{eq},B_{eq}) \label{bulk_appen2}
		\end{align}
		Now, substituting~\eqref{bulk_appen_solve} in~\eqref{bulk_appen2}, we obtain
		\begin{align}
			\Pi_{\parallel}
			= \big< \hat P_{\parallel} \big>_{1}
			+ \big< \hat P_{\parallel} \big>_{2}
			+ \frac{1}{2}\,\big< \hat\varepsilon \big>_{1}^{2}\,
			\frac{\partial^{2}p_{\parallel}}{\partial \varepsilon^{2}}
			+ \frac{1}{2}\,\big< \hat B \big>_{1}^{2}\,
			\frac{\partial^{2}p_{\parallel}}{\partial B^{2}}
			+ \big< \hat\varepsilon \big>_{1} \big<\hat B \big>_{1}\,
			\frac{\partial^{2}p_{\parallel}}{\partial \varepsilon\,\partial B} . \label{fullbulkparallel}
		\end{align}
		Here, we defined 
		\begin{align}
			\hat P_{\parallel} =\hat p_{\parallel} - \hat\varepsilon\,\frac{\partial p_{\parallel}}{\partial \varepsilon}
			- \hat B\,\frac{\partial p_{\parallel}}{\partial B} ,
		\end{align}		
		Similarly, by analogy with Eq.~\eqref{fullbulkparallel}, the second-order expression for $\Pi_\perp$ can be written as
		\begin{align}
			\Pi_{\perp}
			= \big< \hat P_{\perp} \big>_{1}
			+ \big< \hat P_{\perp} \big>_{2}
			+ \frac{1}{2}\,\big< \hat\varepsilon \big>_{1}^{2}\,
			\frac{\partial^{2}p_{\perp}}{\partial \varepsilon^{2}}
			+ \frac{1}{2}\,\big< \hat B \big>_{1}^{2}\,
			\frac{\partial^{2}p_{\perp}}{\partial B^{2}}
			+ \big< \hat\varepsilon \big>_{1} \big<\hat B \big>_{1}\,
			\frac{\partial^{2}p_{\perp}}{\partial \varepsilon\,\partial B} .  \label{fullbulkperp}
		\end{align}
		Here, 
		\begin{align}
			\hat P_{\perp} &=\hat p_{\perp} - \hat\varepsilon\,\frac{\partial p_{\perp}}{\partial \varepsilon}
			- \hat B\,\frac{\partial p_{\perp}}{\partial B}.
		\end{align}

%%%%%%%%%%%%%%%%%%%%%%%%%%%%%%%%%%%%%%%%%%%%%%%%%%%%%%%%%%%%%%%%%%%%%%%%%%%%
%%%%%%%%%%%%%%%%%%%%%%%%%%%%%%%%%%%%%%%%%%%%%%%%%%%%%%%%%%%%%%%%%%%%%%%%%%%

\section{Correlation Function}\label{correlators}
The derivation presented in this section follows closely the analysis given in Ref.~\cite{Hosoya:1983id,Huang:2011dc,HARUTYUNYAN2022168755}. 
Starting from the~\eqref{twopointdef}, we can write
\begin{equation}
	\Big( \hat{X}(\vec{x},t), \hat{Y}(\vec{x}_1,t_1) \Big) = \int_0^1 d\tau \Big< \hat{X}(\vec{x},t) \left[ e^{-\hat{A}\tau} \hat{Y}(\vec{x}_1,t_1)e^{\hat{A}\tau}-\left< e^{-\hat{A}\tau}\hat{Y}(\vec{x}_1,t_1)e^{\hat{A}\tau} \right>_l \right] \Big>_l .\label{corrfun1}
\end{equation}

Referring to Eq.~\eqref{A}, we infer that $\hat{A}=\beta \hat{\mathscr{H}}$,  where $\hat{\mathscr{H}}$ denotes the effective Hamiltonian of the system in the presence of an external magnetic field. Consequently, the imaginary-time evolution of an operator in the Heisenberg picture is given by
\begin{equation}
	\hat{Y}(\vec{x},t+i\tau')=e^{-\hat{\mathscr{H}}\tau'}\hat{Y}(\vec{x},t)e^{\hat{\mathscr{H}}\tau'}.
\end{equation}
This relation immediately implies the Kubo-Martin-Schwinger (KMS) conditions:
\begin{align}
	\left< \hat{Y}(\vec{x},t+i\tau') \right>_l &= \left< \hat{Y}(\vec{x},t) \right>_l, \label{KMS1}\\
	\left< \hat{X}(\vec{x},t)\hat{Y}(\vec{x}_1,t_1+i\beta) \right>_l &= \left< \hat{Y}(\vec{x}_1,t_1) \hat{X}(\vec{x},t)\right>_l.\label{KMS2}
\end{align}
Using~\eqref{KMS1} and~\eqref{KMS2}, in~\eqref{corrfun1} and 
assuming that the correlations vanish at $t\to\infty$, we can 
establish the relation
\begin{equation}
	\Big( \hat{X}(\vec{x},t), \hat{Y}(\vec{x}_1,t_1) \Big) =-\frac{1}{\beta} \int_{-\infty}^{t_1} dt' G^R_{\hat{X}\hat{Y}}(\vec{x}-\vec{x}_1,t-t'),\label{kubo_to_green}
\end{equation}
where
\begin{eqnarray}
	G^R_{\hat{X}\hat{Y}}(\vec{x}-\vec{x}_1,t-t')=-i\theta(t-t')\left< \left[ \hat{X}(\vec{x},t),  \hat{Y}(\vec{x}_1,t') \right] \right>_l.\label{ret_green}
\end{eqnarray}
We now consider the general structure of a frequency-dependent transport coefficient expressed in terms of a two-point correlation function. It is defined as
\begin{eqnarray}
	\mathcal{I}_{\hat{X}\hat{Y}}(\omega) = \beta \int d^3x_1 \int_{-\infty}^t dt_1 e^{\epsilon(t_1-t)} e^{i\omega(t-t_1)}\Big( \hat{X}(\vec{x},t),\hat{Y}(\vec{x}_1,t_1)  \Big).\label{freq_tran}
\end{eqnarray}
The retarded Green's function obtained in~\eqref{ret_green} is 
transitionally invariant both in space as well as in time. We exploit this property to rewrite the integral~\eqref{freq_tran} 
by setting $(\vec{x},t)=(0,0)$. 
Substituting~\eqref{kubo_to_green} in~\eqref{freq_tran}, we obtain
\begin{equation}
	\mathcal{I}_{\hat{X}\hat{Y}}(\omega)=-\int^0_{-\infty} dt_1 e^{(\epsilon-i\omega)t_1} \int_{-\infty}^{t_1}dt'\int d^3 x_1  G_{\hat{X}\hat{Y}}^R (-\vec{x}_1,-t').\label{green_step1}
\end{equation} 
Considering the Fourier transformation 
\begin{equation}
	G^R_{\hat{X}\hat{Y}}(\vec{x}_1,t')=\int \frac{d^3k}{(2\pi)^3} \int_{-\infty}^{\infty}\frac{d\omega'}{2\pi} e^{-i(\omega't'-\vec{k}\cdot\vec{x})} G_{\hat{X}\hat{Y}}^R (\vec{k},\omega'),
\end{equation} 
and using the definition of Dirac delta function, we can write
\begin{equation}
	\int d^3x G_{\hat{X}\hat{Y}}^R (-\vec{x},-t') = \lim_{\vec{k}\to 0} \int_{-\infty}^{\infty}\frac{d\omega'}{2\pi}e^{i\omega't'}G_{\hat{X}\hat{Y}}^R (\vec{k},\omega').\label{integ_space}
\end{equation}
Substituting~\eqref{integ_space} in~\eqref{green_step1} and 
solving the integral over $t'$ by shifting $\omega'\to 
\omega'-i\delta$, where $\delta\to0^+$, we obtain
\begin{equation}
	\mathcal{I}_{\hat{X}\hat{Y}}(\omega)=\lim_{\delta \to 0^+} \frac{i}{\omega+i\epsilon}\oint \frac{d\omega'}{2\pi i}\left[ \frac{1}{\omega'-(\omega+i\epsilon+i\delta)}-\frac{1}{\omega'-i\delta} \right]G^R_{\hat{X}\hat{Y}}(\omega').\label{final_integ_I}
\end{equation}
Here, $G^R_{\hat{X}\hat{Y}}(\omega')\equiv 
\lim_{\vec{k}\to 0}G^R_{\hat{X}\hat{Y}}(\vec{k},\omega')$.

To evaluate~\eqref{final_integ_I}, we first extend $\omega'$ to the complex plane and apply Cauchy’s integral formula together with the residue theorem. From~\eqref{final_integ_I}, we identify two poles at $\omega'=\omega+i\epsilon+i\delta$ and $\omega'=i\delta$. We close the contour in the upper half-plane and assume that $G^R_{\hat{X}\hat{Y}}(\omega')$ decays sufficiently rapidly at infinity compared with ($\frac{1}{\omega'}$). Consequently, we obtain
\begin{equation}
	\mathcal{I}_{\hat{X}\hat{Y}}(\omega)=\lim_{\delta \to 0^+} \frac{i}{\omega+i\epsilon} \left[ G^R_{\hat{X}\hat{Y}}(\omega+i\epsilon+i\delta)-G^R_{\hat{X}\hat{Y}}(i\delta) \right].
\end{equation}
We now take the limits $\delta\to 0^+$ and $\epsilon\to 0^+$, 
where $\epsilon$ denotes the irreversibility parameter that must be 
set to zero at the end of the calculations. In this limit, we obtain
\begin{equation}
	\mathcal{I}_{\hat{X}\hat{Y}}(\omega)= \frac{i}{\omega} \left[ G^R_{\hat{X}\hat{Y}}(\omega)-G^R_{\hat{X}\hat{Y}}(0) \right].\label{kubo_green3}
\end{equation}
For $\omega\to0$,~\eqref{kubo_green3} can be simplified to
\begin{equation}
	\mathcal{I}_{\hat{X}\hat{Y}}(0)= i\lim_{\omega\to0}\frac{d}{d\omega} G^R_{\hat{X}\hat{Y}}(\omega)= i\lim_{\omega\to0}\lim_{\vec{k}\to0}\frac{\partial}{\partial\omega} G^R_{\hat{X}\hat{Y}}(\vec{k},\omega).\label{final_green_formula1}
\end{equation}
Moreover, since the operators $\hat{X}$ and $\hat{Y}$ are Hermitian, 
we can write
\begin{equation}
	\left< \left[ \hat{X}(\vec{x},t), \hat{Y}(\vec{x}_1,t_1) \right] \right>_l^* = -\left< \left[ \hat{X}(\vec{x},t), \hat{Y}(\vec{x}_1,t_1) \right] \right>_l,
\end{equation}
which indicates that $\left< \left[ \hat{X}(\vec{x},t), 
\hat{Y}(\vec{x}_1,t_1) \right] \right>_l$ is a purely imaginary 
quantity.

Using Eq.~\eqref{ret_green}, it follows that the retarded Green’s function 
$G^R_{\hat{X}\hat{Y}}$ and the canonical correlation
$\Big( \hat{X}(\vec{x},t), \hat{Y}(\vec{x}_1,t_1) \Big)$ are both purely real. Consequently,~\eqref{freq_tran} implies that, $\mathcal{I}_{\hat{X}\hat{Y}}(0)$ 
must also be real. Thus, we rewrite~\eqref{final_green_formula1} in the form
\begin{equation}
	\mathcal{I}_{\hat{X}\hat{Y}}(0) = \beta \int d^4 x_1 \Big( \hat{X}(x),\hat{Y}(x_1)\Big)=-\lim_{\omega\to 0}\frac{d}{d\omega}{\rm Im}G^R_{\hat{X}\hat{Y}}(\omega) =-\lim_{\omega\to 0}\lim_{\vec{k}\to 0}\frac{\partial}{\partial\omega}{\rm Im}G^R_{\hat{X}\hat{Y}}(\vec{k},\omega).
\end{equation}
The second type of correlation function that emerges in our analysis is given by
\begin{equation}
	\mathcal{J}^\mu_{\hat{X}\hat{Y}}(\omega) = \beta \int d^4x_1  e^{i\omega(t-t_1)}\Big( \hat{X}(x),\hat{Y}(x_1)  \Big)(x_1-x)^\mu.\label{freq_tran_type2}
\end{equation}
Here, the temporal component is given by
\begin{equation}
	\mathcal{J}^0_{\hat{X}\hat{Y}}(\omega) = i\beta \frac{d}{d\omega} \int d^4x_1 e^{i\omega(t-t_1)}\Big( \hat{X}(x),\hat{Y}(x_1) \Big)=i\frac{d}{d\omega}\mathcal{I}_{\hat{X}\hat{Y}}(\omega).\label{temporal_J} 
\end{equation}
The spatial component of 
$\mathcal{J}^\mu_{\hat{X}\hat{Y}}(\omega)$ vanishes in the local 
rest frame, since the correlator $\Big( \hat{X}(x),\hat{Y}(x_1) 
\Big)$ evaluated in local rest frame depends on 
$|\vec{x}-\vec{x}_1|$, which make 
$\mathcal{J}^i_{\hat{X}\hat{Y}}(\omega)$ an odd function of 
$\vec{x}-\vec{x}_1$.

Furthermore, considering a Taylor expansion 
of $G^R_{\hat{X}\hat{Y}}(\omega)$ around $\omega=0$, we have
\begin{equation}
	G^R_{\hat{X}\hat{Y}}(\omega) =G^R_{\hat{X}\hat{Y}}(0)+\omega \frac{d}{d\omega}G^R_{\hat{X}\hat{Y}}(\omega)\bigg|_{\omega=0}+\frac{\omega^2}{2!}\frac{d^2}{d\omega^2}G^R_{\hat{X}\hat{Y}}(\omega)\bigg|_{\omega=0}+\cdots, \label{Gexpansion}
\end{equation}
Substituting~\eqref{Gexpansion} in~\eqref{kubo_green3}, we obtain
\begin{equation}
	\mathcal{I}_{\hat{X}\hat{Y}}(\omega)=i\frac{d}{d\omega}G^R_{\hat{X}\hat{Y}}(\omega)\bigg|_{\omega=0}+\frac{i\omega}{2!}\frac{d^2}{d\omega^2}G^R_{\hat{X}\hat{Y}}(\omega)\bigg|_{\omega=0}+\cdots . \label{green_expansion}
\end{equation}
Next, upon inserting Eq.~\eqref{green_expansion} into~\eqref{temporal_J} and subsequently taking the limit $\omega\to0$, we find
\begin{align}
	\mathcal{J}^0_{\hat{X}\hat{Y}}(0)&=-\frac{1}{2}\lim_{\omega\to0} \frac{d^2}{d\omega^2}G^R_{\hat{X}\hat{Y}}(\omega)=-\frac{1}{2}\lim_{\omega\to0} \frac{d^2}{d\omega^2}{\rm Re}G^R_{\hat{X}\hat{Y}}(\omega)=-\frac{1}{2}\lim_{\omega\to0}\lim_{\vec{k}\to0} \frac{\partial^2}{\partial\omega^2}{\rm Re}G^R_{\hat{X}\hat{Y}}(\vec{k},\omega).
\end{align}
Finally, we can write
\begin{equation}
	\beta\int d^4x_1 \Big( \hat{X}(x),\hat{Y}(x_1)  \Big)(x_1-x)^\mu =\mathcal{J}^\mu_{\hat{X}\hat{Y}}(0)= u^\mu \mathcal{J}^0_{\hat{X}\hat{Y}}(0) = u^\mu \lim_{\omega \to 0} i\frac{d}{d\omega}\mathcal{I}_{\hat{X}\hat{Y}}(\omega) .\label{freq_dep_trans}
\end{equation}

	\end{appendix}

\end{document}